\documentclass[aps,prb,reprint,twocolumn,floatfix,superscriptaddress,amsmath,amssymb,amsfonts, nofootinbib]{revtex4-2}

\usepackage{url}
\usepackage{bm}
\usepackage{graphicx}
\usepackage{amsmath}
\usepackage{amstext}
\usepackage{amssymb}
\usepackage{amsfonts}
\usepackage{amsbsy}
\usepackage{verbatim}
\usepackage{color}
\usepackage[colorlinks=true, urlcolor=blue, linkcolor=blue, citecolor=blue, pdftex]{hyperref}
\usepackage{floatrow}
\usepackage{float}
\usepackage{tikz}
\usepackage{gensymb}
\usepackage{textcomp}
\usepackage{enumitem}
\usepackage[version=3]{mhchem}
\usepackage{afterpage}
\usepackage{pbox}
\usepackage{dsfont}
\usepackage{siunitx}
\usepackage{stackengine,wasysym,scalerel}
\usepackage{latexsym}
\usepackage{cancel}
\usepackage{array, multirow, bigdelim, makecell, booktabs}
\usepackage[misc]{ifsym}
\usepackage[capitalise]{cleveref}
\usepackage{times}

\usepackage[normalem]{ulem}

\definecolor{darkblue}{HTML}{004D6B}
\definecolor{darkred}{HTML}{8c1515}
\definecolor{darkgreen}{HTML}{006400}
\usepackage{hyperref}
\hypersetup{
	colorlinks=true,
	urlcolor=darkblue,
	citecolor=darkblue,
	linkcolor=darkblue,
	breaklinks
}

\graphicspath{{figures/}}

\begin{document}
\title{Candidate quantum spin liquids on the maple-leaf lattice}
\author{Jonas Sonnenschein}
\thanks{These authors contributed equally.}
\affiliation{Theory of Quantum Matter Unit, Okinawa Institute of Science and Technology, 1919-1 Tancha, Onna-son, Okinawa 904-0495, Japan}
\author{Atanu Maity}
\thanks{These authors contributed equally.}
\affiliation{Institut f\"ur Theoretische Physik und Astrophysik and W\"urzburg-Dresden Cluster of Excellence ct.qmat, Julius-Maximilians-Universit\"at W\"urzburg, Am Hubland Campus S\"ud, W\"urzburg 97074, Germany}
\affiliation{Department of Physics and Quantum Centre of Excellence for Diamond and Emergent Materials (QuCenDiEM), Indian Institute of Technology Madras, Chennai 600036, India}
\author{Chunxiao Liu}
\thanks{These authors contributed equally.}
\affiliation{Department of Physics, University of California, Berkeley, California 94720, USA}
\author{Ronny Thomale}
\affiliation{Institut f\"ur Theoretische Physik und Astrophysik and W\"urzburg-Dresden Cluster of Excellence ct.qmat, Julius-Maximilians-Universit\"at W\"urzburg, Am Hubland Campus S\"ud, W\"urzburg 97074, Germany}
\affiliation{Department of Physics and Quantum Centre of Excellence for Diamond and Emergent Materials (QuCenDiEM), Indian Institute of Technology Madras, Chennai 600036, India}
\author{Francesco Ferrari}
\affiliation{Institut für Theoretische Physik, Goethe Universität Frankfurt, Max-von-Laue-Straße 1, 60438 Frankfurt am Main, Germany}
\affiliation{Department of Physics and Quantum Centre of Excellence for Diamond and Emergent Materials (QuCenDiEM), Indian Institute of Technology Madras, Chennai 600036, India}
\author{Yasir Iqbal}
\email{yiqbal@physics.iitm.ac.in}
\affiliation{Department of Physics and Quantum Centre of Excellence for Diamond and Emergent Materials (QuCenDiEM), Indian Institute of Technology Madras, Chennai 600036, India}

\begin{abstract} 
Motivated by recent numerical studies reporting putative quantum paramagnetic behavior in spin-$1/2$ Heisenberg models on the maple-leaf lattice, we classify Abrikosov fermion mean-field \textit{Ans\"atze} of fully symmetric $U(1)$ and $\mathds{Z2}_{2}$ quantum spin liquids within the framework of projective symmetry groups. We obtain a total of $17$ $U(1)$ and $12$ $\mathds{Z}_{2}$ algebraic PSGs, and, upon restricting their realization via mean-field \textit{Ans\"atze} with nearest-neighbor amplitudes (relevant to the studied models), only 12 $U(1)$ and 8 $\mathds{Z}_{2}$ distinct phases are obtained. We present both singlet and triplet fields for all \textit{Ans\"atze} up to third nearest-neighbor bonds and discuss their spinon dispersions as well as their dynamical spin structure factors. We further assess the effects of Gutzwiller projection on the equal-time spin structure factors, and identify a $U(1)$ Fermi surface spin liquid whose structure factor most closely reproduces the one obtained from pseudo-fermion functional renormalization group calculations.
\end{abstract}

\date{\today}

\maketitle

\section{Introduction}
\begin{figure*}	\includegraphics[width=1.0\linewidth]{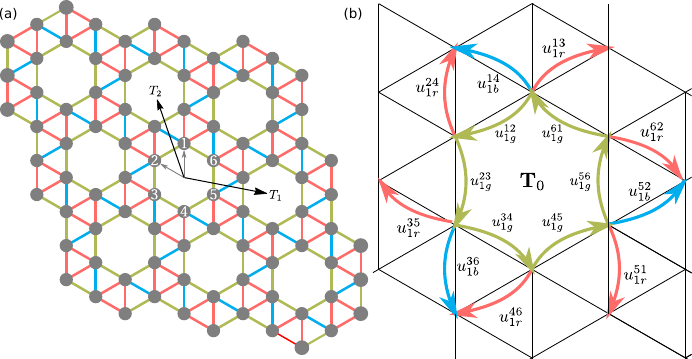}
	\caption{(a) An illustration of the maple-leaf lattice with corresponding lattices vectors $\mathbf{T}_1$ and $\mathbf{T}_2$. The small vectors inside the hexagon correspond to $\mathbf{a}_1$ and $\mathbf{a}_2$ which describe the positions of the site $u$ as explained in the main text. The three different colors stand for three symmetry-inequivalent first nearest-neighbor bonds, whose labelling (within one unit cell) is specified in (b).}
	\label{fig:cell_definitions}
\end{figure*}

Two-dimensional geometrically frustrated lattices have long been a fertile playground for realizing quantum spin liquids (QSLs). Indeed, Heisenberg models on the celebrated kagome and triangular lattices are known to realize a variety of paramagnetic phases, including the exotic $U(1)$ Dirac and chiral QSLs~\cite{Ran-2007,Iqbal-2013,He-2017,Gong-2015,Hu-2015,Iqbal-2016,Hu-2019}. A relatively new entrant in search of QSL behavior is the maple-leaf lattice~\cite{Betts-1995,Misguich-1999,Schulenburg-2000,Schmalfuss-2002,Farnell-2011,Farnell-2014,Farnell-2018,ghosh2022,Gresista-2023,Beck-2024,Ghosh-2024,Ghosh_mag} [see Fig.~\ref{fig:cell_definitions}(a)], where the existence of a putative QSL phase sandwiched between magnetically ordered and a dimerized ground state in the spin $S=1/2$ Heisenberg antiferromagnet has been reported by pseudo-fermion functional renormalization group (pf-FRG) calculations~\cite{Gresista-2023}. A subsequent study of this model employing neural quantum states and density matrix renormalization group (DMRG) approaches also hinted at the possible existence of an intermediate QSL phase~\cite{Beck-2024}. Another recent DMRG exploration of parameter space with ferromagnetic couplings on triangle and dimer bonds [red and blue bonds in Fig.~\ref{fig:cell_definitions}(a), respectively] provided an inkling for an island of spin liquidity surrounded by magnetic and dimer orders~\cite{Ghosh-2024}. The precarious locations of the reported QSL phases on the maple-leaf lattice have naturally led to speculations concerning their possible origin from a nearby deconfined quantum critical point. In similar spirit, it has recently been shown that the inclusion of longer range Heisenberg couplings induces quantum paramagnetic phases~\cite{Gembe-2024}. While the aforementioned approaches provide evidence of quantum paramagnetic behavior, which could putatively be a QSL, they fall short of characterizing its precise microscopic nature, i.e., gapless vs gapped, $U(1)$ vs $\mathbb{Z}_{2}$ gauge structure, etc. A powerful framework to systematically classify quantum spin liquids with different gauge groups is provided for within a parton representation by the method of projective symmetry groups~\cite{Wen-2002,Wenbook}. This framework has been extensively applied on two- and three-dimensional lattices~\cite{Wang-2006,Lawler-2008,Choy-2009,Yang-2010,Lu-2011a,Lu-2011b,Yang-2012,Messio-2013,Bieri-2015,Yang-2016,Lu-2016a,Bieri-2016,Huang-2017,Huang-2018,Lu-2018,Liu-2019,Jin-2020,Sonnenschein-2020,Sahoo-2020,Liu-2021,Chern-2021,Chern-2022,Benedikt-2022,Maity-2023,Chauhan-2023,liu2024schwinger}, and met with wide success in describing the ground state and low-energy behavior of quantum spin models~\cite{Iqbal-2011a,Iqbal-2011b,Iqbal-2012,Iqbal-2013,Iqbal_2014,Iqbal-2016,Iqbal-2018_bk,Iqbal-2021,Ferrari-2019,Ferrari-2023,Hu-2013,Kiese-2023}.

To this end, we employ the projective symmetry group (PSG) framework for fermionic partons to provide a systematic classification of fully symmetric QSL mean-field \textit{Ans\"atze} with different low-energy gauge groups~\cite{Wen-2002}. We find a total of $17$ $U(1)$ and $12$ $\mathbb{Z}_{2}$ distinct algebraic PSGs on the maple-leaf lattice. Upon restricting the (singlet) mean-field \textit{Ans\"atze} to first neighbor amplitudes \emph{only}, a total of 12 $U(1)$ and 8 $\mathds{Z}_{2}$ states can be realized, while, if amplitudes up to third neighbor are included, \emph{all} $U(1)$ and $\mathds{Z}_{2}$ distinct states are realizable. While our treatment principally focuses on singlet QSLs, in general, we also provide the symmetry allowed triplet amplitudes thus enabling for a consideration of competing ferromagnetic and spin-orbit couplings in the original spin system. The Hamiltonians featuring such couplings are likely to be present and potentially relevant in describing natural minerals~\cite{Fennell-2011,Kampf-2013,mills-2014,Iqbal_Spangolite} and synthetic crystals~\cite{Cave-2006,Aliev-2012,Haraguchi-2018,Haraguchi-2021,Makuta-2021} with maple-leaf crystal geometries or distortions thereof. In two dimensions, it is well known that including these triplet fields can lead to a plethora of topologically nontrivial spinon models and possibly spin nematic states~\cite{Reuther-2014,Sonnenschein-2017,Shindou-2009,Dodds-2013,Iqbal-2016_nem}.

The article is organized as follows. In Sec.~\ref{sec2} the projective group approach is explicated and complemented by the symmetry operators on the maple-leaf lattice in Sec.~\ref{sec:lattice_time}. A synopsis of the projective symmetry group results is provided in Sec.~\ref{sec4}, succeeded by a mean field analysis of the PSG candidate states in Sec.~\ref{sec5}. The latter \textit{Ans\"atze} then enable us to compute spinon bands as well as the static and dynamic spin structure factors (Sec.~\ref{sec6}). In Sec.~\ref{sec7}, we conclude that Heisenberg models on the maple-leaf lattice open up an arena of elusive quantum paramagnetic phenomena awaiting further investigation and substantiation.  

\section{Projective Symmetry Group Approach}\label{sec2}
In this section, we review the essential steps of the projective symmetry group (PSG) classification of effective low-energy theories of QSLs~\cite{Wen-2002,Wenbook}. We start from a Heisenberg model defined on a lattice
\begin{align}
    {\hat H}=\sum_{ \mathbf{rr'}}J_{\mathbf{rr'}} \mathbf{\hat S}_{\mathbf{r}}\cdot\mathbf{\hat S}_{\mathbf{r'}},
\end{align}
where $\mathbf{\hat S}_{\mathbf{r}}$ denotes the $SU(2)$ spin-operator acting on the spin-$1/2$ representation on site $\mathbf{r}$. Due to the absence of magnetic order in QSLs, a mean-field treatment must be carried out within a parton representation of spin operators, and here we adopt the representation in terms of two flavors of complex fermions due to Abrikosov~\cite{Abrikosov-1965}
\begin{align}\label{eq:spin_op_fermion}
{\hat S}^\mu_{\mathbf{r}} = \frac{1}{2} \sum_{\alpha\beta} {\hat f}^\dagger_{\mathbf{r}\alpha} \sigma^\mu_{\alpha\beta} {\hat f}_{\mathbf{r\beta}},
\end{align}
where the \emph{spinon} operator ${\hat f}_{\mathbf{r}\alpha}$ annihilates a fermion with spin-$\alpha \in \left\lbrace \uparrow, \downarrow \right\rbrace$ at site $\mathbf{r}$ and $\sigma^\mu$ ($\mu \in \left\lbrace x,y,z \right\rbrace$) are the three Pauli matrices. This mapping artificially enlarges the local Hilbert space from the spin space $\mathbb{C}^{2}$ to the four-dimensional fermionic Fock space. Hence, this mapping reproduces the physical Hilbert space of the spin model only in the subspace of single occupation $n_\mathbf{r}=1$ for all sites, which correspond to $\mathbf{S}^{2}=3/4$, while the unphysical states with $n_\mathbf{r}=0$ or $2$, yield $\mathbf{S}^{2}=0$ which do not correspond to spin states. Therefore, a correct description of the physical model needs to incorporate a constraint which forbids empty and doubly occupied sites. The constrained model leads to a description of the QSL state in terms of a gauge theory~\cite{Baskaran-1988,Affleck-1988}. The Heisenberg Hamiltonian becomes quartic in terms of these fermions and a Hubbard-Stratonovich transformation is employed to further decouple the interacting model. We choose this decoupling to neglect any magnetic terms, which then introduces the auxiliary fields 
\begin{align}
& \chi_{\mathbf{rr'}} \delta_{\alpha\beta} = 2\left\langle {\hat f}^\dagger_{\mathbf{r}\alpha}{\hat f}_{\mathbf{r}\beta} \right\rangle, \notag\\
& \Delta_{\mathbf{rr'}} \epsilon_{\alpha\beta} = -2\left\langle {\hat f}_{\mathbf{r}\alpha}{\hat f}_{\mathbf{r}\beta} \right\rangle.
\end{align}
To make further progress we will assume static fields in a mean-field treatment \cite{Baskaran-1987}. Using Nambu-spinors ${\hat \psi}^\dagger_{\mathbf{r}} = ({\hat f}^\dagger_{\mathbf{r}\uparrow}, {\hat f}_{\mathbf{r}\downarrow})$ one can write the Hamiltonian as
\begin{align}\label{eq:MF-Hamiltonian}
{\hat H}=&\sum_{\mathbf{rr'}}-\frac{3}{8}J_{\mathbf{rr'}}\left[({\hat \psi}^\dagger_{\mathbf{r}}u_{\mathbf{rr'}}{\hat \psi}_{\mathbf{r'}} + {\rm h.c.}) - \frac{1}{2}\text{Tr}[u^\dagger_{\mathbf{rr'}}u_{\mathbf{rr'}}] \right] \notag \\ 
&
+ \sum_{\mathbf{r},\mu} {\hat \psi}^\dagger_{\mathbf{r}} a_{\mu}(\mathbf{r})\sigma^\mu {\hat \psi}_{\mathbf{r}}
\end{align}
with a local multiplier field $a_{\mu}(\mathbf{r})$ which ensures single occupancy on the mean-field level, i.e.,
\begin{align}
\left\langle \sum_{\alpha} {\hat f}^\dagger_{\mathbf{r}\alpha}{\hat f}_{\mathbf{r}\alpha}\right\rangle = 1, \quad \left\langle {\hat f}_{\mathbf{r}\alpha}{\hat f}_{\mathbf{r}\beta}\right\rangle =0 \quad \forall\,\mathbf{r}.
\end{align}
The coupling matrices $u_{\mathbf{rr'}}$ contain the mean-field amplitudes
\begin{align}
u_{\mathbf{rr'}} =
\begin{pmatrix}
\chi^\dagger_{\mathbf{rr'}} & \Delta_{\mathbf{rr'}} \\
\Delta^\dagger_{\mathbf{rr'}} & -\chi_{\mathbf{rr'}}
\end{pmatrix} = \dot\iota \alpha^0_{\mathbf{rr'}}\tau^0 + \sum_{\mu}\alpha^\mu_{\mathbf{rr'}}\tau^\mu,
\end{align}
which can be parameterized by four real coefficients $\alpha^0_{\mathbf{rr'}}, \alpha^\mu_{\mathbf{rr'}} \in \mathbb{R}$. Here, $\tau^0$ is the $2\times 2$ identity matrix and $\tau^\mu$ are the Pauli matrices acting in Nambu space. In the pure mean-field picture (in this context often referred to as zeroth order mean-field \cite{Wenbook}) the Hamiltonian Eq.~\eqref{eq:MF-Hamiltonian} can be readily solved. However, having relaxed the constraint may include some contributions from unphysical states. We will mitigate this problem by constructing particular mean-field \textit{Ans\"atze} which are stable saddle-point solutions beyond the zeroth order mean-field theory. To see which mean-field models are of interest, we consider the type of fluctuations which are expected to induce gapless excitations beyond the pure mean-field picture. Therefore, we re-examine the fermionic representation of the spin operator in Eq.~\eqref{eq:spin_op_fermion} which is invariant under a local $U(1)$ transformation ${\hat f}_{\mathbf{r\alpha}}\rightarrow e^{\dot\iota\theta(\mathbf{r})}{\hat f}_{\mathbf{r}\alpha}$. Within the physical subspace of single occupancy it is further invariant under a particle-hole like transformation ${\hat f}_{\mathbf{r}\alpha}\rightarrow \cos \phi(\mathbf{r}) {\hat f}_{\mathbf{r}\alpha} + sign(\alpha) \sin \phi(\mathbf{r}){\hat f}^\dagger_{\mathbf{r},-\alpha}$, where $-\alpha$ means flipping the spin label and $sign(\uparrow) = +1$,  $sign(\downarrow) = -1$. These two transformations do not commute, as is manifest by considering their action on the spinor ${\hat \psi}_{\mathbf{r}}$. The angles of a successive application of $U(1)$, particle-hole and again $U(1)$ transformations can be regarded as Euler angles parameterizing the group of rotations in 3 dimensional space, $SO(3)$, which is locally isomorphic to $SU(2)$ \cite{Bieri-2016}. It follows that in the fermionic representation the Heisenberg Hamiltonian should be invariant under a local $SU(2)$ transformation~\cite{Affleck-1988}. Note that this local freedom is different from the global spin rotation invariance of the Heisenberg model. The gauge freedom is implemented in our description by an action on the spinors according to ${\hat \psi}_{\mathbf{r}}\rightarrow W_{\mathbf{r}}{\hat \psi}_{\mathbf{r}}$ with $W_{\mathbf{r}} \in SU(2)$. Equivalently, one can act on the coupling matrices of a mean-field \textit{Ansatz} $u_{\mathbf{rr'}}\rightarrow W^\dagger_{\mathbf{r}}u_{\mathbf{rr'}}W_{\mathbf{r'}}$. It is obvious that a generic coupling matrix $u_{\mathbf{rr'}}$ will break this local invariance. However, for a particular choice of the mean-field decoupling there might exist a subgroup $\mathcal{G} \subseteq SU(2)$, called the invariant gauge group (IGG), for which
\begin{align}\label{eq:IGG}
u_{\mathbf{rr'}} = W^\dagger_{\mathbf{r}}u_{\mathbf{rr'}}W_{\mathbf{r'}}, \quad W_{\mathbf{r}} \in \mathcal{G}
\end{align}
is true. Note that such a subgroup always exists since for $\mathds{Z}_2 \subseteq \mathcal{G}$ Eq.~\eqref{eq:IGG} is trivially fulfilled. Different mean-field \textit{Ans\"atze} which are related not only by Eq.~\eqref{eq:IGG} but by a generic $SU(2)$ gauge transformation lead to an equivalent description and, therefore, the elements of $\mathcal{G}$ merely put different labels on the same physical state~\cite{Wenbook}. One can further show that fluctuations over a given mean-field \textit{Ansatz} are generated by elements of its $\mathcal{G}$ \cite{Wen-2002}. In this work, we will consider the scenarios of $\mathcal{G} \simeq U(1)$ and $\mathds{Z}_2$~\footnote{States with $SU(2)$ IGG cannot be realized on the maple-leaf lattice given its nonbipartite nature.}. In the later case, the excitations of the fluctuation-fields are gapped such that at sufficiently low energies the mean-field \textit{Ansatz} leads to a stable saddle-point. In the $U(1)$ case, however, stability arguments are more subtle and need to be considered for each model separately~\cite{Song-2019,Song-2020,Budaraju-2023,Iqbal_2014,Iqbal-2016}.

The emergence of the local invariance has further implications regarding the symmetry properties of the model. Assume that we want to investigate if the system at hand has any underlying symmetry. Such a symmetry would act on a mean-field \textit{Ansatz} as $u_{\mathbf{rr'}}\rightarrow O_{\mathcal{O}(\mathbf{r})}u_{\mathcal{O}(\mathbf{r})\mathcal{O}(\mathbf{r'})}O^\dagger_{\mathcal{O}(\mathbf{r'})}$, where $O$ is a projective representation of the symmetry operation $\mathcal{O}$ acting in Nambu space. Due to the local gauge freedom we say that the system is invariant under a given symmetry operation if we can find a suitable gauge transformation $G_\mathcal{O}$ such that
\begin{align}\label{eq:symmetry_condition}
u_{\mathbf{rr'}} = G_{\mathcal{O}}(\mathcal{O}(\mathbf{r}))u_{\mathcal{O}(\mathbf{r})\mathcal{O}(\mathbf{r'})}G^\dagger_{\mathcal{O}}(\mathcal{O}(\mathbf{r'})), \notag \\
\quad G_{\mathcal{O}}(\mathbf{r}) \in SU(2).
\end{align}
This equation defines the PSG and in mathematical terms it is the extension of the symmetry group (SG) by the IGG
\begin{align}
PSG = IGG \rtimes SG.
\end{align}
PSGs provide a systematic way to classify and construct many possible quantum states. This classification goes beyond the Landau paradigm in the conventional sense \cite{Wen-2002, Wenbook}. We will make use of this method in the following section and construct possible quantum spin liquid states for the maple-leaf lattice.

\begin{table*}
	\caption{The 17 $U(1)$ PSG classes are listed here. $n$ and $n_{\mathcal{I}}$ are integers that can take values $0$ or $1$. }
	\begin{ruledtabular}
		\begin{tabular}{cccccc}
$w_{\mathcal{T}}$&$w_{\mathcal{I}}$&$\theta$&$\Tilde{\theta}_{\mathcal{I}}$&$\rho_{\mathcal{I}}(u)$ & \# of PSG class\\
			\hline	
$0$&$0$&$3\Tilde{\theta}_{\mathcal{I}}$&$\Tilde{\theta}_{\mathcal{I}}$&$0$ & 1\\
$0$&$1$&$n\pi$&$\theta$&$\frac{p_{\mathcal{I}}(u-1)\pi}{3},\;p_\mathcal{I}=0,1,2,3$ & 8\\
$1$&$0$&$n\pi$&$\theta$&$n_{\mathcal{I}}\pi\delta_{\text{mod}(u,2),0}$ & 4\\
$1$&$1$&$n\pi$&$\theta$&$n_{\mathcal{I}}\pi\delta_{\text{mod}(u,2),0}$ & 4\\
		\end{tabular}
	\end{ruledtabular}
	\label{table:u1_psg}
\end{table*}

\section{Lattice and time-reversal symmetries}
\label{sec:lattice_time}

The site coordinates on the maple-leaf lattice can be generically described by $\mathbf{r}=x\mathbf{T}_1 + y\mathbf{T}_2 + \mathbf{u}$. We choose the Bravais lattice vectors in the Cartesian basis as ${\mathbf{T}_1 = \frac{a}{2}(3\sqrt{3},-1)}$ and ${\mathbf{T}_2 = \frac{a}{2}(-\sqrt{3},5)}$, with lattice constant $a$; $\mathbf{u}$ denotes the position of lattice sites within the unit cell. The unit cell is fixed such that its center coincides with the center of a hexagon as depicted in Fig.~\ref{fig:cell_definitions}. Every site $\mathbf{u}$ in the unit cell can be written as $\mathbf{u} = x_u \mathbf{a}_1 + y_u \mathbf{a}_2$ where $\mathbf{a}_1 = a(0,1)$ and $\mathbf{a}_2 = \frac{a}{2}(-\sqrt{3},1)$ where $x_u$, $y_u \in \left\lbrace 0, \pm 1 \right\rbrace$. Using the convention of Fig.~\ref{fig:cell_definitions} we label these sites in a shorthand notation by $u=\left\lbrace 1,\ldots, 6 \right\rbrace$. The underlying symmetry group $\mathcal{S}$ of the maple-leaf lattice is given by the wallpaper group \textit{P6} which can be generated by 4 operations: two translations ($T_1$ and $T_2$), inversion ($\mathcal{I}$) and a $C_3$-rotation ($R$) (we adopt an anticlockwise rotation)~\footnote{Instead of ($\mathcal{I}$) and a $C_3$-rotation, one can equivalently work with only a single six-fold rotation $C_{6}$. Note that this lattice lacks any reflection symmetry.}. These operations act on a lattice site at ($x,y,u$) by
\begin{align}\label{eq:generators}
& T_1(x, y, u) \rightarrow (x+1, y, u),\notag \\
& T_2(x, y, u) \rightarrow (x, y+1, u),\notag \\
& R(x, y, u) \rightarrow (-y, x-y, R(u)), \\
& \mathcal{I}(x, y, u) \rightarrow (-x, -y, \mathcal{I}(u)), \notag\\
& R^{-1}(x, y, u) \rightarrow (y-x, -x, R^{-1}(u)). \notag
\end{align}
Within our convention of the unit cell the operations $R, \mathcal{I}$ only permute elements in $u$. Thus, one can compute their action on the Bravais lattice $(x,y)$ and on one unit cell $u$ separately. The mutual relations of Eq.~\eqref{eq:generators} lead to the following set of algebraic conditions
\begin{align}
& T_1 T_2 = T_2 T_1,  \label{eq:translations} \\
& \mathcal{I}^2 = \mathds{1}, \label{eq:sigma_2}\\
& T_i\mathcal{I} T_i= \mathcal{I}, \label{eq:sigma_T}\\
& R^3 = \mathds{1},  \label{eq:R}\\
& T^{-1}_1 R = R T_1 T_2, \label{eq:R_T1}\\
& T^{-1}_2 R T_1 = R,  \label{eq:R_T2}\\
& R^{-1}\mathcal{I} R \mathcal{I} = \mathds{1}. \label{eq:sigma_R}
\end{align}
These relations fix the underlying group structure. Note that in case of the PSG, the identity $\mathds{1}$ is defined only modulo the invariant gauge group $\mathcal{G}$. 

Besides operations acting on the site index of a spinor, we also include time-reversal symmetry $\mathcal{T}$, which acts on an \textit{Ansatz} according to $\mathcal{T}(u_{\mathbf{rr'}}, a^\mu_{\mathbf{r}})\rightarrow -(u_{\mathbf{rr'}}, a^\mu_{\mathbf{r}})$ \cite{Wenbook}. This leads to the additional conditions
\begin{align}
& \mathcal{T}^2 = \mathds{1}, \label{eq:timerev} \\
& \mathcal{T} \mathcal{S} \mathcal{T}^{-1} \mathcal{S}^{-1}  = \mathds{1} \label{eq:spacetime_comm}. 
\end{align}
In Appendix~\ref{app:genric_gauge_con}, we list all the precise conditions for the corresponding representation matrices $G_{\mathcal{S}}(\mathbf{r})$ and $G_\mathcal{T}(\mathbf{r})$. 

\section{PSG solutions}\label{sec4}
In this section, we present a set of gauge inequivalent representation matrices for the group extensions $\mathcal{G} \simeq U(1)$, $\mathds{Z}_2$. The details of the construction are presented in Appendix~\ref{app:u1_psg_derivation} for the $U(1)$ case, and Appendix~\ref{app:z2_psg_derivation} for the $\mathds{Z}_2$ case.

\label{sec:classification}
\subsection{$U(1)$ PSG solution}
The generic form of $U(1)$ PSG solution for any symmetry operator $\mathcal{O}$ is given by $G_\mathcal{O}(x,y,u)=g_3(\phi_\mathcal{O}(x,y,u))(\dot\iota\tau^1)^{w_\mathcal{O}}$. Here, $g_3(\xi)$ should be read as $e^{\dot\iota\xi\tau^3}$, with $\xi \in [0,2\pi)$, and $w_\mathcal{O}$ is an integer that takes the values $0$ and $1$.
Note that for $\mathcal{O}\in\{T_1,T_2,R\}$ consistent solutions only exist for $w_\mathcal{O}=0$. The solutions for $\phi_\mathcal{O}(x,y,u)$ are 
\begin{align}
 \phi_{T_1}(x,y,u)&=y\theta,\;\phi_{T_2}(x,y,u)=0, \label{eq:g_translation_u}\\
 \phi_{R}(x,y,u)&=[xy-\frac{1}{2}x(x-1)]\theta, \label{eq:g_R_u}\\
 \phi_\mathcal{I}(x,y,u)&=\Tilde{\theta}_{\mathcal{I}}(x+y)+\rho_{\mathcal{I}}(u), \label{eq:g_sigma_u}\\
 \phi_{\mathcal{T}}(x,y,u)&=u\pi\delta_{w_\mathcal{T},0}. 
\label{eq:g_time_u}
\end{align}
The parameters in the above equations are given in Appendix~\ref{app:u1_psg_derivation}.
In total, we enumerated 17 distinct $U(1)$ PSG classes which are listed in Table~\ref{table:u1_psg}.

\subsection{$\mathds{Z}_2$ PSG solution}
In the $\mathds{Z}_2$ case, the $G_\mathcal{O}$ are generic $SU(2)$ matrices. A set of gauge inequivalent solutions is given by
\begin{align}
& G_{T_1}(x,y,u)=\eta^{y}\tau^0,\;G_{T_2}(x,y,u)=\tau^0, \label{eq:g_translation}\\
& G_R(x,y,u)=\eta^{xy-\frac{1}{2}x(x-1)}\tau^0, \label{eq:g_R}\\
& G_\mathcal{I}(x,y,u)=\eta^{x+y}\eta^{u+1}_{\mathcal{I}}\tau^0, \label{eq:g_sigma}\\
& G_{\mathcal{T}}(x,y,u)=\eta^{u+1}_{\mathcal{T}\mathcal{I}}g_{\mathcal{T}},\;\;g_{\mathcal{T}}\in \left\lbrace \tau^0,\dot\iota\tau^2 \right\rbrace .\label{eq:g_time}
\end{align}
Taking only lattice symmetries into account, the number of the gauge inequivalent \textit{Ans\"atze} is classified by the integer (binary) parameters $\eta$, $\eta_{\mathcal{I}}$, yielding $2^2=4$ distinct classes. Inclusion of time-reversal leads to $2^2\times4=16$ PSG classes, as the integer $\eta_{\mathcal{T}\mathcal{I}}$ can take values $0$ or $1$. However, the solutions with $g_{\mathcal{T}}=\tau^0$ and $\eta_{\mathcal{T}\mathcal{I}}=1$ yield vanishing \textit{Ans\"atze}, so we can effectively consider a total of $2^2\times3=12$ PSG classes which lead to fully symmetric QSLs.

\begin{figure}	\includegraphics[width=0.750\linewidth]{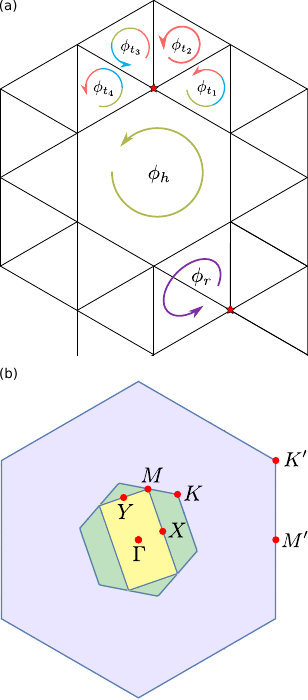}
	\caption{(a) Illustration of $SU(2)$ fluxes on the maple-leaf lattice. Note that all fluxes are defined starting from the same lattice site marked by a red star. The two stars are therefore thought of as equivalent. (b) The green (blue) hexagon depicts the first (extended) Brillouin zone. The extended Brillouin zone is obtained via scaling by a factor of $\sqrt{7}$ and rotation by an angle $\phi=\arccos{\frac{5}{2\sqrt{7}}}$ w.r.t. the first Brillouin zone. The orange rectangular region shows the reduced Brillouin zone corresponding to \textit{Ans\"atze} which double the unit cell along $T_1$. The high symmetry points are $\Gamma(0,0)$, $X(\frac{5\pi}{14\sqrt{3}},\frac{\pi}{14})$, $M(\frac{\pi}{7\sqrt{3}},\frac{3\pi}{7})$, $Y(-\frac{3\pi}{14\sqrt{3}},\frac{5\pi}{14})$, $K(\frac{4\pi}{7\sqrt{3}},\frac{8\pi}{21})$, $K'(\frac{2\pi}{\sqrt{3}},\frac{2\pi}{3})$ and $M'(\frac{2\pi}{\sqrt{3}},0)$. }
	\label{fig:fig2}
\end{figure}

\begin{figure}
\includegraphics[width=1.0\linewidth]{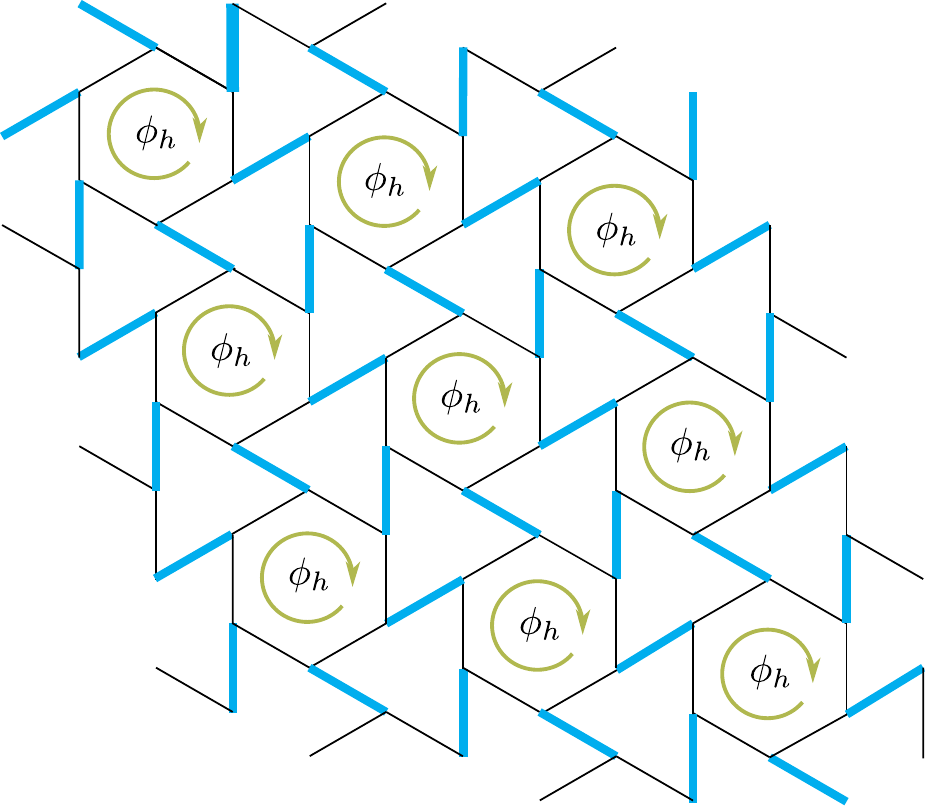}
	\caption{Flux pattern for the \textit{Ansatz} classes UA and UB. These classes share the common property that all red bonds vanish. The flux through the hexagons denoted by $\phi_h$, and the blue bonds differ for specific class members and are further specified in the main text.}\label{fig:flux_uab}
\end{figure}

\section{Short ranged mean-field \textit{Ans\"atze}}\label{sec5}
In this section, we explicitly construct and discuss all mean-field \textit{Ans\"atze} for first nearest neighbors (1NN) amplitudes according to the PSG symmetry conditions derived in the previous section. As shown in Fig.~\ref{fig:cell_definitions}, there are three symmetry inequivalent 1NN bonds, which are referred to as red, blue, and green bonds in the remainder of the paper. 
Inserting the gauge inequivalent PSG representations $G_{\mathcal{O}}$ in Eq.~\eqref{eq:symmetry_condition} enables the construction of the mean-field matrices $u_{\mathbf{rr'}}$ according to the desired symmetry. A similar yet one-site condition can be used for the Lagrange multiplier $a_\mu (\mathbf{r})$. By an \textit{Ansatz}, we refer to the pair $(u_{\mathbf{rr'}}, a_\mu(\mathbf{r}))$. We present the $U(1)$ \textit{Ans\"atze} first, followed by the $\mathds{Z}_2$ states in the later subsection. A concise summary of the general transformation rules Eq.~\eqref{eq:symmetry_condition} for the $\mathds{Z}_2$ states is shown in Fig.~\ref{fig:first_neighbors}. Appendix~\ref{sec:u1_ansatze_3NN} contains the results for mean-field models up to third nearest neighbors.

\subsection{$U(1)$ mean-field \textit{Ans\"atze}}
\label{sec:u1_ansatze}
We divide the $U(1)$ \textit{Ans\"atze} in four classes labeled by UA, UB, UC and UD corresponding to the different PSG labels $(w_\mathcal{T},w_\mathcal{I})= (0,0)$, $(0,1)$, $(1,0)$ and $(1,1)$, respectively.
\subsubsection{$w_\mathcal{I}=0$ and $w_\mathcal{T}=0$ (class UA)}
The \textit{Ansatz} matrices for this class are given by
\begin{equation}\label{eq:u00_ansatze_1NN}
\left.\begin{aligned}
&u^{12}_{1g}=u^{34}_{1g}=u^{56}_{1g}=u^{23}_{1g}=u^{45}_{1g}=u^{61}_{1g}=\dot\iota\chi^0_{1g}\tau^0+\chi^3_{1g}\tau^3,\\
&u^{14}_{1b}=u^{36}_{1b}=u^{52}_{1b}=\dot\iota\chi^0_{1b}\tau^0+\chi^3_{1b}\tau^3,\\
& u^{uu'}_{1r}=0,\quad a_\mu(u)=0,\\
&2\tan^{-1}\left(-\frac{\chi^3_{1b}}{\chi^0_{1b}}\right)+\Tilde{\theta}_{\mathcal{I}}=\pi.
\end{aligned}\right.
\end{equation}
Applying $N$ successive translations along $T_1$ will modify $u^{14}_{1b}$ and $u^{36}_{1b}$ according to
\begin{equation}
\left.\begin{aligned}
&u^{14}_{1b}\equiv u_{(x,y,1),(x,y+1,4)}\\
& \overset{T_1^{N}}{\rightarrow} g_3(-3 N \Tilde{\theta}_{\mathcal{I}})u_{(x+N,y,1),(x+N,y+1,4)}\\
& =g_3(-3 N \Tilde{\theta}_{\mathcal{I}})u^{14}_{1b},\\
&u^{36}_{1b}\equiv u_{(x,y,3),(x-1,y-1,6)}\\
& \overset{T_1^N}{\rightarrow} g_3(3N\Tilde{\theta}_{\mathcal{I}})u_{(x+N,y,3),(x+N-1,y-1,6)}\\
& = g_3(3N\Tilde{\theta}_{\mathcal{I}})u^{36}_{1b}
\end{aligned}\right.    
\end{equation}
while all other bonds remain invariant. The realization of such an \textit{Ansatz} on a finite lattice needs the implementation of the following constraint 
\begin{equation}
\Tilde{\theta}_{\mathcal{I}}=\frac{m}{n}\pi\text{, with }m,n\in \mathds{Z}.
\end{equation}

\begin{figure}
\includegraphics[width=1.0\linewidth]{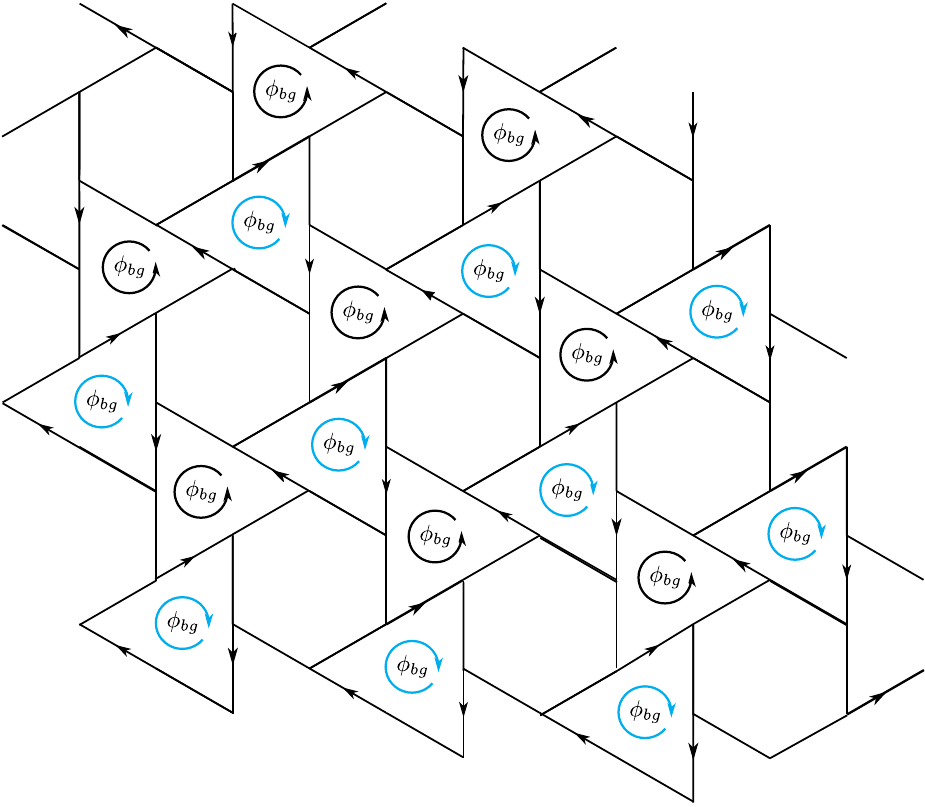}
	\caption{Flux pattern for the \textit{Ansatz} UB03. The arrows on the lines denote the orientation of the bonds. The triangles are pierced by nontrivial flux $\phi_{bg}$ which is defined for an anticlockwise rotation. Triangles for which the circulation goes clockwise carry negative flux $-\phi_{bg}$ and here drawn in blue.}\label{fig:flux_ub03}
\end{figure}

Members of this class will be labeled as
\begin{equation} 
\text{UA}[m,n].
\end{equation}
Since $u^{uu'}_{1r}=0$, all the red bonds vanish for 1NNs and the only $SU(2)$ flux $\phi_h$ [see Fig.~\ref{fig:fig2}(a) for the definition of the different fluxes] piercing through the central hexagon as depicted in Fig.~\ref{fig:flux_uab}. The precise flux value is determined by the mean-field parameters $\chi^{0/3}_{1g}$.

\subsubsection{$w_\mathcal{I}=1$ and $w_\mathcal{T}=0$ (class UB)}
We label this class as 
\begin{equation}
    \text{UB}np_\mathcal{I},
\end{equation}
where the classifying quantum numbers are $n \in \left\lbrace 0,1 \right\rbrace$, and $p_\mathcal{I} \in \left\lbrace 0,1,2,3 \right\rbrace$. The coupling matrices are
\begin{equation}\label{eq:u10_ansatze_1NN}
\left.\begin{aligned}
&u^{12}_{1g}=u^{34}_{1g}=u^{56}_{1g}=\dot\iota\chi^0_{1g}\tau^0+\chi^3_{1g}\tau^3,\\
&u^{23}_{1g}=u^{45}_{1g}=u^{61}_{1g}=-g_3(p_\mathcal{I}\pi/3)(u^{12}_{1g})^\dagger,\\
&u^{14}_{1b}=u^{36}_{1b}=u^{52}_{1b}=\dot\iota\chi^0_{1b}\tau^0+\chi^3_{1b}\tau^3,\\
& u^{14}_{1b}=-\eta g_3(p_\mathcal{I}\pi)u^{14}_{1b},\\
& u^{uu'}_{1r}=0,\quad a_\mu(u)=0.
\end{aligned}\right.
\end{equation}
where we denote $g_3(n\pi)=\eta$. The spatial dependence of  $u^{14}_{1b}$ and $u^{36}_{1b}$ is given by
\begin{equation}
\left.\begin{aligned}
& \eta^{N}u^{14}_{1b}(x+N, y) = u^{14}_{1b}(x,y),\\
& \eta^{N}u^{36}_{1b}(x+N, y) = u^{36}_{1b}(x,y).
\end{aligned}\right.
\end{equation}

\begin{figure}
\includegraphics[width=1.0\linewidth]{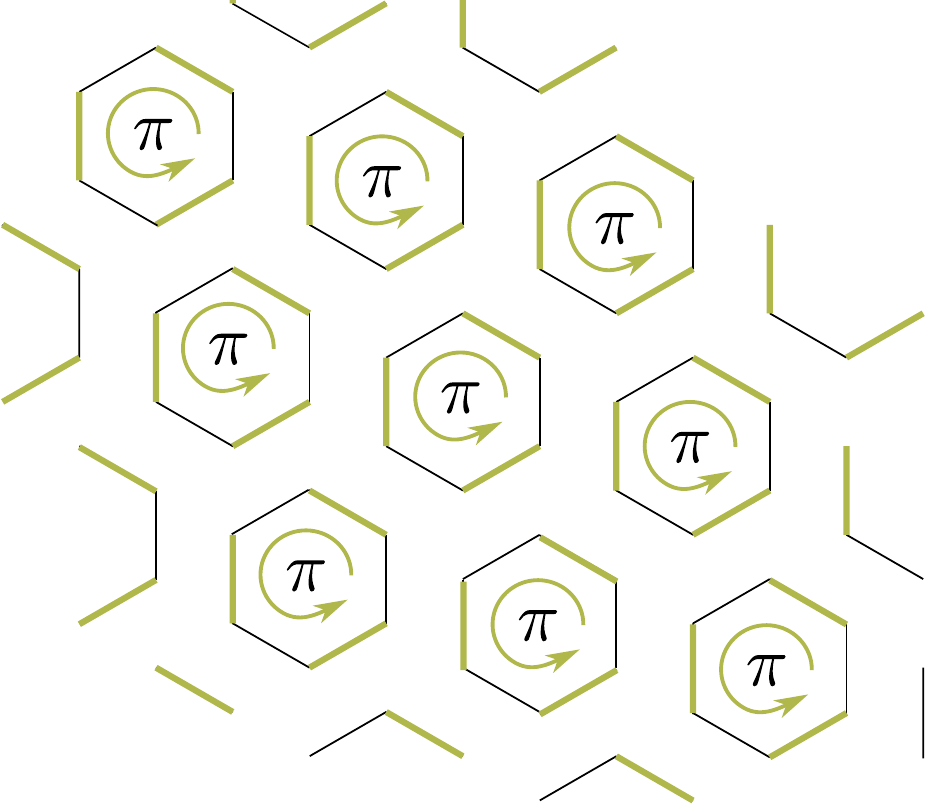}
	\caption{Flux pattern for the \textit{Ansatz} UB00. The green lines denote bonds for which the coupling matrices are multiplied by $-1$.}\label{fig:flux_ub00}
\end{figure}

Therefore, the realization of an \textit{Ansatz} with $\eta=-1$ requires a doubling of the unit cell. Like in the UA class, only the loop operator corresponding to the hexagons is finite with
\begin{equation}
    P_{\phi_h}\propto g_3((p_\mathcal{I}+1)\pi).
\end{equation}
The gauge inequivalent \textit{Ans\"atze} in this class are as follows:

For $\eta = +1$ with $\phi_h=0$ an \textit{Ansatz} has to obey
\begin{equation}\label{eq:ub03}
\left.\begin{aligned}
&u^{12}_{1g}=u^{34}_{1g}=u^{56}_{1g}=u^{23}_{1g}=u^{45}_{1g}=u^{61}_{1g}=\chi_{1g}\tau^3,\\
&u^{14}_{1b}=u^{36}_{1b}=u^{52}_{1b}=\dot\iota\chi^0_{1b}\tau^0+\chi^3_{1b}\tau^3,\\
& u^{uu'}_{1r}=0.
\end{aligned}\right.
\end{equation}
An appropriate gauge transformation sets $\chi^0_{1g}=0$ and we redefine $\chi^3_{1g}=\chi_{1g}$. This state is then labeled as UB03 and shown in Fig.~\ref{fig:flux_ub03}. Note that for 1NN $p_\mathcal{I}=3$ and $p_\mathcal{I}=1$ lead to the same result.

For $\eta = +1$ with $\phi_h=\pi$ the \textit{Ans\"atze} are
\begin{equation}\label{eq:hex_pi}
\left.\begin{aligned}
&u^{12}_{1g}=u^{34}_{1g}=u^{56}_{1g}=-u^{23}_{1g}=-u^{45}_{1g}=-u^{61}_{1g}=\chi_{1g}\tau^3,\\
&u^{uu'}_{1b}=0,\quad u^{uu'}_{1r}=0.
\end{aligned}\right.
\end{equation}
We label this state as UB00 where $p_\mathcal{I}=0$. Notice that this configuration corresponds to the $\pi$-flux hexagonal plaquette singlet state as shown in Fig.~\ref{fig:flux_ub00}. $p_\mathcal{I}=2$ gives the same state for 1NN. 

\begin{figure}
\includegraphics[width=1.0\linewidth]{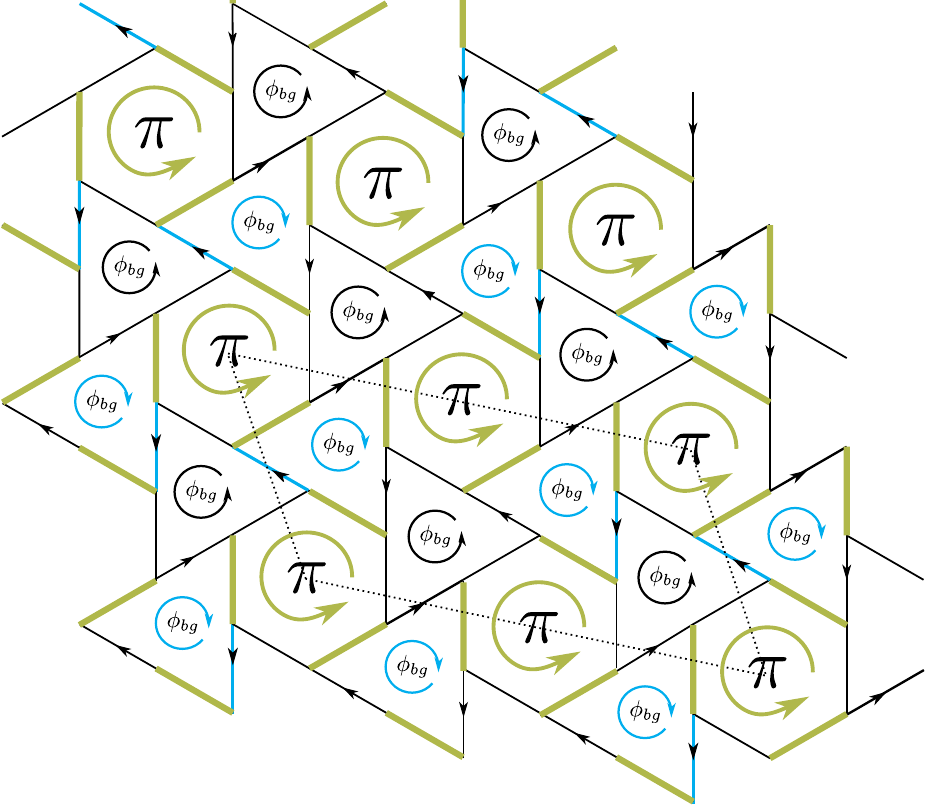}
	\caption{Flux pattern for the \textit{Ansatz} UB10. The green and blue lines denote where the coupling matrices are multiplied by a sign factor $-1$. The arrows on the lines denote the orientation of the bonds. With $\eta = -1$ one has a doubling of the unit cell which is enclosed by the dashed lines. The triangles are pierced by nontrivial flux $\phi_{bg}$ which is defined for an anticlockwise rotation. Triangles for which the circulation goes clockwise carry negative flux $-\phi_{bg}$ and here drawn in blue. The hexagons carry a $\pi$ flux.}\label{fig:flux_ub10}
\end{figure}

For $\eta = -1$ an \textit{Ansatz} with $\phi_h=\pi$ appears for $p_\mathcal{I}=0$ labelled by UB10 and is given by
\begin{equation}\label{eq:ubpi0}
\left.\begin{aligned}
&u^{12}_{1g}=u^{34}_{1g}=u^{56}_{1g}=-u^{23}_{1g}=-u^{45}_{1g}=-u^{61}_{1g}=\chi_{1g}\tau^3,\\
&u^{14}_{1b}=u^{36}_{1b}=u^{52}_{1b}=\dot\iota\chi^0_{1b}\tau^0+\chi^3_{1b}\tau^3, \quad u^{uu'}_{1r}=0,
\end{aligned}\right.
\end{equation}
and shown in Fig.~\ref{fig:flux_ub10}. An \textit{Ansatz} with $p_\mathcal{I}=2$ is gauge equivalent to UB10 for 1NN.

\textit{Ans\"atze} for $\eta = -1$ with $\phi_h=0$ appear for $p_\mathcal{I}=3$ and are labelled by UB13. The matrices are given by
\begin{equation}\label{eq:hex_zero}
\left.\begin{aligned}
&u^{12}_{1g}=u^{34}_{1g}=u^{56}_{1g}=u^{23}_{1g}=u^{45}_{1g}=u^{61}_{1g}=\chi_{1g}\tau^3,\\
&u^{uu'}_{1b}=0, \quad u^{uu'}_{1r}=0.
\end{aligned}\right.
\end{equation}
Such an \textit{Ansatz} corresponds to the $0$-flux hexagonal plaquette singlet state. $p_\mathcal{I}=1$ yields the same state for 1NN. This state is effectively the same as for $\eta =+1$ as the mean-field amplitudes on the blue bonds vanish. Notice that for both, $0$- and $\pi$-flux, hexagonal singlet plaquette \textit{Ans\"atze} the IGG is $SU(2)$.

\subsubsection{$w_\mathcal{I}=0$ and $w_\mathcal{T}=1$ (class UC)}
This class is labeled as
\begin{equation}
  \label{eq:uc_label}  \text{UC}nn_\mathcal{I}
\end{equation}
with $n, n_\mathcal{I} \in \left\lbrace 0,1 \right\rbrace$. We use the shorthand notation $\eta=g_3(n\pi)$ and $\eta_\mathcal{I}=g_3(n_\mathcal{I}\pi)$. The coupling matrices are determined as
\begin{equation}\label{eq:u01_ansatze_1NN}
\left.\begin{aligned}
&u^{12}_{1g}=\eta_\mathcal{I}u^{23}_{1g}=u^{34}_{1g}=\eta_\mathcal{I}u^{45}_{1g}=u^{56}_{1g}=\eta_\mathcal{I}u^{61}_{1g}=\chi_{1g}\tau^3,\\
&u^{14}_{1b}=u^{36}_{1b}=u^{52}_{1b}=\chi_{1b}\tau^3, \quad u^{14}_{1b}=\eta\eta_\mathcal{I}(u^{14}_{1b})^\dagger,\\
&u^{13}_{1r}=u^{24}_{1r}=\eta u^{35}_{1r}=u^{46}_{1r}=\eta u^{51}_{1r}=u^{62}_{1r}=\chi_{1r}\tau^3,\\
&a_3(u)\neq 0, \quad a_1, a_2 = 0.
\end{aligned}\right.
\end{equation}

\begin{figure}	\includegraphics[width=1.0\linewidth]{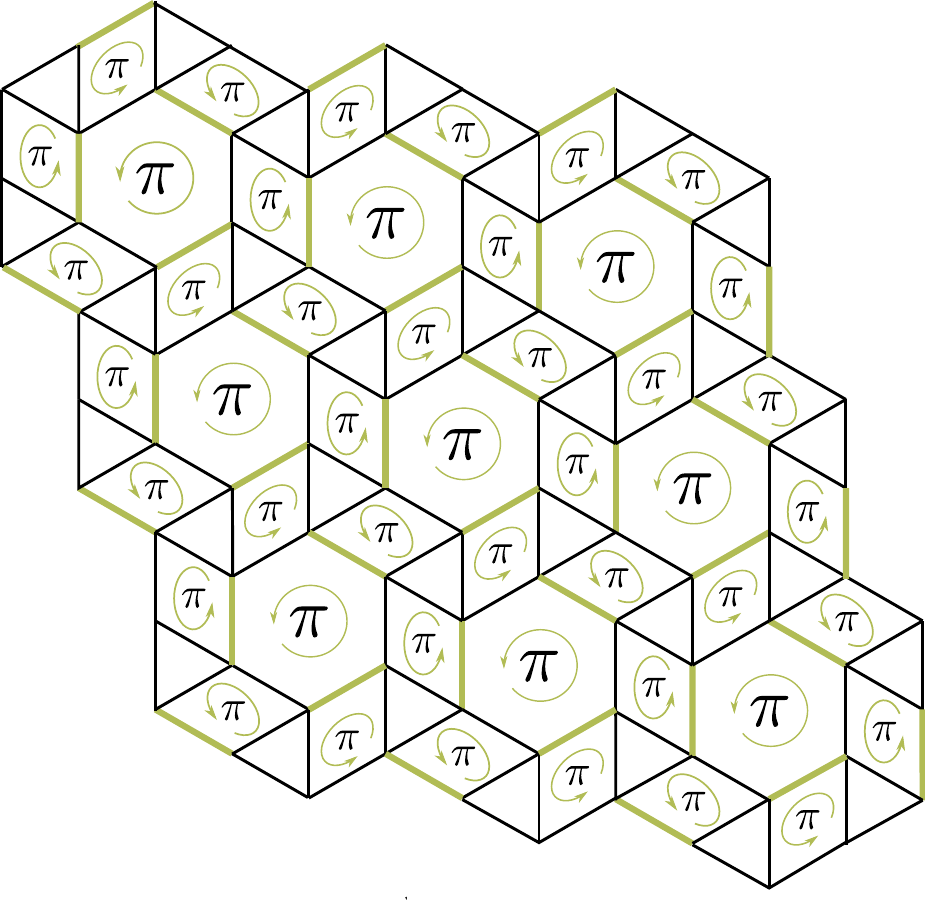}
	\caption{Flux pattern ($\pi,*,0,*,*,\pi$) for the \textit{Ansatz} class UC01 given by Eq.~\eqref{eq:uc0pi}. The green lines denote where the coupling matrices are multiplied by $\eta_\mathcal{I} = -1$.}
	\label{fig:flux_uc01}
\end{figure}

The spatial dependence is given by
\begin{equation}\label{eq:u1_spatial_dependence}
\left.\begin{aligned}
&\eta^N u^{14}_{1b}(x+N, y) = u^{14}_{1b}(x,y),\\
&\eta^N u^{36}_{1b}(x+N, y) = u^{36}_{1b}(x,y),\\
&\eta^N u^{24}_{1r}(x+N, y) = u^{24}_{1r}(x,y),\\
&\eta^N u^{13}_{1r}(x+N, y) = u^{13}_{1r}(x,y),\\
&\eta^N u^{46}_{1r}(x+N, y) = u^{46}_{1r}(x,y),\\
&\eta^N u^{51}_{1r}(x+N, y) = u^{51}_{1r}(x,y),
\end{aligned}\right.    
\end{equation}
while all other bonds are translation invariant. The \textit{Ans\"atze} for $\eta = +1$ and $\eta_\mathcal{I} = +1$ are
\begin{equation}\label{eq:uc00}
\left.\begin{aligned}
&u^{12}_{1g}=u^{23}_{1g}=u^{34}_{1g}=u^{45}_{1g}=u^{56}_{1g}=u^{61}_{1g}=\chi_{1g}\tau^3,\\
&u^{14}_{1b}=u^{36}_{1b}=u^{52}_{1b}=\chi_{1b}\tau^3,\\
&u^{13}_{1r}=u^{24}_{1r}=u^{35}_{1r}=u^{46}_{1r}=u^{51}_{1r}=u^{62}_{1r}=\chi_{1r}\tau^3,\\
&a_3(u)\neq 0, \quad a_1, a_2 = 0,
\end{aligned}\right.
\end{equation}
while for $\eta_\mathcal{I} = -1$ 
\begin{equation}\label{eq:uc0pi}
\left.\begin{aligned}
&u^{12}_{1g}=-u^{23}_{1g}=u^{34}_{1g}=-u^{45}_{1g}=u^{56}_{1g}=-u^{61}_{1g}=\chi_{1g}\tau^3,\\
&u^{uu}_{1b}=0,\\
&u^{13}_{1r}=u^{24}_{1r}=u^{35}_{1r}=u^{46}_{1r}=u^{51}_{1r}=u^{62}_{1r}=\chi_{1r}\tau^3,\\
& a_3(u)\neq 0, \quad a_1, a_2 = 0.
\end{aligned}\right.
\end{equation}

In the notation of Eq.~\eqref{eq:uc_label}, the above two states are labelled as UC00 and UC01, respectively. Another way of describing these states is by specifying their flux structures ($\phi_h,\phi_{t_1},\phi_{t_2},\phi_{t_3},\phi_{t_4},\phi_r$). The definition of these different fluxes is shown in Fig.~\ref{fig:fig2}. For the two previously discussed states this alternative notation leads to ($0,0,0,0,0,0$) and ($\pi,*,0,*,*,\pi$), respectively. The `$*$'-symbol indicates the absence of a well defined flux operator of the associated loop. Figure~\ref{fig:flux_uc01} shows the flux pattern of the UC01 class.

\begin{figure}	\includegraphics[width=1.0\linewidth]{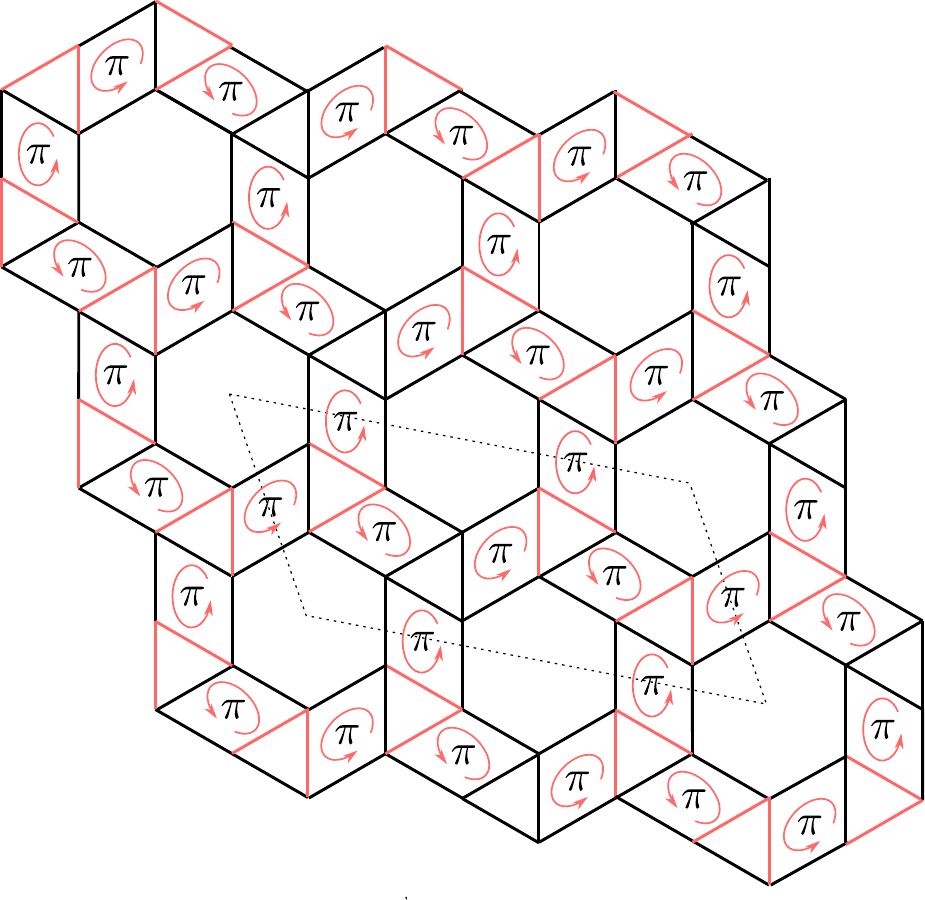}
	\caption{Flux pattern ($0,*,0,*,*,\pi$) for the \textit{Ansatz} class UC10 given by Eq.~\eqref{eq:ucpi0}. The red lines denote where the coupling matrices are multiplied by the sign factor $\eta = -1$. This sign factor leads to a doubled unit cell which is enclosed by the dashed lines.}
	\label{fig:flux_uc10}
\end{figure}

\begin{figure}	\includegraphics[width=1.0\linewidth]{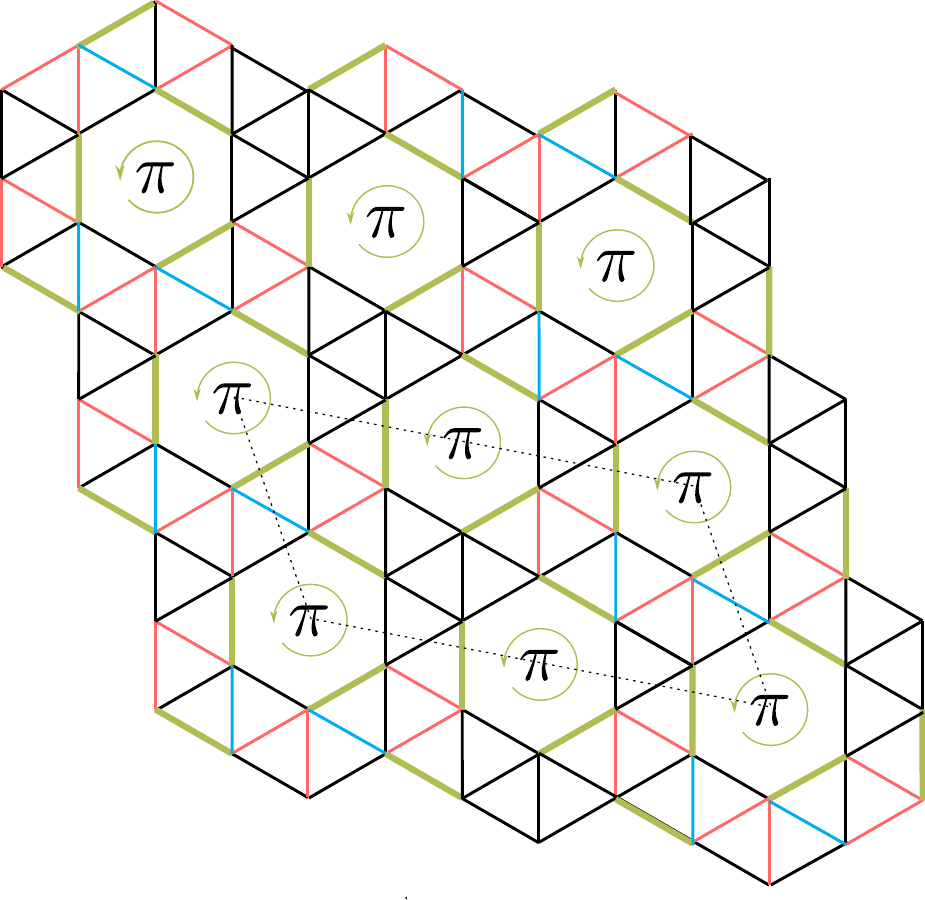}
	\caption{Flux pattern ($\pi,0,0,0,0,0$) for the \textit{Ansatz} class UC11 given by Eq.~\eqref{eq:ucpipi}. The red, blue and green lines denote the bonds in which the coupling matrices are multiplied by a factor $-1$. The $\eta = -1$ factor leads to a doubled unit cell which is enclosed by the dashed lines.}
	\label{fig:flux_uc11}
\end{figure}

The mean-field amplitudes for the $\eta=-1$ \textit{Ansatz} are given by
\begin{equation}\label{eq:ucpi0}
\left.\begin{aligned}
&u^{12}_{1g}=u^{23}_{1g}=u^{34}_{1g}=u^{45}_{1g}=u^{56}_{1g}=u^{61}_{1g}=\chi_{1g}\tau^3,\\
&u^{13}_{1r}=u^{24}_{1r}=-u^{35}_{1r}=u^{46}_{1r}=-u^{51}_{1r}=u^{62}_{1r}=\chi_{1r}\tau^3,\\
&u^{uu}_{1b}=0,\quad a_3(u)\neq 0, \quad a_1, a_2 = 0,
\end{aligned}\right.
\end{equation}
for $\eta_\mathcal{I} = +1$. This state is classified as UC10 with the associated flux pattern ($0,*,0,*,*,\pi$) which is shown in Fig.~\ref{fig:flux_uc10}. In the case that $\eta_\mathcal{I} = -1$ one finds
\begin{equation}\label{eq:ucpipi}
\left.\begin{aligned}
&u^{12}_{1g}=-u^{23}_{1g}=u^{34}_{1g}=-u^{45}_{1g}=u^{56}_{1g}=-u^{61}_{1g}=\chi_{1g}\tau^3,\\
&u^{14}_{1b}=u^{36}_{1b}=u^{52}_{1b}=\chi_{1b}\tau^3,\\
&u^{13}_{1r}=u^{24}_{1r}=-u^{35}_{1r}=u^{46}_{1r}=-u^{51}_{1r}=u^{62}_{1r}=\chi_{1r}\tau^3,\\
&a_3(u)\neq 0, \quad a_1, a_2 = 0.
\end{aligned}\right.
\end{equation}
We will call this \textit{Ansatz} UC11. It features the flux pattern ($\pi,0,0,0,0,0$), as depicted in Fig.~\ref{fig:flux_uc11}.

\begin{figure}	\includegraphics[width=1.0\linewidth]{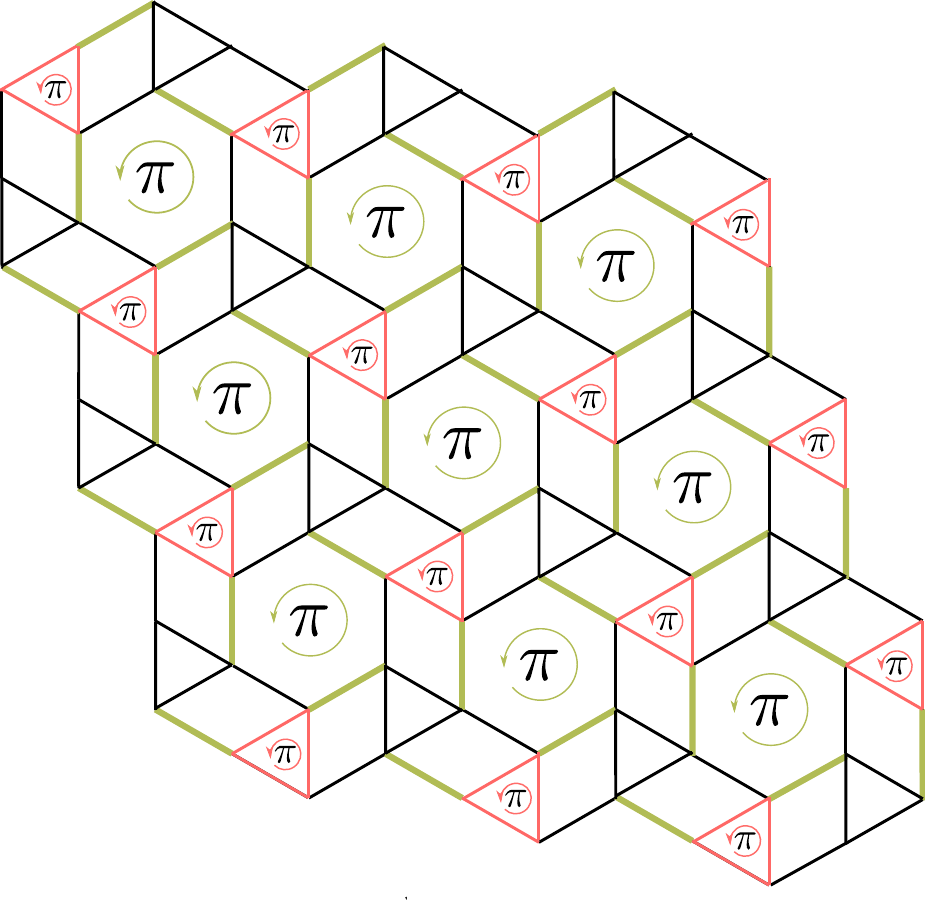}
	\caption{Flux pattern for the \textit{Ansatz} class UD00 given by Eq.~\eqref{eq:ud00}. The red and green lines denote where the coupling matrices are multiplied by sign factors.}
	\label{fig:flux_ud00}
\end{figure}
\begin{figure}	\includegraphics[width=1.0\linewidth]{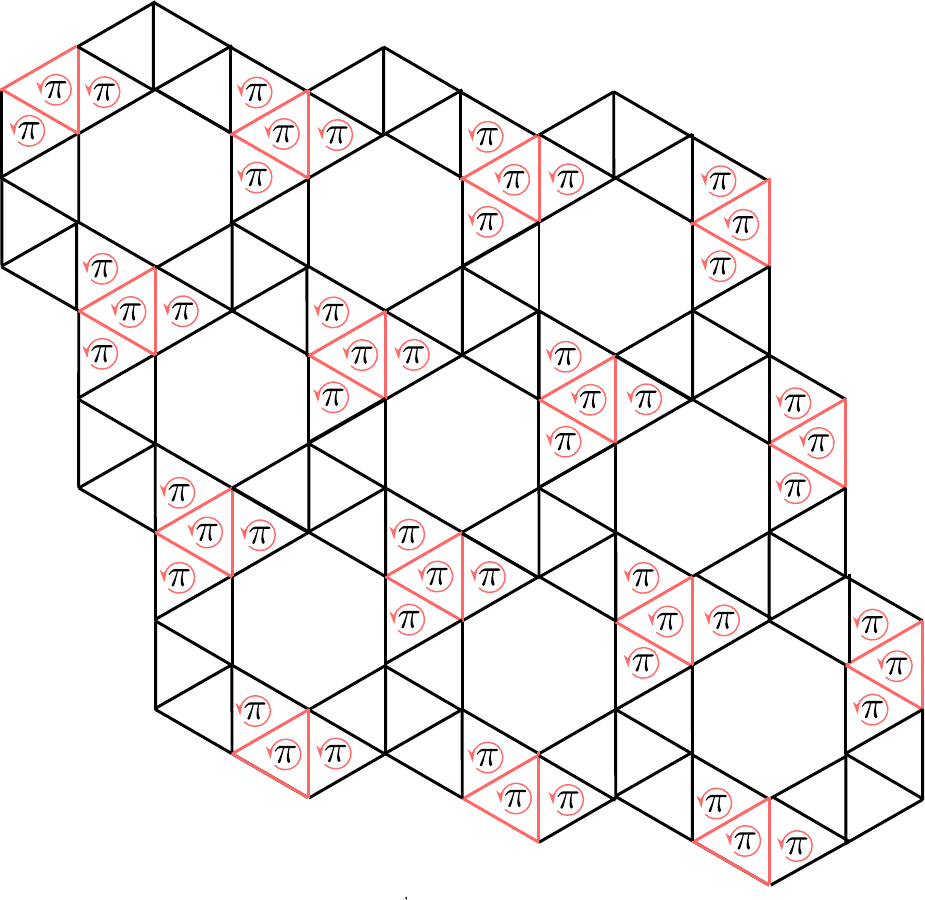}
	\caption{Flux pattern for the \textit{Ansatz} class UD01 given by Eq.~\eqref{eq:ud03}. The red lines denote where the coupling matrices are multiplied by $-1$.}
	\label{fig:flux_ud01}
\end{figure}

\begin{figure}	\includegraphics[width=1.0\linewidth]{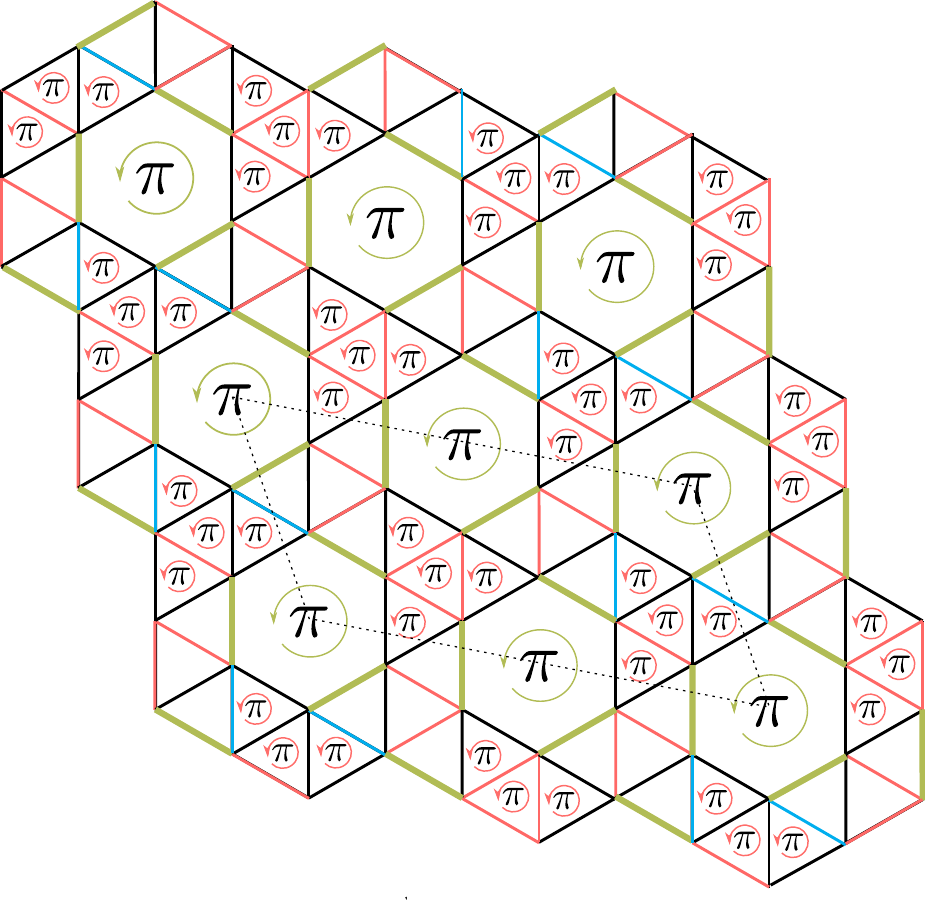}
	\caption{Flux pattern for the \textit{Ansatz} class UD10 given by Eq.~\eqref{eq:udpi0}. The red and green lines denote where the coupling matrices are multiplied by sign factors. The dashed lines enclose the doubled unit cell.}
	\label{fig:flux_ud10}
\end{figure}
\begin{figure}	\includegraphics[width=1.0\linewidth]{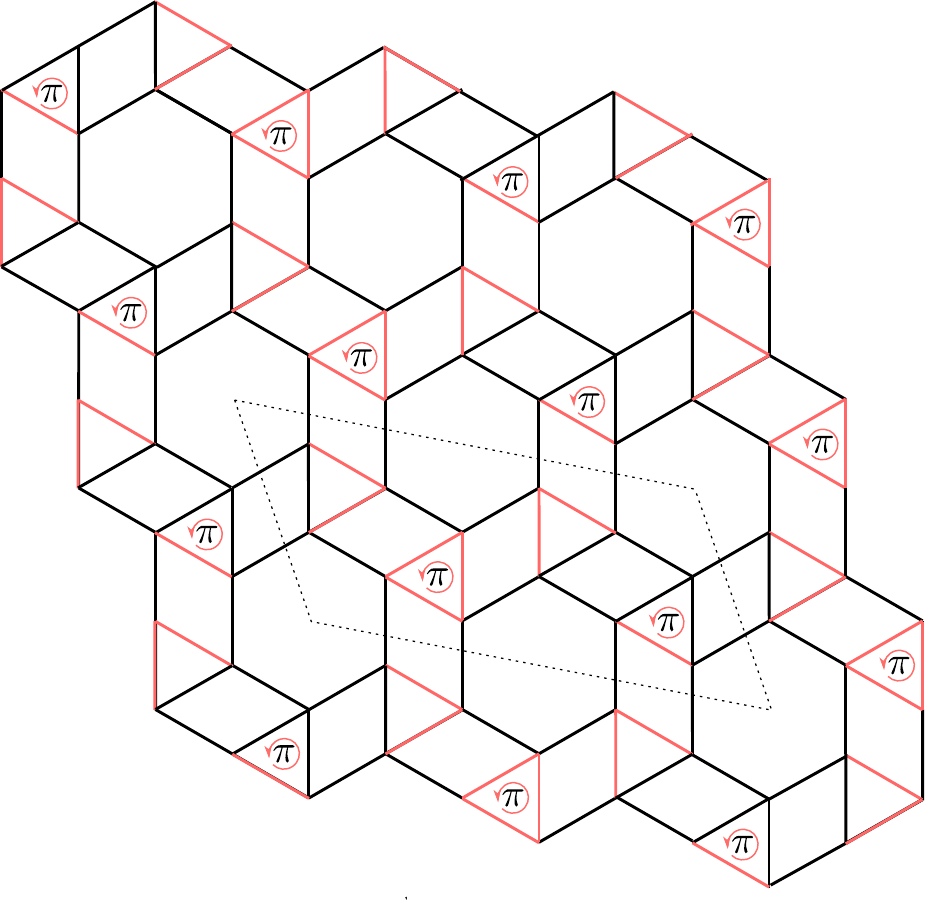}
	\caption{Flux pattern for the \textit{Ansatz} class UD11 given by Eq.~\eqref{eq:udpi3}. The red and green lines denote where the coupling matrices are multiplied by sign factors. The dashed lines enclose the enlarged unit cell.}
	\label{fig:flux_ud11}
\end{figure}

\subsubsection{$w_\mathcal{I}=1$ and $w_\mathcal{T}=1$ (class UD)}
Members of this class are labeled as
\begin{equation}
    \text{UD}nn_\mathcal{I}.
\end{equation}
The transformation properties of the mean-field amplitudes are 
\begin{equation}\label{eq:u11_ansatze_1NN}
\left.\begin{aligned}
&u^{12}_{1g}=-\eta_\mathcal{I}u^{23}_{1g}=u^{34}_{1g}=-\eta_\mathcal{I}u^{45}_{1g}=u^{56}_{1g}=-\eta_\mathcal{I}u^{61}_{1g}=\chi_{1g}\tau^3,\\
&u^{14}_{1b}=u^{36}_{1b}=u^{52}_{1b}=\chi_{1b}\tau^3,\quad u^{14}_{1b}=-\eta\eta_\mathcal{I}u^{14}_{1b},\\
&u^{13}_{1r}=-u^{24}_{1r}=\eta u^{35}_{1r}=-u^{46}_{1r}=\eta u^{51}_{1r}=-u^{62}_{1r}=\chi_{1r}\tau^3,\\
&a_\mu(u) =0.
\end{aligned}\right.
\end{equation}
The spatial dependence is the same as that of the UC-class [see Eq.~\eqref{eq:u1_spatial_dependence}]. The following \textit{Ans\"atze} are for $\eta=+1$ and  $\eta_\mathcal{I} = +1$
\begin{equation}\label{eq:ud00}
\left.\begin{aligned}
&u^{12}_{1g}=-u^{23}_{1g}=u^{34}_{1g}=-u^{45}_{1g}=u^{56}_{1g}=-u^{61}_{1g}=\chi_{1g}\tau^3,\\
&u^{uu'}_{1b}=0,\\
&u^{13}_{1r}=-u^{24}_{1r}=u^{35}_{1r}=-u^{46}_{1r}=u^{51}_{1r}=-u^{62}_{1r}=\chi_{1r}\tau^3.\\
\end{aligned}\right.
\end{equation}
The ones corresponding to $\eta_\mathcal{I} = -1$ are
\begin{equation}\label{eq:ud03}
\left.\begin{aligned}
&u^{12}_{1g}=u^{23}_{1g}=u^{34}_{1g}=u^{45}_{1g}=u^{56}_{1g}=u^{61}_{1g}=\chi_{1g}\tau^3,\\
&u^{14}_{1b}=u^{36}_{1b}=u^{52}_{1b}=\chi_{1b}\tau^3,\\
&u^{13}_{1r}=-u^{24}_{1r}=u^{35}_{1r}=-u^{46}_{1r}=u^{51}_{1r}=-u^{62}_{1r}=\chi_{1r}\tau^3.\\
\end{aligned}\right.
\end{equation}
UD00 and UD01 denote the two states above with their associated flux structures presented in Fig.~\ref{fig:flux_ud00} and Fig.~\ref{fig:flux_ud01}, respectively. 

In the case of $\eta=-1$ the \textit{Ans\"atze} for $\eta_\mathcal{I} = +1$ are 
\begin{equation}\label{eq:udpi0}
\left.\begin{aligned}
&u^{12}_{1g}=-u^{23}_{1g}=u^{34}_{1g}=-u^{45}_{1g}=u^{56}_{1g}=-u^{61}_{1g}=\chi_{1g}\tau^3,\\
&u^{14}_{1b}=u^{36}_{1b}=u^{52}_{1b}=\chi_{1b}\tau^3,\\
&u^{13}_{1r}=-u^{24}_{1r}=-u^{35}_{1r}=-u^{46}_{1r}=-u^{51}_{1r}=-u^{62}_{1r}=\chi_{1r}\tau^3,\\
\end{aligned}\right.
\end{equation}
while those belonging to $\eta_\mathcal{I} = -1$ are
\begin{equation}\label{eq:udpi3}
\left.\begin{aligned}
&u^{12}_{1g}=u^{23}_{1g}=u^{34}_{1g}=u^{45}_{1g}=u^{56}_{1g}=u^{61}_{1g}=\chi_{1g}\tau^3,\\
&u^{uu'}_{1b}=0,\\
&u^{13}_{1r}=-u^{24}_{1r}=-u^{35}_{1r}=-u^{46}_{1r}=-u^{51}_{1r}=-u^{62}_{1r}=\chi_{1r}\tau^3.\\
\end{aligned}\right.
\end{equation}
These states are labelled as UD10 (flux structure shown in Fig.~\ref{fig:flux_ud10}) and UD11 (flux structure shown in Fig.~\ref{fig:flux_ud11}), respectively.

\begin{table}[b]
\caption{First nearest-neighbor symmetric $\mathds{Z}_2$ mean-field \textit{Ans\"atze} as they are classified in the main text. The allowed mean-field amplitudes for the three symmetry inequivalent bonds are listed for each class, together with the allowed onsite terms.}
\begin{ruledtabular}
\begin{tabular}{ccccc}
\multirow{2}{*}{Label} & \multicolumn{3}{c}{1NN}  & \multirow{2}{*}{Onsite}\\
\cline{2-4}
& $u_{1g}$ & $u_{1b}$ & $u_{1r}$ & \\
\hline
$Z0002$ & $\chi_{1g}\tau^{3}+\Delta_{1g}\tau^{1}$ & $\chi_{1b}\tau^3+\Delta_{1b}\tau^{1}$ & $\chi_{1r}\tau^3+\Delta_{1r}\tau^{1}$ & $\tau^{3}$ \\
$Z0102$ & $\chi_{1g}\tau^{3}+\Delta_{1g}\tau^{1}$ & $0$ & $\chi_{1r}\tau^3+\Delta_{1r}\tau^{1}$ & $\tau^{3}$ \\
$Z1002$ & $\chi_{1g}\tau^{3}+\Delta_{1g}\tau^{1}$ & $0$ & $\chi_{1r}\tau^3+\Delta_{1r}\tau^{1}$ & $\tau^{3}$ \\
$Z1102$ & $\chi_{1g}\tau^{3}+\Delta_{1g}\tau^{1}$ & $\chi_{1b}\tau^3+\Delta_{1b}\tau^{1}$ & $\chi_{1r}\tau^3+\Delta_{1r}\tau^{1}$ & $\tau^{3}$ \\
\hline
$Z0012$ & $\dot\iota\chi_{1g}\tau^0+\Delta_{1g}\tau^{2}$ & $\Delta_{1b}\tau^{2}$ & $\chi_{1r}\tau^{3}+\Delta_{1r}\tau^{1}$ & $\tau^{3}$ \\
$Z0112$ & $\dot\iota\chi_{1g}\tau^0+\Delta_{1g}\tau^{2}$ & $\dot\iota\chi_{1b}\tau^0$ & $\chi_{1r}\tau^{3}+\Delta_{1r}\tau^{1}$ & $\tau^{3}$ \\
$Z1012$ & $\dot\iota\chi_{1g}\tau^0+\Delta_{1g}\tau^{2}$ & $\dot\iota\chi_{1b}\tau^0$ & $\chi_{1r}\tau^{3}+\Delta_{1r}\tau^{1}$ & $\tau^{3}$ \\
$Z1112$ & $\dot\iota\chi_{1g}\tau^0+\Delta_{1g}\tau^{2}$ & $\Delta_{1b}\tau^{2}$ & $\chi_{1r}\tau^{3}+\Delta_{1r}\tau^{1}$ & $\tau^{3}$ \\
\end{tabular}
\end{ruledtabular}
\label{table:z2_ansatze_1nn}
\end{table}

\subsection{$\mathds{Z}_2$ mean-field \textit{Ans\"atze}}
\label{sec:z2_ansatze}

\begin{figure*}
    \centering
    \includegraphics{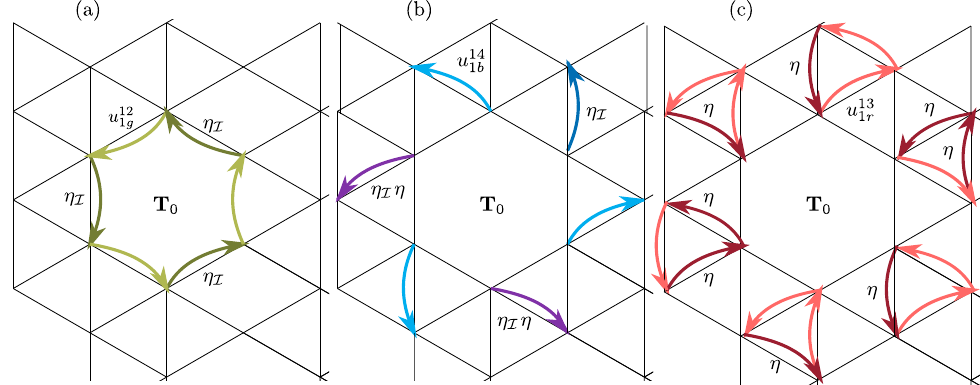}
    \caption{Transformation pattern for the first neighbor bonds. The labeling convention is that site 1 sits at the twelve o'clock position and increases counterclockwise (cfr Fig.~\ref{fig:cell_definitions}). The form of the independent \textit{Ansatz} matrices $u^{12}_{1g}, u^{14}_{1b}$ and $u^{13}_{1r}$ for the $\mathds{Z}_2$ states is given in Table~\ref{table:z2_ansatz} and their transformation rules are depicted in (a), (b), and (c), respectively. Different shadings of the color encode the different phase factors $\eta, \eta_\mathcal{I}$ and their products.} 
    \label{fig:first_neighbors}
\end{figure*}

All the \textit{Ans\"atze} with IGG $\simeq \mathds{Z}_2$ have been listed in Table~\ref{table:z2_ansatze_1nn}. The corresponding sign structures, depicted in Fig.~\ref{fig:first_neighbors}, are given as
\begin{equation}\label{eq:z2_ansatze_1NN}
\left.\begin{aligned}
&u^{12}_{1g}=\eta_\mathcal{I}u^{23}_{1g}=u^{34}_{1g}=\eta_\mathcal{I}u^{45}_{1g}=u^{56}_{1g}=\eta_\mathcal{I}u^{61}_{1g}=u_{1g}\\
&u^{14}_{1b}=u^{36}_{1b}=u^{52}_{1b}=u_{1b}\\
&u^{13}_{1r}=u^{24}_{1r}=\eta u^{35}_{1r}=u^{46}_{1r}=\eta u^{51}_{1r}=u^{62}_{1r}=u_{1r}.\\
\end{aligned}\right.
\end{equation}
The spatial dependence is the same as for the UC and UD classes [see  Eq.~\eqref{eq:u1_spatial_dependence}]. We adopt the following labelling scheme for the $\mathds{Z}_2$ \textit{Ans\"atze}:
\begin{equation}
Z\eta\eta_\mathcal{I}\eta_\mathcal{TI}\gamma.
\end{equation}
Here, we denote the positive signs of the $\eta$-parameters by `0' and the negative by `1' accordingly. The $\gamma$ label corresponds to the representation of time-reversal, i.e., $g_\mathcal{T}\propto\tau^\gamma$.

\begin{figure}
    \includegraphics[scale=1]{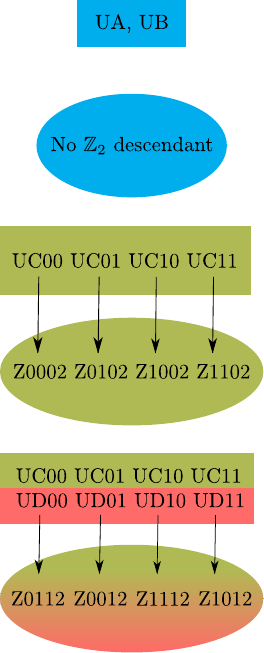}
    \caption{Different PSG classes and the connection between the parent $U(1)$ and their descendent $\mathds{Z}_2$ states.} 
    \label{fig:flowchart}
\end{figure}

The first four \textit{Ans\"atze} in the Table~\ref{table:z2_ansatze_1nn} correspond to a homogeneous representation of time-reversal $G_\mathcal{T}(x,y,u)=\dot\iota\tau^2$. Notice that the $\mathds{Z}_2$ state labelled by $Z0002$ belongs to the non-projective class, i.e., the linear representation of the space group with uniform real hopping and $s$-wave pairing. From the given structures of these four \textit{Ans\"atze} one sees the connection to their parent $U(1)$ states which is highlighted in Fig.~\ref{fig:flowchart}. Z0002, Z0102, Z1002 and Z1102 appear in the vicinity of the $U(1)$-\textit{Ans\"atze} labelled by UC00, UC01, UC10 and UC11, respectively.

\begin{figure}	\includegraphics[width=1.0\linewidth]{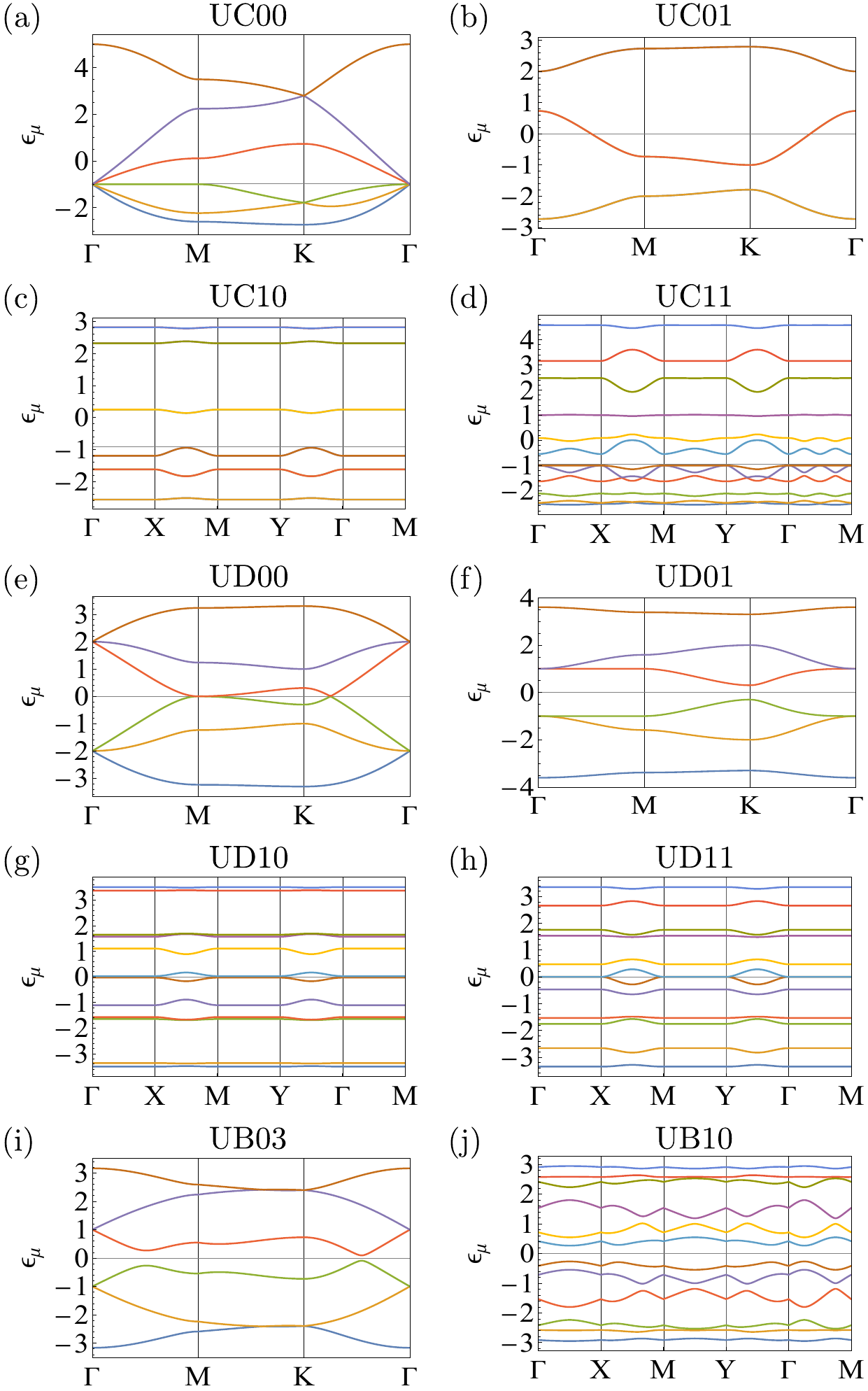}
	\caption{Spinon band structures for the different $U(1)$ \textit{Ans\"atze} with nearest neighbor hoppings only, corresponding to the gauge shown in the manuscript figures. The magnitude of the symmetry allowed hoppings is set to one. The gray line marks the Fermi level.}
	\label{fig:fig13}
\end{figure}

\begin{figure*}	\includegraphics[width=0.95\linewidth]{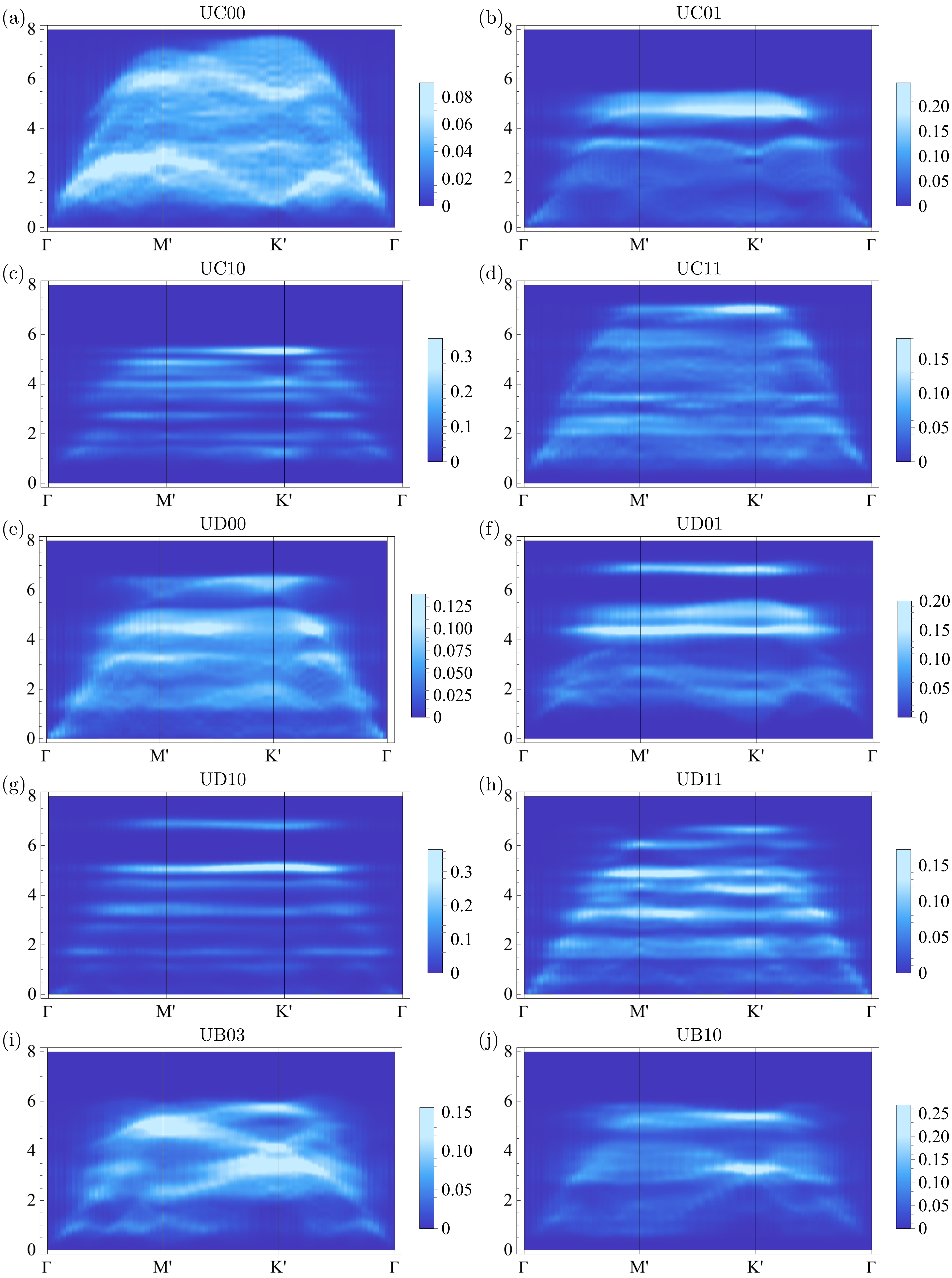}
\caption{Dynamical structure factor plotted along the high-symmetry path [see Fig.~\ref{fig:fig2}(b)] in the extended Brillouin zone for a system size of $14\times14\times6$ sites, where all symmetry allowed first-neighbor hopping amplitudes are set to one, and farther neighbors to zero.} 
\label{fig:fig14}
\end{figure*}

\begin{figure*}	\includegraphics[width=\linewidth]{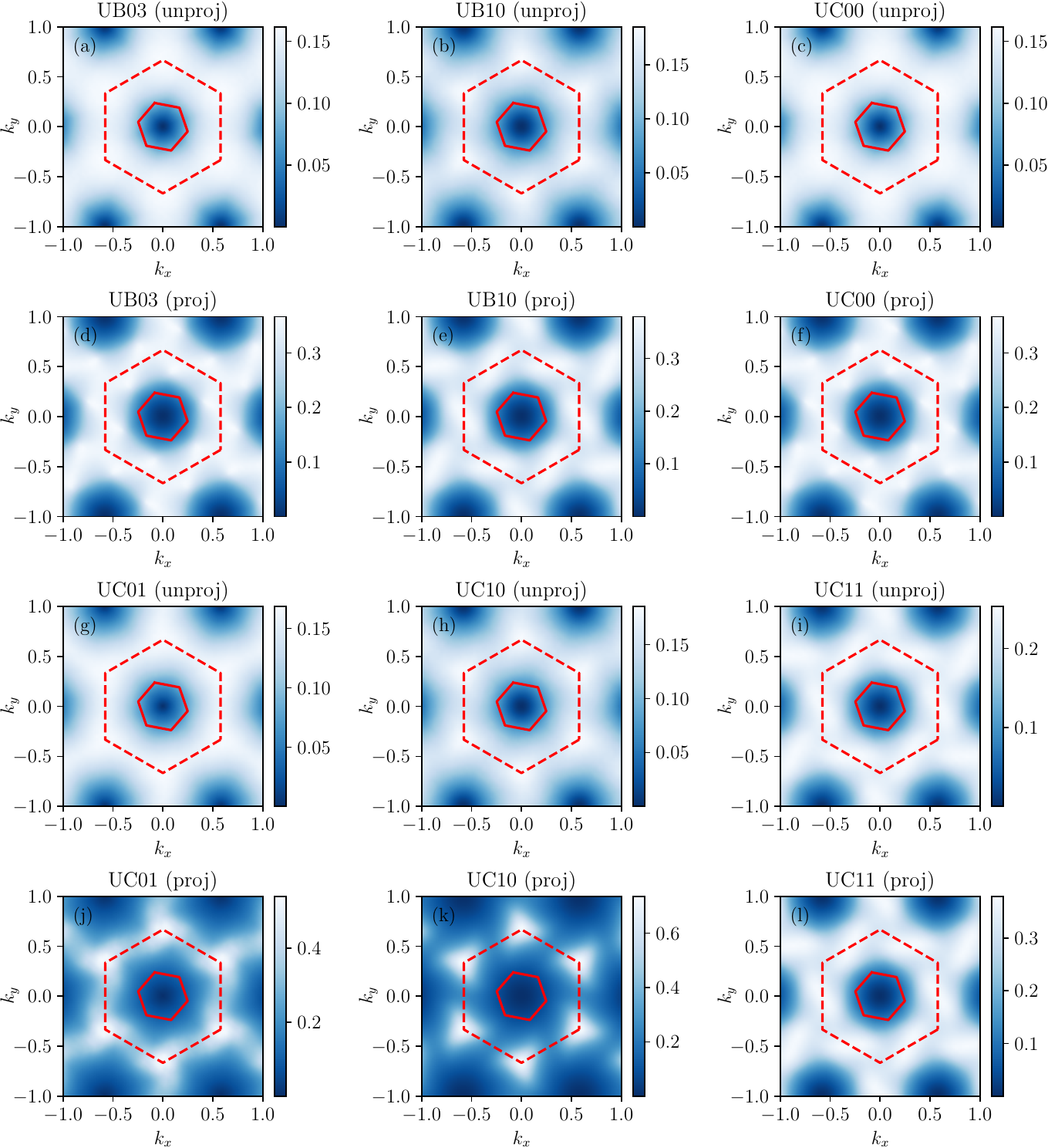}
\caption{Equal-time spin structure factors of the UB and UC states (with all symmetry allowed hopping amplitudes set to one), as obtained with the unprojected and projected fermionic wave functions within VMC. The color
plot shows the isotropic structure factor S$({\bf k})$ in the $\boldsymbol{k}_x - \boldsymbol{k}_y$ plane. The momenta are in units of $2\pi$. The results have been obtained on a $6\times12\times12~(=864)$-site finite cluster with all the symmetries of the lattice. The red hexagons with solid (dashed) lines delimits the first (extended) Brillouin zones.} 
\label{fig:sq_ubuc}
\end{figure*}

\begin{figure*}	\includegraphics[width=0.7\linewidth]{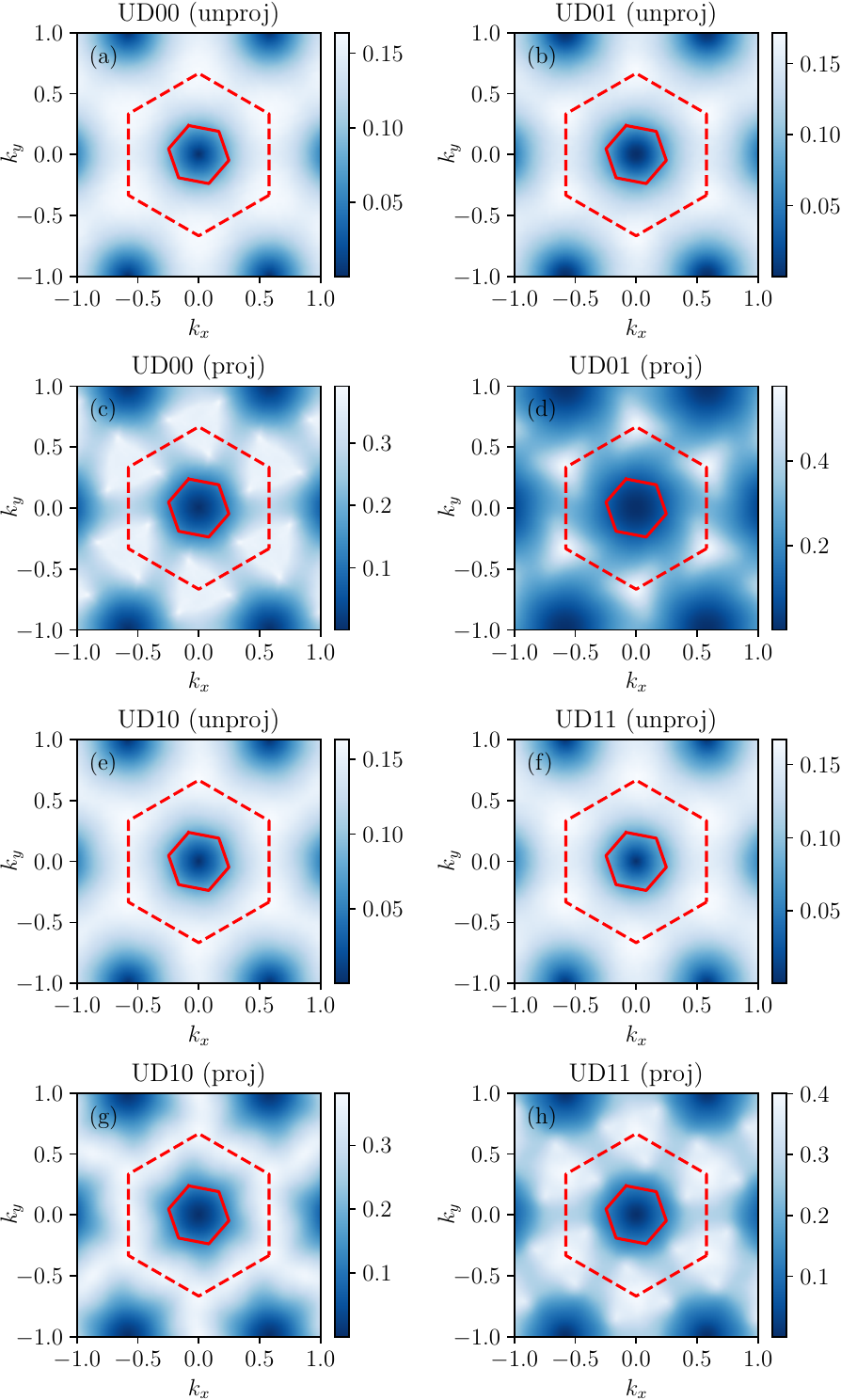}
\caption{Equal-time spin structure factors of the UD states (with all symmetry allowed hopping amplitudes are set to one), as obtained with the unprojected and projected fermionic wave functions within VMC. The color plot shows the isotropic structure factor S$({\bf k})$ in the $\boldsymbol{k}_x - \boldsymbol{k}_y$ plane. The momenta are in units of $2\pi$. The results have been obtained on a $6\times12\times12~(=864)$-site finite cluster with all the symmetries of the lattice. The red hexagons with solid (dashed) lines delimits the first (extended) Brillouin zones.} 
\label{fig:sq_ud}
\end{figure*}

The next four $\mathds{Z}_2$ \textit{Ans\"atze} contain a site dependent representation of time-reversal $G_\mathcal{T}(x,y,u)=(-1)^{u+1}\dot\iota\tau^2$. The connection to the parent $U(1)$ states is not apparent here but it can be established by using appropriate gauge transformations. Let us consider the \textit{Ansatz} labelled by Z0012. First all pairing terms have to be set equal to zero $\Delta_{1g}=0$, $\Delta_{1b}=0$ and $\Delta_{1r}=0$ which restores the continuous $U(1)$ symmetry. Then, using a gauge transformation of the form $W(x,y,u)=-\dot\iota\tau^3\delta_{\text{mod}(u,2),0}$ transforms it into UC01. On the other hand, it transforms into UD01 if one uses a gauge transformation of the form $W(x,y,u)=\dot\iota\tau^1\delta_{\text{mod}(u,2),0}$ which sets $\chi_{1g}=0$. Therefore, UC01 and UD01 share the same $\mathds{Z}_2$ descendant given by Z0012. Similarly, one can verify that UC00 and UD00 descent to Z0112, UC11 and UD11 to Z1012, and UC10 and UD10 to Z1112. Finally, for $G_\mathcal{T}(x,y,u)=(-1)^{u+1}\tau^0$, $\mathds{Z}_2$ \textit{Ans\"atze} with first nearest neighbour amplitudes cannot be realized. 

\section{Spinon bands, Dynamical and Static Spin Structure Factors}\label{sec6}
In this section, we summarize the properties of the spinon excitation spectrum and dynamical structure factors of different $U(1)$ \textit{Ans\"atze}. We adopt the gauge choice given in Sec.~\ref{sec:u1_ansatze}. In Fig.~\ref{fig:fig13}, we present the spinon spectrum obtained upon fixing the magnitude of the symmetry allowed first-neighbor hoppings equal to unity, and all further neighbor hoppings fixed to zero. We plot the energy along the high symmetry path, which is $\Gamma\rightarrow M\rightarrow K\rightarrow\Gamma$ of the first Brillouin zone [green hexagon in Fig.~\ref{fig:fig2}(b)] for the \textit{Ans\"atze} realizable in a single unit cell and $\Gamma\rightarrow X\rightarrow M\rightarrow Y\rightarrow\Gamma\rightarrow M$ of the reduced Brillouin zone [yellow rectangle in Fig.~\ref{fig:fig2}(b)] for the \textit{Ans\"atze} realizable in a doubled unit cell.
\begin{itemize}
\item UC00: This state is described by uniform hopping pattern and thus its spectrum is that of the maple-leaf lattice band structure~\cite{Schmalfuss-2002}. This state has a Dirac point at the center of the Brillouin zone [Fig.~\ref{fig:fig13}(a)]. However, the presence of the Dirac like dispersion is an artifact of fixing all first-neighbor hoppings to one. In general, for other choices of hoppings the spectrum can be gapped or feature a Fermi surface.
\item UC01: The spectrum consists of three doubly degenerate bands [Fig.\ref{fig:fig13}(b)]. The Fermi level is such that the lower half of the middle band is filled which gives rise to a Fermi surface.
\item UD00: A nodal Fermi surface is observed in this \textit{Ansatz} [Fig.~\ref{fig:fig13}(e)]. However, this is not a robust generic property as it gets gapped out upon varying the hopping amplitudes.
\item UD11: This state consists of quasi-flat bands and features a nodal Fermi surface [Fig.~\ref{fig:fig13}(h)] for the given choice of parameters. It can nonetheless be completely gapped out for other choices of hopping amplitudes.
\item UC10, UC11, UD01, UD10, UB03, and UB10: These states comprise of gapped excitations for a generic choices of hopping parameters, however, among these the UB03 \textit{Ansatz} shows Dirac dispersion for some choices of hopping parameters.
\end{itemize}

Further characterization can be made by studying the spin-spin correlations via the spin structure factors. Here, we consider both the dynamical and equal-time structure factors, and for the latter also assess the impact of Gutzwiller projection. We first discuss the dynamical structure factor (DSF) as defined in Eq.~\eqref{eq:dsf} [see Appendix~\ref{app:calculation_dsf} for details]. In Fig.~\ref{fig:fig14}, we show the DSF for different $U(1)$ \textit{Ans\"atze} along the high symmetry directions $\Gamma\rightarrow M'\rightarrow K'\rightarrow\Gamma$ of the extended Brillouin zone. For the UC00 state [see Fig.~\ref{fig:fig14}(a)], we notice the appearance of low-energy intensity around the $\Gamma$-point which is expected due to the presence of a Dirac point. In addition, there appear two principal dispersive variations in intensity. Among these, the dispersive continuum at lower energy occurs due to the scattering process from the three filled bands to the first empty band, while the intensity variation at higher energy is due the scattering from the filled bands to the dispersive uppermost empty band. For the UC01 state [see Fig.~\ref{fig:fig14}(b)], the dome like variation at lower energies is due to the contribution from excitations near the Fermi surface. Besides, there appears a flat strong intensity ontinuum at higher energies on a diffuse low intensity background, and this can be ascribed to scattering processes between the lowest (filled) mode and the highest (empty) mode. As the UC10 state consists of all quasi-flat modes, this reflects in the observed flat continuum [see Fig.~\ref{fig:fig14}(c)]. Similar inferences can be drawn for UC11, UD10 and UD11, shown in Figs.~\ref{fig:fig14}(d), (g) and (h), respectively. However, due to the presence of few dispersive modes above and below the Fermi level, a very low intensity diffusive background can be noted for UC11. For UD11, the remnant finite intensity down to zero energy at the $\Gamma$ point is due to the excitations around the Fermi surface. Similar consequence of the nodal Fermi surface can also be found in case of UD00 state [see Fig.~\ref{fig:fig14}(e)]. Here, the horizontal intensities at $\omega\approx3,4.5,6$, are due to the excitations between the quasi-flat bands in the segment $\overline{MK}$. 

While the calculation of the DSF is performed at the mean-field level given the numerical complexities involved, one can still assess the effects of gauge fluctuations beyond mean-field in the equal-time (frequency integrated) structure factor which is given by
\begin{equation}
S(\mathbf{k})=\frac{1}{N_s}\sum_{i,\,j} e^{\dot{\iota} \mathbf{k}\cdot (\mathbf{r}_i-\mathbf{r}_j)} \langle \Psi| \hat{\mathbf{S}}_i \cdot \hat{\mathbf{S}}_j |\Psi\rangle\,,
\end{equation}
Here, the wave function $|\Psi\rangle$ is defined by Gutzwiller-projecting the fermionic wave function to spin space, i.e., enforcing single fermionic occupation of each lattice sites. The Gutzwiller projection is treated numerically by means of a suitable Monte Carlo framework~\cite{Becca_Sorella_2017}. The results obtained with and without projection are compared in Fig.~\ref{fig:sq_ubuc} and~\ref{fig:sq_ud} for the various $U(1)$ \textit{Ans\"atze}. 
The unprojected $S(\mathbf{k})$ show a rather featureless ring of intensity encircling the extended Brillouin zone. The effects of gauge fluctuations introduced by the Gutzwiller projection are pronounced and one observes the appearance of well-defined momentum modulated features in the projected structure factors. These are triangulated patterns around the $K'$ points, featuring either a homogeneous intensity distribution or soft maxima at the triangular vertices. This pattern is qualitatively similar to that of the dimerized hexagonal singlet state of Ref.~\cite{Ghosh-2024}. It is interesting to note that the projected $S(\mathbf{k})$ of the UC01 state most closely resembles that obtained from a recent pf-FRG calculation~\cite{Gresista-2023} within the QSL phase, possibly hinting at the UC01 Fermi surface state providing a description of the spin liquid nature. Other states with similar structure factors are gapped $U(1)$ QSLs and therefore unstable~\cite{Polyakov-1977}. A generic feature of all \textit{Ans\"atze} is the absence of pinch points in their projected $S(\mathbf{k})$ in contrast to the Dirac spin liquid on the kagome lattice~\cite{Kiese-2023}.

\section{Discussion and Outlook}\label{sec7}
In this work we have performed a projective symmetry group classification of spin-$1/2$ symmetric quantum spin liquids with different gauge groups on the maple-leaf lattice. Employing the Abrikosov fermion representation we obtain 17 $U(1)$ and 12 $\mathds{Z}_{2}$ distinct PSGs. The restriction of mean-field \textit{Ans\"atze} to short-range (first-neighbor) \emph{singlet} amplitudes, of relevance to concerned models, leads to only 12 $U(1)$ and 8 $\mathds{Z}_{2}$ distinct phases. In light of recent numerical studies pointing to QSL ground states in extended $S=1/2$ Heisenberg models on the maple-leaf lattice, our classification thus sets the stage for future works aimed at characterizing their precise microscopic nature. The Gutzwiller projected static structure factors for the different variational states could be compared to those obtained from unconstrained numerical approaches to narrow down and identify promising candidate ground states. Subsequently, it would be worthwhile to perform a variational Monte Carlo study towards optimizing the corresponding Gutzwiller projected wave functions and assess the energetic competitiveness of the $U(1)$ and $\mathds{Z}_{2}$ states for Hamiltonian parameter regimes displaying QSL ground states. The evidence of a $U(1)$ Dirac spin liquid ground state on the triangular lattice~\cite{Iqbal-2016,Hu-2019} and on its $1/4$-site depleted version, the kagome lattice~\cite{Iqbal-2013,He-2017}, poses the interesting question concerning the potential stability of the Dirac state under a periodic site depletion. Viewed from this perspective, it would be interesting to gauge its stability on the maple-leaf lattice which is an intermediate depletion density, being a $1/5$-site depletion of the triangular lattice. An alternate treatment of these \textit{Ans\"atze} would be their analysis within the pseudo-fermion functional renormalization group framework~\cite{Mueller-2024} by using the low-energy effective vertex functions (instead of the bare couplings) within a self-consistent Fock-like mean-field scheme to compute low-energy theories for emergent
spinon excitations~\cite{Hering-2019,Hering-2022}. Within the parameter space of nearest-neighbor couplings, a QSL has been located between magnetic and dimer orders, which fuels the speculation of its possible origin from a proximate deconfined quantum critical point, and the scenario of a gapless spin liquid as a plausible candidate. Furthermore, since Ref.~\cite{Gembe-2024} reports nonmagnetic behavior arising from quantum melting of noncoplanar orders in a $S=1/2$ $J_{1}$-$J_{2}$-$J_{3}$ Heisenberg model, it would be important to extend the current analysis to classify chiral spin liquids. Given the recent reporting of QSL behavior in a model featuring mixed ferro- and antiferromagnetic couplings~\cite{Ghosh-2024}, the incorporation of symmetry allowed triplet fields in the projected wave functions would prove essential to accurately capture the ground state behavior. Finally, it would be interesting to identify the respective parent QSLs whose potential instabilities yield the plethora of dimer orders that have been reported in the generalized parameter space of the nearest-neighbor Heisenberg model. 

\section{Acknowledgments}
Y.I. thanks S. Bhattacharjee, M. Gemb\'e, P. Ghosh, L. Gresista, C. Hickey, T. Müller, J. Naumann, K. Penc, R. Samajdar, H.-J. Schmidt, P. Schmoll, S. Trebst, and A. Wietek for helpful discussions and collaboration on related projects. J.S. received financial support from the Theory of Quantum Matter Unit of the Okinawa Institute of Science and Technology Graduate University (OIST). The work of Y.I. was performed, in part, at the Aspen Center for Physics, which is supported by National Science Foundation Grant No. PHY-2210452. The participation of Y.I. at the Aspen Center for Physics was supported by the Simons Foundation. The research of Y.I. was carried out, in part, at the Kavli Institute for Theoretical Physics in Santa Barbara during the ``A New Spin on Quantum Magnets" program in summer 2023, supported by the National Science Foundation under Grant No.~NSF PHY-1748958. Y.I. acknowledges support from the ICTP through the Associates Programme and from the Simons Foundation through Grant No. 284558FY19, IIT Madras through the Institute of Eminence (IoE) program for establishing QuCenDiEM (Project No. SP22231244CPETWOQCDHOC), and the International Centre for Theoretical Sciences (ICTS), Bengaluru, India during a visit for participating in the program Frustrated Metals and Insulators (Code No. ICTS/frumi2022/9). Y.I. acknowledges the use of the computing resources at HPCE, IIT Madras. The work in W\"urzburg was supported by DFG Grant No. 258499086-SFB 1170 and the W\"urzburg-Dresden Cluster of Excellence on Complexity and Topology in Quantum Matter, Grant No. 390858490-EXC 2147. F. F. and R. T. thank IIT Madras for a Visiting Researcher position under the IoE program which facilitated the completion of this work. F.F. acknowledges support by the Deutsche Forschungsgemeinschaft (DFG, German Research Foundation) 
for funding through TRR 288 -- 422213477.

\appendix

\section{Generic gauge conditions}
\label{app:genric_gauge_con}
The algebraic conditions Eqs.~\eqref{eq:translations}\textendash\eqref{eq:spacetime_comm} written in terms of PSG representations yield

\begin{align}
& G_{T_1}(x,y,u) G_{T_{2}}(x-1,y,u)\notag\\
&G^{-1}_{T_1}(x,y-1,u)G^{-1}_{T_{2}}(x,y,u)=e^{\dot\iota\theta\tau^3}/\eta\tau^0  \label{eq:gauge_translations} \\
& G_\mathcal{I}(x,y,u)G_\mathcal{I}(-x,-y,\mathcal{I}(u)) =e^{\dot\iota\theta_\mathcal{I}\tau^3}/ \eta_{\mathcal{I}}\tau^0 \label{eq:gauge_sigma}\\
& G^{-1}_{T_1}(x+1,y,u) G_{\mathcal{I}}(x+1,y,u)\notag\\
&G^{-1}_{T_1}(-x,-y,\mathcal{I}(u))G^{-1}_{\mathcal{I}}(x,y,u)=e^{\dot\iota\Tilde{\theta}_{\mathcal{I}}\tau^3}/\eta_{\mathcal{I}_x}\tau^0  \label{eq:gauge_sigma_T1} \\
& G^{-1}_{T_2}(x,y+1,u) G_{\mathcal{I}}(x,y+1,u)\notag\\
&G^{-1}_{T_2}(-x,-y,\mathcal{I}(u))G^{-1}_{\mathcal{I}}(x,y,u)=e^{\dot\iota\theta_{\mathcal{I}_y}\tau^3}/\eta_{\mathcal{I}_y}\tau^0  \label{eq:gauge_sigma_T2} \\
& G_R(x,y,u)G_R(y-x,-x,R^2(u))\notag\\
&G_R(-y,x-y,R(u))=e^{\dot\iota\theta_{R}\tau^3}/\eta_R\tau^0  \label{eq:gauge_R}\\
& G^{-1}_R(-y,x-y,R(u))G_{T_1}(-y,x-y,R(u))\notag\\
&G_R(-y-1,x-y,R(u))\notag\\&G_{T_1}(x+1,y+1,u)G_{T_2}(x,y+1,u)=e^{\dot\iota\theta_{R_x}\tau^3}/\eta_{R_x}\tau^0  \label{eq:gauge_R_T1}\\
& G^{-1}_R(-y,x-y,R(u))G^{-1}_{T_2}(-y,x-y+1,R(u))\notag\\
&G_R(-y,x-y+1,R(u))G_{T_1}(x+1,y,u)=e^{\dot\iota\theta_{R_y}\tau^3}/\eta_{R_y}\tau^0  \label{eq:gauge_R_T2}\\
& G^{-1}_R(-y,x-y,R(u))G_{\mathcal{I}}(-y,x-y,R(u))\notag\\
&G_R(y,y-x,R\mathcal{I}(u))G_{\mathcal{I}}(-x.-y,\mathcal{I}(u))=e^{\dot\iota\theta_{R\mathcal{I}}\tau^3}/\eta_{R\mathcal{I}}\tau^0  \label{eq:gauge_sigma_R}\\
&G_\mathcal{T}(x,y,u)G_\mathcal{O}(x,y,u)\notag\\&G^{-1}_\mathcal{T}(O^{-1}(x,y,u))G^{-1}_\mathcal{O}(x,y,u)=e^{\dot\iota\theta_{\mathcal{T}\mathcal{O}}\tau^3}/\eta_{\mathcal{T}\mathcal{O}}\tau^0\label{eq:gauge_time_O}\\
&[G_\mathcal{T}(x,y,u)]^2=e^{\dot\iota\theta_{\mathcal{T}}\tau^3}/\eta_{\mathcal{T}}\tau^0  \label{eq:gauge_time},
\end{align}
where on the right-hand side of the equations the entry corresponds to the $U(1)$ extension and the second entry denotes the $\mathds{Z}_2$ case.

\section{$U(1)$ PSG}
\label{app:u1_psg_derivation}
The canonical form of a $U(1)$ \textit{Ansatz} is given by,
\begin{equation}
    \label{eq:canonical_u1}
    u_{ij}=\dot\iota\text{Im}\chi_{ij} \tau^0 + \text{Re}\chi_{ij} \tau^3
\end{equation}
Correspondingly, the loop operators take the form $P_C=u_{ij}u_{jk}\ldots u_{li}=e^{\dot\iota\xi\tau^3}\equiv g_3(\xi)$. The structure of the gauge transformation which keeps the canonical form intact is
\begin{equation}
    \label{eq:canonical_u1_gauge structrure}
    G_\mathcal{O}(x,y,u)=g_3(\phi_\mathcal{O}(x,y,u))(\dot\iota\tau^1)^{w_\mathcal{O}},
\end{equation}
where $w_\mathcal{O}$ can take values $0,1$ and $\mathcal{O}\in\{T_1,T_2,R,\mathcal{I},\mathcal{T}\}$.

\subsection{Lattice Symmetries}
For $\mathcal{O}\in\{T_1,T_2\}$, there are three cases (i): $(w_{T_1},w_{T_2})=(0,0)$, (ii): $(w_{T_1},w_{T_2})=(1,0)$ and (iii): $(w_{T_1},w_{T_2})=(1,1)$. The cases (ii) and (iii) can not satisfy Eq.~\eqref{eq:gauge_R_T2} and Eq.~\eqref{eq:gauge_R_T1}, respectively. Therefore, we need to consider only case (i), i.e., $w_{T_1}=w_{T_2}=0$. Using the local gauge freedom one can choose
\begin{equation}
    \phi_{T_1}(x,0,u)=\phi_{T_2}(x,y,u)=0.
\end{equation}
Using this, Eq.~\eqref{eq:gauge_translations} gives
\begin{equation}\label{eq:tran_1_u1}
 G_{T_1}=g_3(y\theta),\; G_{T_2}=\tau^0.
\end{equation}
Let us now define $\Delta_i\phi_{\mathcal{O}}(x,y,u)=\phi_{\mathcal{O}}(x,y,u)-\phi_{\mathcal{O}}[T^{-1}_i(x,y,u)]$. With this definition and Eq.~\eqref{eq:tran_1_u1} we can recast Eq.~\eqref{eq:gauge_sigma_T1} and Eq.~\eqref{eq:gauge_sigma_T2} as
\begin{equation}\label{eq:sigma_tran_u}
	\left.\begin{aligned}
&\Delta_1\phi_\mathcal{I}(x,y,u)=\Tilde{\theta}_{\mathcal{I}}+(1-(-1)^{w_{\mathcal{I}}})y\theta,\\
&\Delta_2\phi_\mathcal{I}(x,y,u)=\theta_{\mathcal{I}_y}.
\end{aligned}\right.
\end{equation}
Generally, all solutions must obey the following consistency relation
\begin{equation}
\label{eq:consistent}
\left.\begin{aligned}
\Delta_1\phi_\mathcal{O}(x,y,u)&+\Delta_{2}\phi_{\mathcal{O}}[T^{-1}_1(x,y,u)]\\
&=\Delta_{2}\phi_\mathcal{O}(x,y,u)+\Delta_1\phi_{\mathcal{O}}[T^{-1}_{2}(x,y,u)].
\end{aligned}\right.
\end{equation}
For $\mathcal{O} =\mathcal{I}$ and insertion of Eq.~\eqref{eq:sigma_tran_u} in the above relation yields
\begin{equation}\label{eq:sigma_consistent}
    (1-(-1)^{w_{\mathcal{I}}})\theta=0,
\end{equation}
which means for $w_\mathcal{I}=1$ we have $2\theta=0$. Note that there is no constraint on $\theta$ for $w_\mathcal{I}=0$. Substituting this back in Eq.~\eqref{eq:sigma_tran_u}, we obtain the following solution for $\phi_\mathcal{I}$
\begin{equation}\label{eq:sigma_sol_u_1}
   \phi_\mathcal{I}(x,y,u)=x\Tilde{\theta}_{\mathcal{I}}+y\theta_{\mathcal{I}_y}+\rho_{\mathcal{I}}(u)
\end{equation}
where $\rho_{\mathcal{I}}(u)=\phi_\mathcal{I}(0,0,u)$. Using this solution~\eqref{eq:sigma_sol_u_1} in Eq.~\eqref{eq:gauge_sigma} gives
\begin{equation}
\label{eq:cyclic_sigma_u}
\left.\begin{aligned}
&\rho_{\mathcal{I}}(u)+(-1)^{w_\mathcal{I}}\rho_{\mathcal{I}}(\mathcal{I}(u))=\theta_\mathcal{I},\\
&(1+(-1)^{w_\mathcal{I}})\Tilde{\theta}_{\mathcal{I}}=(1+(-1)^{w_\mathcal{I}})\theta_{\mathcal{I}_y}=0,
\end{aligned}\right.
\end{equation}
which implies $2\Tilde{\theta}_{\mathcal{I}}=2\theta_{\mathcal{I}_y}=0$ for $w_{\mathcal{I}}=1$, while there is no constraint on $\theta$ for $w_{\mathcal{I}}=0$.
From Eq.~\eqref{eq:gauge_R_T1} and Eq.~\eqref{eq:gauge_R_T2} one obtains relations similar to Eq.~\eqref{eq:sigma_tran_u}
\begin{equation}\label{eq:R_tran_u}
	\left.\begin{aligned}
&\Delta_1\phi_R(x,y,u)=y\theta+(-1)^{w_R}((1-x)\theta-\theta_{R_x}),\\
&\Delta_2\phi_R(x,y,u)=(-1)^{w_R}(x\theta+\theta_{R_y}).
\end{aligned}\right.
\end{equation}
The consistency condition~\eqref{eq:consistent} for $\mathcal{O}=R$ imposes the following restriction on $\theta$
\begin{equation}\label{eq:R_consistent}
    (1-(-1)^{w_R})\theta=0,
\end{equation}
which means for $w_R=1$, $2\theta=0$, while there is again no constraint on $\theta$ for $w_R=0$. Substituting this in Eq.~\eqref{eq:R_tran_u}, yields the following solution for $\phi_R$
\begin{equation}\label{eq:R_sol_u_1}
\left.\begin{aligned}
   \phi_R(x,y,u)&=x\left[y-\frac{(-1)^{w_R}}{2}(x-1)\right]\theta\\
   &-(-1)^{w_R}(x\theta_{R_x}-y\theta_{R_y})+\rho_{R}(u).
\end{aligned}\right.
\end{equation}
Notice that $w_R=1$ does not satisfy Eq.~\eqref{eq:gauge_R}. Furthermore, Eq.~\eqref{eq:gauge_R} gives the following condition for $w_R=0$
\begin{equation}\label{eq:cyclic_R_u}
\rho_{R}(u)+\rho_{R}(R^2(u))+\rho_{R}(R(u))=\theta_R.
\end{equation}
Under a local gauge transformation $W(x,y,u)$, the projective representation $G_\mathcal{O}$ transforms as $G_\mathcal{O}(x,y,u)\rightarrow W^\dagger(x,y,u)G_\mathcal{O}(x,y,u)W[\mathcal{O}^{-1}(x,y,u)]$. A local gauge transformation of the form
\begin{equation}
\label{eq:local}
    W(x,y,u)=g_3(x\xi_x+y\xi_y)
\end{equation}
does not change the structure of the $G_{T_i}$ besides a negligible global phase which has no consequence on the \textit{Ans\"atze}. It will, however, modify the phases $\theta_{R_x}$, $\theta_{R_y}$, $\Tilde{\theta}_{\mathcal{I}}$ and $\theta_{\mathcal{I}_y}$ such that we can use a suitable choice to set
\begin{equation}\label{eq:global_fixing_u1}
    \theta_{R_x}=\theta_{R_y}=0.
\end{equation}
Then Eq.~\eqref{eq:gauge_sigma_R} yields
\begin{align}
& \text{for $w_{\mathcal{I}}=0$: } \notag\\
&\Tilde{\theta}_{\mathcal{I}}=\theta_{\mathcal{I}y}=\frac{1}{3}(\theta+2\pi p)\label{eq:wi0_ti},\;p=0,1,2. \\
& \text{and for $w_{\mathcal{I}}=1$: }\notag\\ 
&\Tilde{\theta}_{\mathcal{I}}=\theta_{\mathcal{I}y}=\theta, \label{eq:wi1_ti}\\
& \rho_{\mathcal{I}}(u)-\rho_{R}(u)+(-1)^{w_\mathcal{I}}(\rho_{\mathcal{I}}(\mathcal{I}R^{-1}(u))+\rho_{R}(\mathcal{I}(u)))=\theta_{R\mathcal{I}}. \label{eq:con_sigma_R}
\end{align}
Inserting Eq.~\eqref{eq:wi0_ti} and Eq.~\eqref{eq:wi1_ti}, the solution for $\phi_\mathcal{I}$ can be rewritten as
\begin{equation}\label{eq:sigma_sol_u_2}
   \phi_\mathcal{I}(x,y,u)=\frac{1}{3}(\theta+2\pi p)(x+y)\delta_{w_\mathcal{I},0}+\theta(x+y)\delta_{w_\mathcal{I},1}+\rho_{\mathcal{I}}(u).
\end{equation}
Furthermore, we are left with a sublattice-dependent gauge transformation of the form
\begin{equation}
    W(x,y,u)=g_3(\xi_u).
\end{equation}
Under such a transformation $\rho_{R,u}$ transforms for $u\in \left\lbrace 1,3,5 \right\rbrace$ as
\begin{equation}\label{eq:u1_fixing_1}
\left.\begin{aligned}
   &\Tilde{\rho}_{R}(1)=-\xi_1+\rho_{R}(1)+\xi_5,\\
   &\Tilde{\rho}_{R}(3)=-\xi_3+\rho_{R}(3)+\xi_1,\\
   &\Tilde{\rho}_{R}(5)=-\xi_5+\rho_{R}(5)+\xi_3.
\end{aligned}\right.
\end{equation}
We choose $\xi_1$, $\xi_3$ and $\xi_5$ such that $\Tilde{\rho}_{R}(1)=\Tilde{\rho}_{R}(3)=\Tilde{\rho}_{R}(5)=\rho_R$. This requires
\begin{equation}\label{eq:u1_fixing_2}
\left.\begin{aligned}
   &\xi_1=\rho_{R}(1)+\xi_5-\rho_R,\\
   &\xi_3=\rho_{R}(1)+\rho_{R}(3)+\xi_5-2\rho_R,\\
   &3\rho_R=\rho_{R}(1)+\rho_{R}(3)+\rho_{R}(5).\\
\end{aligned}\right.
\end{equation}
Substituting Eq.~\eqref{eq:cyclic_R_u} in the last condition of the above equations, gives $3\rho_R=\theta_R$ which implies
\begin{equation}\label{eq:u1_fixing_3}
\rho_R=\frac{1}{3}(\theta_R+2\pi p_R),\text{ with } p_R=0,1,2.
\end{equation}
One can fix $\Tilde{\rho}_{R}(2)=\Tilde{\rho}_{R}(4)=\Tilde{\rho}_{R}(6)=\rho_R$ similarly by the following choices
\begin{equation}\label{eq:u1_fixing_4}
\left.\begin{aligned}
   &\xi_2=\rho_{R}(2)+\xi_6-\rho_R,\\
   &\xi_4=\rho_{R}(4)+\rho_{R}(2)+\xi_6-2\rho_R.
\end{aligned}\right.
\end{equation}
Note that we are still left with unfixed $\xi_5$ and $\xi_6$. With a suitable gauge choice, one can set $\rho_{\mathcal{I}}(1)=0$. Let us fix the other $\rho_{\mathcal{I}}(u)$ separately for $w_\mathcal{I}=0$ and $w_\mathcal{I}=1$:

\subsubsection{$w_\mathcal{I}=0$}
In this case, using Eq.~\eqref{eq:u1_fixing_3}, the relations given in Eq.~\eqref{eq:cyclic_sigma_u} and Eq.~\eqref{eq:con_sigma_R} are rewritten as
\begin{equation}\label{eq:u1_sigma_cyclic_R_sigma_wi0}
\left.\begin{aligned}
   &\rho_{\mathcal{I}}(u)+\rho_{\mathcal{I}}(\mathcal{I}(u))=\theta_{\mathcal{I}},\\
   &\rho_{\mathcal{I}}(u)+\rho_{\mathcal{I}}(\mathcal{I}R^{-1}(u))=\theta_{R\mathcal{I}}.
\end{aligned}\right.
\end{equation}
These two relations lead to
\begin{equation}\label{eq:u1_fixing_5}
\left.\begin{aligned}
   &\rho_{\mathcal{I}}(1)=\rho_{\mathcal{I}}(3)=\rho_{\mathcal{I}}(5)=0,\;\\   &\rho_{\mathcal{I}}(2)=\rho_{\mathcal{I}}(4)=\rho_{\mathcal{I}}(6)=\theta_\mathcal{I}.
\end{aligned}\right.
\end{equation}

\subsubsection{$w_\mathcal{I}=1$}
In this case, using Eq.~\eqref{eq:u1_fixing_3}, the relations given by Eq.~\eqref{eq:cyclic_sigma_u} and Eq.~\eqref{eq:con_sigma_R} are rewritten as
\begin{equation}\label{eq:u1_sigma_cyclic_R_sigma_wi1}
\left.\begin{aligned}
   &\rho_{\mathcal{I}}(u)-\rho_{\mathcal{I}}(\mathcal{I}(u))=\theta_{\mathcal{I}},\\
   &\rho_{\mathcal{I}}(u)-\rho_{\mathcal{I}}(\mathcal{I}R^{-1}(u))=\theta_{R\mathcal{I}}+2\rho_R.\\
\end{aligned}\right.
\end{equation}
leading to
\begin{equation}
    \rho_{\mathcal{I}}(u)=\frac{p_{\mathcal{I}}(u-1)\pi}{3},\;\text{with }p_\mathcal{I}=1,2,3.
\end{equation}
This completes the gauge fixing procedure of the lattice group operations.

\subsection{Time-reversal}
We proceed in finding the PSG solutions for time-reversal symmetry. Using Eq.~\eqref{eq:gauge_time_O} with $\mathcal{O} \in \left\lbrace T_1,T_2 \right\rbrace$, one gets
\begin{equation}
\label{eq:time_u1_sol_1}
\left.\begin{aligned}
&\Delta_1\phi_{\mathcal{T}}(x,y,u)=\theta_{\mathcal{T}_x}+[1-(-1)^{w_{\mathcal{T}}}]y\theta\\
&\Delta_2\phi_{\mathcal{T}}(x,y,u)=\theta_{\mathcal{T}_y}.
\end{aligned}\right.
\end{equation}
The consistency condition~\eqref{eq:consistent} for $\mathcal{O}=\mathcal{T}$ gives
\begin{equation}
    [1-(-1)^{w_{\mathcal{T}}}]\theta=0,
\end{equation}
which implies for $w_{\mathcal{T}}=1$ that $2\theta=0$. A solution for $G_\mathcal{T}$ can be obtained from Eq.~\eqref{eq:time_u1_sol_1} as
\begin{equation}\label{eq:time_u1_sol_2}
    \phi_\mathcal{T}(x,y,u)=x\theta_{\mathcal{T}_x}+y\theta_{\mathcal{T}_y}+\rho_{\mathcal{T}}(u)\, .
\end{equation}
Let us consider the remaining conditions for $w_\mathcal{T}=0$ and $w_\mathcal{T}=1$ separately:

\subsubsection{$w_\mathcal{T}=0$}
In this case, Eq.~\eqref{eq:gauge_time} yields
\begin{align}
\label{eq:utime_fixing_1}
&2\theta_{\mathcal{T}_x}=2\theta_{\mathcal{T}_y}=0,\quad \rho_{\mathcal{T}}(u)=\frac{\theta_{\mathcal{T}}}{2}+\pi n_{\mathcal{T}}(u),\\
&\text{with }n_{\mathcal{T}}(u)=0,1. \notag    
\end{align}
From Eq~\eqref{eq:gauge_time_O} with $\mathcal{O}= R$ we obtain
\begin{align}
& \theta_{\mathcal{T}_x}=\theta_{\mathcal{T}_y},\quad  3\theta_{\mathcal{T}_x}=0 \label{eq:utime_fixing_2a}, \\
& \rho_{\mathcal{T}}(u)-\rho_{\mathcal{T}}(R^{-1}(u))=\theta_{\mathcal{T}R}. \label{eq:utime_fixing_2b}
\end{align}
From Eq.~\eqref{eq:utime_fixing_1} and Eq.~\eqref{eq:utime_fixing_2a} it follows
\begin{equation}\label{eq:utime_fixing_3}    
\theta_{\mathcal{T}_x}=\theta_{\mathcal{T}_y}=0.
\end{equation}
Finally, Eq~\eqref{eq:gauge_time_O} with $\mathcal{O}=\mathcal{I}$ gives
\begin{equation}\label{eq:utime_fixing_4}    
\rho_{\mathcal{T}}(u)-(-1)^{w_\mathcal{I}}\rho_{\mathcal{T}}(\mathcal{I}(u))=\theta_{\mathcal{T}\mathcal{I}}.
\end{equation}
Using Eq.~\eqref{eq:utime_fixing_1}, Eq.~\eqref{eq:utime_fixing_2b} and Eq.~\eqref{eq:utime_fixing_4}, we can fix $\rho_{\mathcal{T}}(u)$ as 
\begin{align}\label{eq:utime_fixing_5} 
&\rho_{\mathcal{T}}(1)=\rho_{\mathcal{T}}(3)=\rho_{\mathcal{T}}(5)=0,\\
&\rho_{\mathcal{T}}(2)=\rho_{\mathcal{T}}(4)=\rho_{\mathcal{T}}(6)=\pi .  
\end{align}
If $w_\mathcal{I}=0$, one can set $\rho_{\mathcal{I}}(u)=\theta_\mathcal{I}/2$, i.e., independent of the sublattice $u$ by using a sublattice dependent gauge transformation of the form $W(x,y,u)=g_3(\theta_\mathcal{I}/2)\delta_{\text{mod}(u,2),0}$ without altering our previous results. As in the case of the lattice symmetries, a global phase has no impact and we can conveniently set $\rho_{\mathcal{I}}(u)=0$.

\subsubsection{$w_\mathcal{T}=1$}
Here, Eq.~\eqref{eq:gauge_time} does not yield any constraint. From Eq~\eqref{eq:gauge_time_O} with $\mathcal{O} = R$, and using the fact that $2\theta=0$ for $w_\mathcal{T}=1$, we obtain
\begin{align}
& \theta_{\mathcal{T}_x}=\theta_{\mathcal{T}_y},\; 3\theta_{\mathcal{T}_x}=0 \label{eq:utime_fixing_6a}, \\
& \rho_{\mathcal{T}}(u)-\rho_{\mathcal{T}}(R^{-1}(u))=\theta_{\mathcal{T}R}+2\rho_{R}(u). \label{eq:utime_fixing_6b}
\end{align}
Furthermore, Eq~\eqref{eq:gauge_time_O} with  $\mathcal{O}=\mathcal{I}$ for $w_\mathcal{I}=0$ yields
\begin{align}
& 2\theta_{\mathcal{T}_x}=\frac{2}{3}(\theta+2p\pi)\implies \theta_{\mathcal{T}_x}=\frac{1}{3}(\theta+2p\pi)+n\pi,\label{eq:utime_fixing_7a} \\
& \rho_{\mathcal{T}}(u)-\rho_{\mathcal{T}}(\mathcal{I}(u))=\theta_{\mathcal{T}\mathcal{I}}+2\rho_{\mathcal{I}}(u). \label{eq:utime_fixing_7b}
\end{align}
Eq.~\eqref{eq:utime_fixing_6a} and Eq.~\eqref{eq:utime_fixing_7a} require $\theta=n\pi$. Using Eq.~\eqref{eq:utime_fixing_6b} and Eq.~\eqref{eq:utime_fixing_7b}, we can fix $\rho_{\mathcal{T}}(u)$ for $w_\mathcal{I}=0$ as 
\begin{align}
& \rho_{\mathcal{T}}(u) \in \{0,\theta_{\mathcal{TI}}-\theta'_{\mathcal{T}R},\theta'_{\mathcal{T}R},\theta_{\mathcal{TI}},-\theta'_{\mathcal{T}R},\theta_{\mathcal{TI}}+\theta'_{\mathcal{T}R}\} \label{eq:utime_fixing_9a}, \\
& \theta'_{\mathcal{T}R}=\theta'_{\mathcal{T}R}+2\rho_R=\frac{2p'_{\mathcal{T}R}\pi}{3} \label{eq:utime_fixing_9b},\;p'_{\mathcal{T}R}=0,1,2 \\
& 2\theta_{\mathcal{TI}}+2\theta_{\mathcal{I}}=0\implies\theta_{\mathcal{TI}}=-\theta_{\mathcal{I}}+n_{\mathcal{I}}\pi,\;n_{\mathcal{I}}=0,1 . \label{eq:utime_fixing_9c}
\end{align}
Eq.~\eqref{eq:gauge_time_O} with  $\mathcal{O}=\mathcal{I}$ for $w_\mathcal{I}=1$ yields
\begin{equation}\label{eq:utime_fixing_10} \rho_{\mathcal{T}}(u)+\rho_{\mathcal{T}}(\mathcal{I}(u))=\theta_{\mathcal{T}\mathcal{I}}+2\rho_{\mathcal{I}}(u)=\theta_{\mathcal{T}\mathcal{I}}+\frac{2p_{\mathcal{I}}(u-1)\pi}{3}. 
\end{equation}
This together with Eq.~\eqref{eq:utime_fixing_7b} gives
\begin{equation}\label{eq:utime_fixing_11}
\rho_{\mathcal{T}}(u)\in \{\rho_1,\theta_{\mathcal{TI}}-\rho_1,\rho_1,\theta_{\mathcal{TI}}-\rho_1,-\rho_1,\theta_{\mathcal{TI}}-\rho_1\}  
\end{equation}
Notice that the $w_\mathcal{T}=1$ solution for $G_\mathcal{T}$ has a spatial dependence. To remove this, similar to the $w_\mathcal{T}=0$ case, we use a gauge transformation of the form $W(x,y,u)=g_3(\xi(x,y,u))$ with
\begin{equation}\label{eq:local_gauge_u1_time}
    \xi(x,y,u)=\frac{\theta_{\mathcal{T}_x}(x+y)}{2},
\end{equation}
which yields the following
\begin{align}
&\Tilde{\phi}_{\mathcal{T}}(x,y,u)=\rho_{\mathcal{T}}(u) \label{eq:utime_fixing_11a}, \\
&\Tilde{\phi}_{R}(x,y,u)=[xy-\frac{1}{2}x(x-1)]n\pi+\rho_{R}(u)+n_{\mathcal{T}_x}\pi x\label{eq:utime_fixing_11b}, \\
&\Tilde{\phi}_{\mathcal{I}}(x,y,u)=n\pi(x+y)+\rho_{\mathcal{I}}(u), \quad  n_{\mathcal{T}_x}=0,1. \label{eq:utime_fixing_11c}
\end{align}
This can be set to zero using a gauge transformation of the form $W(x,y,u)=g_3((x+y)n_{\mathcal{T}_x}\pi)$.
Furthermore, for $w_\mathcal{T}=1$ we can choose a sublattice-dependent gauge transformation of the form $W(x,y,u)=g_3(\xi(u))$ with
\begin{equation}\label{eq:local_gauge_sub_u1_time}
    \xi(u)=\frac{\rho_{\mathcal{T}}(u)}{2},
\end{equation}
so that $\rho_{\mathcal{T}}(u)=0$. The advantage of such a choice is that the mean-field amplitudes contain only real hopping values. For $w_\mathcal{I}=0$, in the new gauge we obtain
\begin{align}
&\rho_{\mathcal{T}}(u)=0,\;\rho_{R}(u)=\rho_R-\frac{\theta'_{\mathcal{T}R}}{2} \label{eq:u_fixing_13a}, \\
&\rho_{\mathcal{I}}(u)=\theta_\mathcal{I}\delta_{\text{mod}(u,2),0}-(-1)^{\text{mod}(u,2)}\frac{\theta_{\mathcal{TI}}}{2}. \label{eq:u_fixing_13b}
\end{align}
Using Eq.~\eqref{eq:utime_fixing_9c} one can rewrite Eq.~\eqref{eq:u_fixing_13b} as
\begin{equation}
\left.\begin{aligned}
 \rho_{\mathcal{I}}(u)&=\left(\frac{-\theta_\mathcal{I}+n_{\mathcal{I}}\pi}{2}\right)\delta_{\text{mod}(u,2),1}\\
 &+\left(\frac{3\theta_\mathcal{I}-n_{\mathcal{I}}\pi}{2}\right)\delta_{\text{mod}(u,2),0}.  \\
 \end{aligned}\right.
\end{equation}
For $w_\mathcal{I}=1$, in the this gauge we obtain,
\begin{align}
&\rho_{\mathcal{T}}(u)=0,\;\rho_{R}(u)=\rho_R \label{eq:u_fixing_14a}, \\
&\rho_{\mathcal{I}}(u)=\frac{p_{\mathcal{I}}(u-1)\pi}{3}+\frac{\theta_{\mathcal{TI}}}{2}. \label{eq:u_fixing_14b}
\end{align}
As the global phases do not have any impact on the \textit{Ans\"atze} we can discard them. For example, in Eq.~\eqref{eq:u_fixing_13a} and Eq.~\eqref{eq:u_fixing_14b} we could set $\rho_R-\frac{\theta'_{\mathcal{T}R}}{2}=0$ and $\frac{\theta_{\mathcal{TI}}}{2}=0$, respectively. As a summary, all the gauge inequivalent choices are listed in Table~\ref{table:u1_psg}.

\section{$\mathbf{\mathds{Z}_2}$ PSG}
\label{app:z2_psg_derivation}
\subsection{Lattice symmetry}
Using the local gauge redundancy, the relation Eq.~\eqref{eq:gauge_translations} leads to the solution for the projective gauge matrices for $\mathcal{O}\in\{T_1,T_2\}$ as follows
\begin{equation}
	\label{eq:tran_1_z2}
	\left.\begin{aligned}
		&G_{T_1}(x,y,u)=\eta^{y}\tau^0,\quad G_{T_2}(x,y,u)=\tau^0
	\end{aligned}\right.
\end{equation}
Using Eq.~\eqref{eq:gauge_sigma_T1} and Eq.~\eqref{eq:gauge_sigma_T2} gives
\begin{equation}
	\label{eq:sigma_sol_1}		
 G_\mathcal{I}(x,y,u)=\eta^x_{\mathcal{I}_x}\eta^y_{\mathcal{I}_y}g_\mathcal{I}(u).
\end{equation}
The cyclic condition given by Eq.~\eqref{eq:gauge_sigma} for $\mathcal{I}$ gives
\begin{equation}\label{eq:sigma_constraint}		
 g_\mathcal{I}(u)g_\mathcal{I}(\mathcal{I}(u))=\eta_\mathcal{I}\tau^0.
\end{equation}
After coordinate transformation $(x,y,u)\rightarrow R^{-1}(x,y,u)$ Eq.~\eqref{eq:gauge_R_T1} and Eq.~\eqref{eq:gauge_R_T2} can be written as
\begin{equation}
	\left.\begin{aligned}
		&G_R(x,y,u)=\eta_{R_x}\eta^{y-x+1}G_R(x-1,y,u),\\
		&G_R(x,y,u)=\eta_{R_y}\eta^{x}G_R(x,y-1,u).
	\end{aligned}\right.
\end{equation}
These relations yield the solution for $G_R$ as follows
\begin{equation}\label{eq:R_sol_1}		
G_R(x,y,u)=\eta^x_{R_x}\eta^y_{R_y}\eta^{xy-\frac{1}{2}x(x-1)}g_R(u).
\end{equation}
The cyclic condition Eq.~\eqref{eq:gauge_R} for $R$ yields
\begin{equation}\label{eq:R_constraint}		
 g_R(u)g_R(R^2(u))g_R(R(u))=\eta_R\tau^0.
\end{equation}
Exploiting Eq.~\eqref{eq:gauge_sigma_R} leads us to the following constraints
\begin{equation}\label{eq:R_sigma_constraint}
	\left.\begin{aligned}
    &\eta_{\mathcal{I}_x}=\eta_{\mathcal{I}_y}=\eta,\\
    &g_\mathcal{I}(R(u))g_R(R\mathcal{I}(u))g_\mathcal{I}(\mathcal{I}(u))=\eta_{R\mathcal{I}}g_R(R(u))\\
    &\implies g_\mathcal{I}(u)g_R(\mathcal{I}(u))g_\mathcal{I}(\mathcal{I} R^{-1}(u))=\eta_{R\mathcal{I}}g_R(u).
	\end{aligned}\right.
\end{equation}
Further simplification can be obtained if we consider the gauge
\begin{equation}\label{eq:local_z2}
	W(x,y,u)=\eta^{x}_x\eta^{y}_y\tau^0.
\end{equation}
We find that the above transformation does not change the structure of the translational gauges except for a global sign modification. This sign can yet be absorbed by a redefinition including these modified signs. It can further be seen that the gauge transformation Eq.~\eqref{eq:local_z2} modulates $G_R$ as
\begin{equation}
\left.\begin{aligned}
\Tilde{G}_R(x,y,u)&=W^\dagger(x,y,u)G_R(x,y,u)W_{R^{-1}(x,y,u)}\\
&=(\eta_x\eta_y\eta_{R_x})^x(\eta_y\eta_{R_y})^y\eta^{xy-\frac{1}{2}x(x-1)}g_R(u).\\
\end{aligned}\right.    
\end{equation}
Setting $\eta_y=\eta_{R_y}$ and $\eta_x=\eta_y\eta_{R_x}$, the result becomes
\begin{equation}\label{eq:R_sol_2}		
\Tilde{G}_R(x,y,u)=\eta^{xy-\frac{1}{2}x(x-1)}g_R(u).
\end{equation}
Hereafter, we shall omit the tilde symbol. The representation $G_\mathcal{I}$ does not get modified by the gauge transformation. We can also exploit a sublattice dependent gauge transformation $W(x,y,u)=W(u)$ for further fixing the $g$ matrices as follows
\begin{equation}\label{eq:g_fixing_1}
g_R(u)=\eta_R\tau^0.
\end{equation}
In the above, we have used Eq.~\eqref{eq:R_constraint}. The remaining sign factor $\eta_R$ which is global can be neglected. The $g_\mathcal{I}$-matrices can be fixed using the above equation and Eq.~\eqref{eq:sigma_constraint} and Eq.~\eqref{eq:R_sigma_constraint} as follows.
\begin{align}\label{eq:g_fixing_2}
& g_\mathcal{I}(1)=g_\mathcal{I}(3)=g_\mathcal{I}(5)=\tau^0,\\
& g_\mathcal{I}(2)=g_\mathcal{I}(4)=g_\mathcal{I}(6)=\eta_\mathcal{I}\tau^0.
\end{align}
This completes the gauge fixing of the lattice symmetry operations in the $\mathds{Z}_2$ case.

\subsection{Time-reversal}
Here, we find a PSG representation for time-reversal symmetry. Using Eq.~\eqref{eq:gauge_time_O} for $\mathcal{O}\in \left\lbrace T_1, T_2 \right\rbrace$ the solution for $G_{\mathcal{T}}(x,y,u)$ can be written as
\begin{equation}\label{eq:T_sol_1}
	G_{\mathcal{T}}(x,y,u)=\eta^{x}_{\mathcal{T}_x}\eta^{y}_{\mathcal{T}_y}g_\mathcal{T}(u).
\end{equation} 
Eq.~\eqref{eq:gauge_time} leads to the following condition
\begin{equation}\label{eq:T_constraint}		
[g_\mathcal{T}(u)]^2=\eta_\mathcal{T}\tau^0.
\end{equation}
From Eq.~\eqref{eq:gauge_time_O} for $\mathcal{O}=\mathcal{I}$ and substituting Eq.~\eqref{eq:sigma_sol_1} and Eq.~\eqref{eq:g_fixing_2} we have
	\begin{align}\label{eq:constraint_time_sigma}
		&G_{\mathcal{T}}(x,y,u)G_{\mathcal{I}}(x,y,u) \; \notag \\
		&G^{-1}_{\mathcal{T}}(-x,-y,\mathcal{I}(u))G^{-1}_{\mathcal{I}}(x,y,u)=\eta_{\mathcal{T}\mathcal{I}}\tau^0 \; \notag \\
		&\implies g_\mathcal{T}(\mathcal{I}(u))=\eta_{\mathcal{T}\mathcal{I}}g_\mathcal{T}(u) \;.
	\end{align}
Similarly for $\mathcal{O}= R$ and substituting Eq.~\eqref{eq:R_sol_2} we find
	\begin{align}\label{eq:constraint_time_R}
		&G_{\mathcal{T}}(x,y,u)G_{R}(x,y,u) \notag \\
		&G^{-1}_{\mathcal{T}}(y-x,-x,R^{-1}(u))G^{-1}_{R}(x,y,u)=\eta_{\mathcal{T}R}\tau^0,\notag \\
  &\implies\eta^{y}_{\mathcal{T}_x}\eta^{y-x}_{\mathcal{T}_x}g_{\mathcal{T}}(u)g_{R}(u)g^{-1}_{\mathcal{T}}(R^{-1}(u))g^{-1}_{R}(u)=\eta_{\mathcal{T}R}\tau^0, \notag \\
		&\implies  \eta_{\mathcal{T}_x}=\eta_{\mathcal{T}_x}=1,\\
  &\text{and } g_\mathcal{T}(R^{-1}(u))=\eta_{\mathcal{T}R}g_\mathcal{T}(u). 
	\end{align}
Further simplification using Eq.~\eqref{eq:constraint_time_sigma} and Eq.~\eqref{eq:constraint_time_sigma} yields the concise solution for PSG representation of time-reversal symmetry
\begin{equation}\label{eq:sol_time_2}
G_{\mathcal{T}}(x,y,u)=\eta^{u+1}_{\mathcal{T}\mathcal{I}}g_{\mathcal{T}}.   
\end{equation}

\begin{figure*}
    \centering
    \includegraphics[scale=0.7]{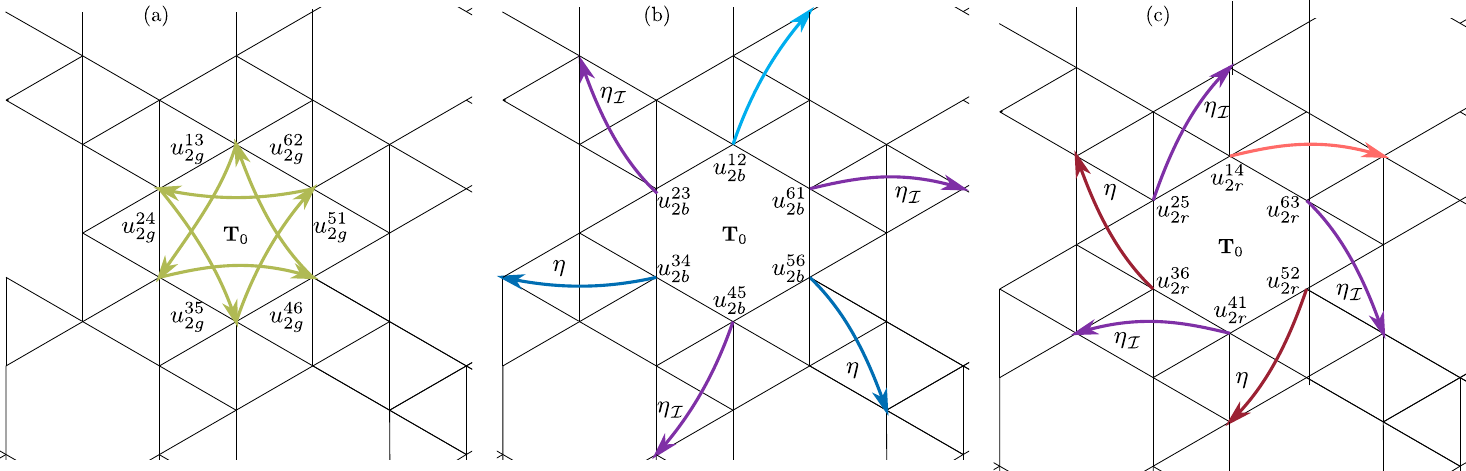}
    \caption{Symmetry inequivalent \textit{Ansatz} matrices $u^{13}_{2g}, u^{12}_{2b}$ and $u^{14}_{2r}$ [see Table~\ref{table:z2_ansatz} for the $\mathds{Z}_2$ states] for the second neighbor bonds transform as depicted in (a), (b), and (c), respectively. Different shadings of the colors encode the different phase factors $\eta, \eta_\mathcal{I}$ and their products.} 
    \label{fig:second_neighbors}
\end{figure*}

\begin{figure*}
    \centering
    \includegraphics[scale=0.7]{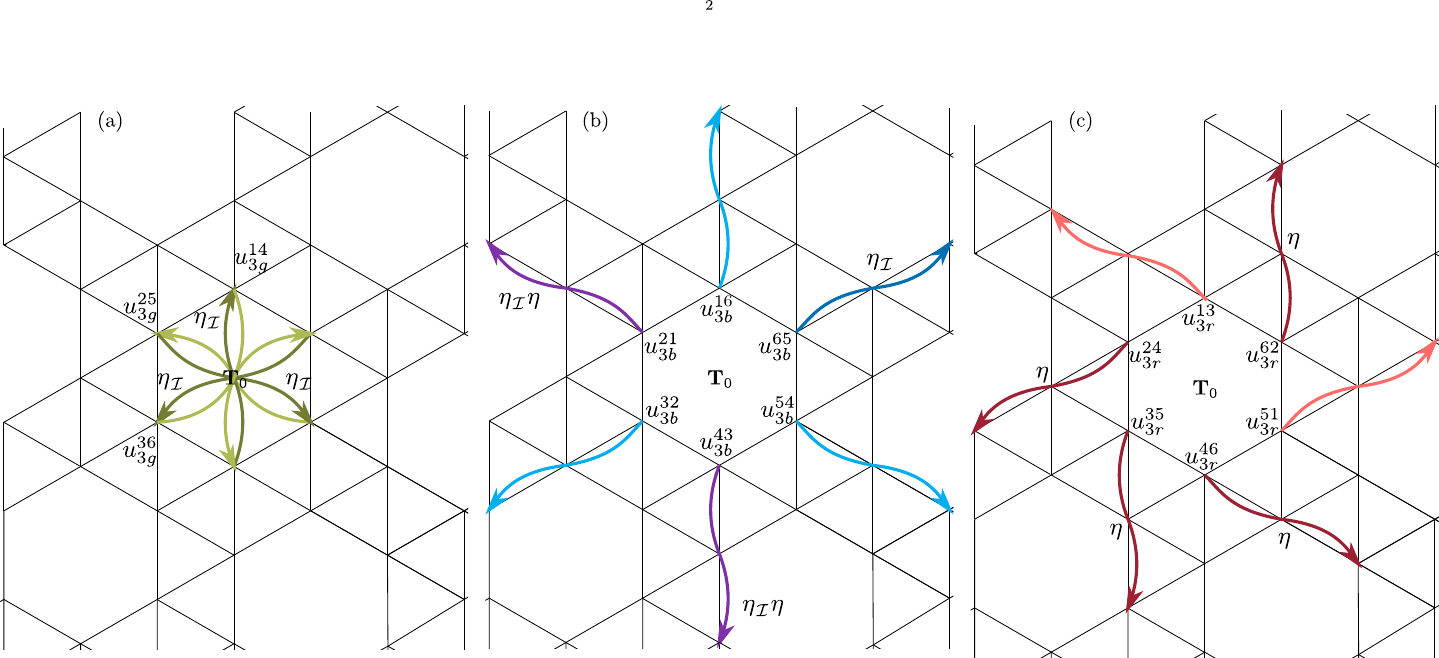}
    \caption{Symmetry inequivalent \textit{Ansatz} matrices $u^{14}_{3g}, u^{16}_{3b}$ and $u^{13}_{3r}$ [see Table~\ref{table:z2_ansatz} for the $\mathds{Z}_2$ states] for the third neighbor bonds transform as depicted in (a), (b), and (c),  respectively. Different shadings of the colors encode the different phase factors $\eta, \eta_\mathcal{I}$ and their products.} 
    \label{fig:third_neighbors}
\end{figure*}

\section{$\mathbf{\mathds{Z}_2}$ \textit{Ans\"atze} before time-reversal}
This appendix contains all different symmetry relations that are needed to construct up to third nearest neighbor mean-field Hamiltonians. The definitions of the different $u-$matrices are shown in Fig.~\ref{fig:cell_definitions} for the first neighbor bonds, and for the second and third neighbor bonds in Fig.~\ref{fig:second_neighbors} and Fig.~\ref{fig:third_neighbors}, respectively. For every range of neighbor bonds we find three symmetry inequivalent bonds that are colored green, red and blue. For each color we fix one \textit{Ansatz} whereas the allowed form varies for different PSGs and can be found in Table~\ref{table:z2_ansatz}. All the other are related by the underlying symmetries of the lattice.
\subsection{1NN}
\subsubsection{Green 1NN bonds ($u_{1g}$)}
For the green colored bonds we fix the \textit{Ansatz} matrix $u_{1g}^{12}$. Applying a chain of symmetry operations induces all symmetry related bonds  
\begin{equation}
	\left.\begin{aligned}
		&u^{12}_{1g}\xrightarrow[]{R}u^{34}_{1g}\xrightarrow[]{R}u^{56}_{1g}\xrightarrow[]{\mathcal{I} R}u^{45}_{1g}\xrightarrow[]{R}u^{61}_{1g}\xrightarrow[]{R}u^{23}_{1g}.
	\end{aligned}\right.   
\end{equation}
Using the PSG representations yields
\begin{equation}
	\left.\begin{aligned}
		&u^{12}_{1g}=u^{34}_{1g}=u^{56}_{1g}=\eta_\mathcal{I} u^{45}_{1g}=\eta_\mathcal{I} u^{61}_{1g}=\eta_\mathcal{I} u^{23}_{1g}.
	\end{aligned}\right.   
\end{equation}
\subsubsection{Blue 1NN bonds ($u_{1b}$)}
Similarly to the green bonds, operating with the chain of transformations yields for the blue bonds
\begin{equation}
	\left.\begin{aligned}
		&u^{14}_{1b}\xrightarrow[]{R}u^{36}_{1b}\xrightarrow[]{R}u^{52}_{1b}\xrightarrow[]{\mathcal{I} R}u^{25}_{1b}\xrightarrow[]{R}u^{41}_{1b}\xrightarrow[]{R}u^{63}_{1b},\\
  &u^{14}_{1b}\xrightarrow[]{T_2\mathcal{I}}(u^{14}_{1b})^\dagger.
	\end{aligned}\right.   
\end{equation}
This results in
\begin{equation}
	\left.\begin{aligned}
		&u^{14}_{1b}=u^{36}_{1b}=u^{52}_{1b}=\eta\eta_\mathcal{I} u^{25}_{1b}=\eta\eta_\mathcal{I} u^{41}_{1b}=\eta_\mathcal{I} u^{63}_{1b}\\
  &u^{14}_{1b}=\eta\eta_\mathcal{I}(u^{14}_{1b})^\dagger
	\end{aligned}\right.   
\end{equation}
\subsubsection{Red 1NN bonds ($u_{1r}$)}
For the red bonds we have
\begin{equation}
	\left.\begin{aligned}
		&u^{13}_{1r}\xrightarrow[]{R}u^{35}_{1r}\xrightarrow[]{R}u^{51}_{1r}\xrightarrow[]{\mathcal{I} R}u^{46}_{1r}\xrightarrow[]{R}u^{62}_{1r}\xrightarrow[]{R}u^{24}_{1r}\\
		&u^{15}_{1r}\xrightarrow[]{R}u^{31}_{1r}\xrightarrow[]{R}u^{53}_{1r}\xrightarrow[]{\mathcal{I} R}u^{42}_{1r}\xrightarrow[]{R}u^{64}_{1r}\xrightarrow[]{R}u^{26}_{1r}\\
  &u^{135}_{1r}\xrightarrow[]{R}u^{351}_{1r}\xrightarrow[]{R}u^{513}_{1r}\xrightarrow[]{\mathcal{I} R}u^{462}_{1r}\xrightarrow[]{R}u^{624}_{1r}\xrightarrow[]{R}u^{246}_{1r}.
	\end{aligned}\right.   
\end{equation}
This yields,
\begin{equation}
	\left.\begin{aligned}		
 &u^{13}_{1r}=\eta  u^{35}_{1r}=\eta u^{51}_{1r}=u^{46}_{1r}=u^{62}_{1r}=u^{24}_{1r}\\		&u^{15}_{1r}=u^{31}_{1r}=u^{53}_{1r}=\eta u^{42}_{1r}=u^{64}_{1r}=\eta u^{26}_{1r}\\  &u^{135}_{1r}=\eta u^{351}_{1r}=\eta u^{513}_{1r}=\eta u^{462}_{1r}=u^{624}_{1r}=\eta u^{246}_{1r}\\
	\end{aligned}\right.   
\end{equation}
On the other hand inclusion of translations gives the following constraints
\begin{equation}
	\left.\begin{aligned}
		&u^{35}_{1r}\xrightarrow[]{T_1}(u^{51}_{1r})^\dagger\xrightarrow[]{T_2}u^{135}_{1r}\\
		&u^{13}_{1r}\xrightarrow[]{T^{-1}_2}u^{513}_{1r}\xrightarrow[]{T^{-1}_1}(u^{31}_{1r})^\dagger\\
  &u^{351}_{1r}\xrightarrow[]{T^{-1}_2}u^{51}_{1r}\xrightarrow[]{T^{-1}_1}(u^{15}_{1r})^\dagger\\
	\end{aligned}\right.   
\end{equation}
yielding
\begin{equation}
    u^{135}_{1r}=\eta u^{13}_{1r},\;u^{15}_{1r}=\eta (u^{13}_{1r})^\dagger.
\end{equation}
The results of this appendix are also summarized in Fig.~\ref{fig:first_neighbors}.
\subsection{2NN and 3NN}
Instead of showing the explicit symmetry relations we only state the results for the second and third neighbors after inserting the PSG representations. A summary of these results are shown in Fig.~\ref{fig:second_neighbors} for the second neighbors and in Fig.~\ref{fig:third_neighbors} for the third neighbors. The second neighbors matrices are given by
\begin{equation}\label{eq:signs_2nn_z2}
  	\left.\begin{aligned}
	&u^{13}_{2g} = u^{35}_{2g} = u^{51}_{2g}	= u^{46}_{2g} = u^{62}_{2g} = u^{24}_{2g}, \\
    &u^{12}_{2b} = \eta u^{34}_{2b} = \eta u^{56}_{2b} = \eta_{\mathcal{I}} u^{45}_{2b} = \eta_{\mathcal{I}} u^{61}_{2b} = \eta_{\mathcal{I}} u^{23}_{2b}, \\
    &u^{14}_{2r} = \eta u^{36}_{2r} = \eta u^{52}_{2r} = \eta_{\mathcal{I}} u^{41}_{2r} = \eta_{\mathcal{I}} u^{63}_{2r} = u^{25}_{2r}.
	\end{aligned}\right.   
\end{equation}
The third neighbor matrices are 
\begin{equation}\label{eq:signs_3nn_z2}
  	\left.\begin{aligned}
	&u^{14}_{3g} = u^{36}_{3g} = u^{52}_{3g} = \eta_{\mathcal{I}} u^{41}_{3g} = \eta_{\mathcal{I}} u^{63}_{3g} = \eta_{\mathcal{I}} u^{25}_{3g} \\
    &u^{16}_{3b} = u^{32}_{3b} = u^{54}_{3b} = \eta_{\mathcal{I}} \eta u^{43}_{3b} = \eta_{\mathcal{I}} u^{65}_{3b} = \eta_{\mathcal{I}} \eta u^{21}_{3b} \\
    &u^{15}_{3r} = u^{31}_{3r} = \eta u^{53}_{2r} = \eta u^{42}_{3r} = \eta u^{64}_{3r} = \eta u^{26}_{3r} \\
	\end{aligned}\right.   
\end{equation}

\begin{table*}
\caption{Symmetric $U(1)$ mean-field \textit{Ans\"atze} up to third nearest neighbours. See Appendix~\ref{sec:u1_ansatze_3NN} for the sign structure.}
\label{table:u1_ansatz_3nn}
\begin{ruledtabular}
\begin{tabular}{cccccccccccc}
\multirow{2}{*}{$\{w_\mathcal{I},w_\mathcal{T}\}$} & \multirow{2}{*}{$\{n,p_\mathcal{I}\}$} & \multicolumn{3}{c}{1NN} & %
    \multicolumn{3}{c}{2NN} & \multicolumn{3}{c}{3NN} & \multirow{2}{*}{Onsite}\\
\cline{3-5}
\cline{6-8}
\cline{9-11}
& & $u_{1g}$ & $u_{1b}$ & $u_{1r}$ & $u_{2g}$ & $u_{2b}$ & $u_{2r}$ & $u_{3g}$ & $u_{3b}$ & $u_{3r}$ & \\
\hline
$\{1,0\}$ & $\{0,0\}$ & $\dot\iota\tau^0,\tau^{3}$ & $0$ & $0$ & $0$ & $\dot\iota\tau^0,\tau^{3}$ & $0$ & $0$ & $\dot\iota\tau^0,\tau^{3}$ & $0$  & $0$ \\
$\{1,0\}$ & $\{0,1\}$ & $\dot\iota\tau^0,\tau^{3}$ & $\dot\iota\tau^0,\tau^{3}$ & $0$ & $0$ & $\dot\iota\tau^0,\tau^{3}$ & $\dot\iota\tau^0,\tau^{3}$ & $\dot\iota\tau^0,\tau^{3}$ & $\dot\iota\tau^0,\tau^{3}$ & $0$  & $0$ \\
$\{1,0\}$ & $\{0,2\}$ & $\dot\iota\tau^0,\tau^{3}$ & $0$ & $0$ & $0$ & $\dot\iota\tau^0,\tau^{3}$ & $0$ & $0$ & $\dot\iota\tau^0,\tau^{3}$ & $0$  & $0$ \\
$\{1,0\}$ & $\{0,3\}$ & $\dot\iota\tau^0,\tau^{3}$ & $\dot\iota\tau^0,\tau^{3}$ & $0$ & $0$ & $\dot\iota\tau^0,\tau^{3}$ & $\dot\iota\tau^0,\tau^{3}$ & $\dot\iota\tau^0,\tau^{3}$ & $\dot\iota\tau^0,\tau^{3}$ & $0$  & $0$ \\
$\{1,0\}$ & $\{1,0\}$ & $\dot\iota\tau^0,\tau^{3}$ & $\dot\iota\tau^0,\tau^{3}$ & $0$ & $0$ & $\dot\iota\tau^0,\tau^{3}$ & $\dot\iota\tau^0,\tau^{3}$ & $0$ & $\dot\iota\tau^0,\tau^{3}$ & $0$  & $0$ \\
$\{1,0\}$ & $\{1,1\}$ & $\dot\iota\tau^0,\tau^{3}$ & $0$ & $0$ & $0$ & $\dot\iota\tau^0,\tau^{3}$ & $0$ & $\dot\iota\tau^0,\tau^{3}$ & $\dot\iota\tau^0,\tau^{3}$ & $0$  & $0$ \\
$\{1,0\}$ & $\{1,2\}$ & $\dot\iota\tau^0,\tau^{3}$ & $\dot\iota\tau^0,\tau^{3}$ & $0$ & $0$ & $\dot\iota\tau^0,\tau^{3}$ & $\dot\iota\tau^0,\tau^{3}$ & $0$ & $\dot\iota\tau^0,\tau^{3}$ & $0$  & $0$ \\
$\{1,0\}$ & $\{1,3\}$ & $\dot\iota\tau^0,\tau^{3}$ & $0$ & $0$ & $0$ & $\dot\iota\tau^0,\tau^{3}$ & $0$ & $\dot\iota\tau^0,\tau^{3}$ & $\dot\iota\tau^0,\tau^{3}$ & $0$  & $0$ \\
\hline
$\{w_\mathcal{I},w_\mathcal{T}\}$ & $\{n,n_\mathcal{I}\}$ & $u_{1g}$ & $u_{1b}$ & $u_{1r}$ & $u_{2g}$ & $u_{2b}$ & $u_{2r}$ & $u_{3g}$ & $u_{3b}$ & $u_{3r}$ & Onsite\\
\hline
$\{0,1\}$ & $\{0,0\}$ & $\tau^{3}$ & $\tau^{3}$ & $\tau^{3}$ & $\tau^{3}$ & $\tau^{3}$ & $\tau^{3}$ & $\tau^{3}$ & $\tau^{3}$ & $\tau^{3}$  & $\tau^{3}$ \\
$\{0,1\}$ & $\{0,1\}$ & $\tau^{3}$ & $0$ & $\tau^{3}$ & $\tau^{3}$ & $\tau^{3}$ & $0$ & $0$ & $\tau^{3}$ & $\tau^{3}$ & $\tau^{3}$ \\
$\{0,1\}$ & $\{1,0\}$ & $\tau^{3}$ & $0$ & $\tau^{3}$ & $\tau^{3}$ & $\tau^{3}$ & $0$ & $\tau^{3}$ & $\tau^{3}$ & $\tau^{3}$  & $\tau^{3}$ \\
$\{0,1\}$ & $\{1,1\}$ & $\tau^{3}$ & $\tau^{3}$ & $\tau^{3}$ & $\tau^{3}$ & $\tau^{3}$ & $\tau^{3}$ & $0$ & $\tau^{3}$ & $\tau^{3}$  & $\tau^{3}$ \\
\hline
$\{1,1\}$ & $\{0,0\}$ & $\tau^{3}$ & $0$ & $\tau^{3}$ & $\tau^{3}$ & $\tau^{3}$ & $0$ & $0$ & $\tau^{3}$ & $\tau^{3}$  & $\tau^{3}$ \\
$\{1,1\}$ & $\{0,0\}$ & $\tau^{3}$ & $\tau^3$ & $\tau^{3}$ & $\tau^{3}$ & $\tau^{3}$ & $\tau^3$ & $\tau^3$ & $\tau^{3}$ & $\tau^{3}$  & $\tau^{3}$ \\
$\{1,1\}$ & $\{0,0\}$ & $\tau^{3}$ & $\tau^3$ & $\tau^{3}$ & $\tau^{3}$ & $\tau^{3}$ & $\tau^3$ & $0$ & $\tau^{3}$ & $\tau^{3}$  & $\tau^{3}$ \\
$\{1,1\}$ & $\{0,0\}$ & $\tau^{3}$ & $0$ & $\tau^{3}$ & $\tau^{3}$ & $0$ & $\tau^3$ & $\tau^3$ & $\tau^{3}$ & $\tau^{3}$  & $\tau^{3}$ \\
\hline
$\{w_\mathcal{I},w_\mathcal{T}\}$ & $\{\Tilde{\theta}_{\mathcal{I}}\}$ & $u_{1g}$ & $u_{1b}$ & $u_{1r}$ & $u_{2g}$ & $u_{2b}$ & $u_{2r}$ & $u_{3g}$ & $u_{3b}$ & $u_{3r}$ & Onsite \\
\hline
$\{0,0\}$ & $\{m\pi/n\}$ & $\dot\iota\tau^0,\tau^{3}$ & $\dot\iota\tau^0,\tau^{3}$ & $0$ & $0$ & $\dot\iota\tau^0,\tau^{3}$ & $\dot\iota\tau^0,\tau^{3}$ & $\tau^{3}$ & $\dot\iota\tau^0,\tau^{3}$ & $0$  & $0$ \\
\end{tabular}
\end{ruledtabular}
\end{table*}

\begin{table*}
\caption{Symmetric $\mathbf{\mathds{Z}_2}$ mean-field \textit{Ans\"atze} up to third nearest neighbours. The sign configurations on the second and third nearest neighbour bonds can be found in Eq.~\eqref{eq:signs_2nn_z2} and Eq.~\eqref{eq:signs_3nn_z2}. }
\label{table:z2_ansatz}
\begin{ruledtabular}
\begin{tabular}{cccccccccccc}
\multirow{2}{*}{$\{\eta_\mathcal{T},g_\mathcal{T}\}$} & \multirow{2}{*}{$\{\eta,\eta_\mathcal{I}\}$} & \multicolumn{3}{c}{1NN} & %
    \multicolumn{3}{c}{2NN} & \multicolumn{3}{c}{3NN} & \multirow{2}{*}{Onsite}\\
\cline{3-5}
\cline{6-8}
\cline{9-11}
& & $u_{1g}$ & $u_{1b}$ & $u_{1r}$ & $u_{2g}$ & $u_{2b}$ & $u_{2r}$ & $u_{3g}$ & $u_{3b}$ & $u_{3r}$ & \\
\hline
$\{+,\dot\iota\tau^2\}$ & $\{+,+\}$ & $\tau^{1,3}$ & $\tau^{1,3}$ & $\tau^{1,3}$ & $\tau^{1,3}$ & $\tau^{1,3}$ & $\tau^{1,3}$ & $\tau^{1,3}$ & $\tau^{1,3}$ & $\tau^{1,3}$  & $\tau^{3}$ \\
$\{+,\dot\iota\tau^2\}$ & $\{+,-\}$ & $\tau^{1,3}$ & $0$ & $\tau^{1,3}$ & $\tau^{1,3}$ & $\tau^{1,3}$ & $0$ & $0$ & $\tau^{1,3}$ & $\tau^{1,3}$ & $\tau^{3}$ \\
$\{+,\dot\iota\tau^2\}$ & $\{-,+\}$ & $\tau^{1,3}$ & $0$ & $\tau^{1,3}$ & $\tau^{1,3}$ & $\tau^{1,3}$ & $0$ & $\tau^{1,3}$ & $\tau^{1,3}$ & $\tau^{1,3}$  & $\tau^{3}$ \\
$\{+,\dot\iota\tau^2\}$ & $\{-,-\}$ & $\tau^{1,3}$ & $\tau^{1,3}$ & $\tau^{1,3}$ & $\tau^{1,3}$ & $\tau^{1,3}$ & $\tau^{1,3}$ & $0$ & $\tau^{1,3}$ & $\tau^{1,3}$  & $\tau^{3}$ \\
\hline
$\{-,\dot\iota\tau^2\}$ & $\{+,+\}$ & $\dot\iota\tau^0,\tau^{2}$ & $\tau^{2}$ & $\tau^{1,3}$ & $\tau^{1,3}$ & $\dot\iota\tau^0,\tau^{2}$ & $\tau^{2}$ & $\tau^{2}$ & $\dot\iota\tau^0,\tau^{2}$ & $\tau^{1,3}$  & $\tau^{3}$ \\
$\{-,\dot\iota\tau^2\}$ & $\{+,-\}$ & $\dot\iota\tau^0,\tau^{2}$ & $\dot\iota\tau^0$ & $\tau^{1,3}$ & $\tau^{1,3}$ & $\dot\iota\tau^0,\tau^{2}$ & $\dot\iota\tau^0$ & $\dot\iota\tau^0$ & $\dot\iota\tau^0,\tau^{2}$ & $\tau^{1,3}$  & $\tau^{3}$ \\
$\{-,\dot\iota\tau^2\}$ & $\{-,+\}$ & $\dot\iota\tau^0,\tau^{2}$ & $\dot\iota\tau^0$ & $\tau^{1,3}$ & $\tau^{1,3}$ & $\dot\iota\tau^0,\tau^{2}$ & $\dot\iota\tau^0$ & $\tau^{2}$ & $\dot\iota\tau^0,\tau^{2}$ & $\tau^{1,3}$  & $\tau^{3}$ \\
$\{-,\dot\iota\tau^2\}$ & $\{-,-\}$ & $\dot\iota\tau^0,\tau^{2}$ & $\tau^{2}$ & $\tau^{1,3}$ & $\tau^{1,3}$ & $\dot\iota\tau^0,\tau^{2}$ & $\tau^{2}$ & $\dot\iota\tau^0$ & $\dot\iota\tau^0,\tau^{2}$ & $\tau^{1,3}$  & $\tau^{3}$ \\
\hline
$\{-,\tau^0\}$ & $\{+,+\}$ & $\dot\iota\tau^0,\tau^{2}$ & $\tau^{3}$ & $0$ & $0$ & $\dot\iota\tau^0,\tau^{1,2,3}$ & $\tau^{1,2,3}$ & $\tau^{1,2,3}$ & $\dot\iota\tau^0,\tau^{1,2,3}$ & $0$ & $0$ \\
$\{-,\tau^0\}$ & $\{+,-\}$ & $\dot\iota\tau^0,\tau^{2}$ & $\dot\iota\tau^0$ & $0$ & $0$ & $\dot\iota\tau^0,\tau^{1,2,3}$ & $\dot\iota\tau^0$ & $\dot\iota\tau^0$ & $\dot\iota\tau^0,\tau^{1,2,3}$ & $0$ & $0$ \\
$\{-,\tau^0\}$ & $\{-,+\}$ & $\dot\iota\tau^0,\tau^{2}$ & $\dot\iota\tau^0$ & $0$ & $0$ & $\dot\iota\tau^0,\tau^{1,2,3}$ & $\dot\iota\tau^0$ & $\tau^{1,2,3}$ & $\dot\iota\tau^0,\tau^{1,2,3}$ & $0$ & $0$ \\
$\{-,\tau^0\}$ & $\{-,-\}$ & $\dot\iota\tau^0,\tau^{2}$ & $\tau^{3}$ & $0$ & $0$ & $\dot\iota\tau^0,\tau^{1,2,3}$ & $\tau^{1,2,3}$ & $\dot\iota\tau^0$ & $\dot\iota\tau^0,\tau^{1,2,3}$ & $0$ & $0$ \\
\end{tabular}
\end{ruledtabular}
\end{table*}

\begin{table*}
\caption{Definition for bond parameters for the representative bonds. The last column lists the spatial symmetry operation (either none or the twofold rotation $\mathcal{I}$) that maps the bond to itself up to translation.}
\label{table:bond_convention}
\begin{ruledtabular}
\begin{tabular}{cccc}
Bond type & Representative bond $\mathbf{r}_\alpha \leftarrow \mathbf{r}'_\beta$ & Parameters  for the bond $u^{\alpha}_{\mathbf{r}_\alpha,\mathbf{r}'_\beta}$ ($\alpha=0,x,y,z$) &Stabilizer\\
\hline
{Onsite bond } & $(0,0,1)\leftarrow(0,0,1)$ & $(\alpha_h,0,0,0,0,\beta_p,\gamma_p,\delta_p)$\\
\hline
NN bond type ``$g$''
&$(0,0,1)\leftarrow(0,0,2)$& $(a_{1g,h},b_{1g,h},c_{1g,h},d_{1g,h},a_{1g,p},b_{1g,p},c_{1g,p},d_{1g,p})$&None\\
NN bond type ``$b$'' &$(0,0,1)\leftarrow(0,1,4)$&$(a_{1b,h},b_{1b,h},c_{1b,h},d_{1b,h},a_{1b,p},b_{1b,p},c_{1b,p},d_{1b,p})$ & $\mathcal{I}$ \\
NN bond type ``$r$'' & $(0,0,1)_0\leftarrow(0,1,5)$&$(a_{1r,h},b_{1r,h},c_{1r,h},d_{1r,h},a_{1r,p},b_{1r,p},c_{1r,p},d_{1r,p})$ &None\\
\hline
{2nd NN bond type ``$g$'' }&$(0,0,1)\leftarrow(0,0,3)$&$(a_{2g,h},b_{2g,h},c_{2g,h},d_{2g,h},a_{2g,p},b_{2g,p},c_{2g,p},d_{2g,p})$&None\\
{2nd NN bond type ``$b$'' }& $(0,0,1)\leftarrow(-1,0,6)$&
$(a_{2b,h},b_{2b,h},c_{2b,h},d_{2b,h},a_{2b,p},b_{2b,p},c_{2b,p},d_{2b,p})$&None\\
{2nd NN bond type ``$r$'' }& $(0,0,1)\leftarrow(1,1,4)$&
$(a_{2r,h},b_{2r,h},c_{2r,h},d_{2r,h},a_{2r,p},b_{2r,p},c_{2r,p},d_{2r,p})$&$\mathcal{I}$\\
\hline
{3nd NN bond type ``$g$'' }& $(0,0,1)\leftarrow(0,0,4)$&$ (a_{3g,h},b_{3g,h},c_{3g,h},d_{3g,h},a_{3g,p},b_{3g,p},c_{3g,p},d_{3g,p})$ & $\mathcal{I}$\\
{3nd NN bond type ``$b$'' }& $(0,0,1)\leftarrow(0,1,6)$&$(a_{3b,h},b_{3b,h},c_{3b,h},d_{3b,h},a_{3b,p},b_{3b,p},c_{3b,p},d_{3b,p})$&None\\
{3nd NN bond type ``$r$'' }& $(0,0,1)\leftarrow(-1,0,5)$&$(a_{3r,h},b_{3r,h},c_{3r,h},d_{3r,h},a_{3r,p},b_{3r,p},c_{3r,p},d_{3r,p})$&None\\
\end{tabular}
\end{ruledtabular}
\end{table*}

\section{$U(1)$ mean-field \textit{Ans\"atze} up to 3NN}
\label{sec:u1_ansatze_3NN}
In analogy to the previous appendix we present here the $U(1)$ mean-field models up to third nearest neighbors. The initial form of one of the \textit{Ans\"atze} is determined by the underlying PSG.

\subsubsection{$w_\mathcal{I}=0$ and $w_\mathcal{T}=0$ (class UA)}
\begin{equation}\label{eq:u00_ans_1NN}
\left.\begin{aligned}
&u^{12}_{1g}=u^{34}_{1g}=u^{56}_{1g}=u^{23}_{1g}=u^{45}_{1g}=u^{61}_{1g}=\dot\iota\chi^0_{1g}\tau^0+\chi^3_{1g}\tau^3\\
&u^{14}_{1b}=u^{36}_{1b}=u^{52}_{1b}=\dot\iota\chi^0_{1b}\tau^0+\chi^3_{1b}\tau^3\\
&u^{uu'}_{1r}=0,\;\;u^{14}_{1b}= g_3(-\Tilde{\theta}_{\mathcal{I}})(u^{14}_{1b})^\dagger.\\
\end{aligned}\right.
\end{equation}

\begin{equation}\label{eq:u00_ans_2NN}
\left.\begin{aligned}
&u^{uu'}_{2g}=0\\
&u^{12}_{2b}=g_3(-3\Tilde{\theta}_{\mathcal{I}})u^{34}_{2b}=g_3(-3\Tilde{\theta}_{\mathcal{I}}) u^{56}_{2b}=\dot\iota\chi^0_{2b}\tau^0+\chi^3_{2b}\tau^3\\
&u^{23}_{2b}=u^{45}_{2b}=u^{61}_{2b}=g_3(2\Tilde{\theta}_{\mathcal{I}})u^{12}_{2b}\\
&u^{14}_{2r}=g_3(-3\Tilde{\theta}_{\mathcal{I}}) u^{36}_{2r}=g_3(-3\Tilde{\theta}_{\mathcal{I}}) u^{52}_{2r}=\dot\iota\chi^0_{2r}\tau^0+\chi^3_{2r}\tau^3\\
&u^{14}_{2r}= g_3(\Tilde{\theta}_{\mathcal{I}})(u^{14}_{2r})^\dagger.\\
\end{aligned}\right.
\end{equation}

\begin{equation}\label{eq:u00_ans_3NN}
\left.\begin{aligned}
&u^{14}_{3g}=u^{36}_{3g}=(u^{25}_{3g})^\dagger=\chi_{3g}\tau^3r\\
&u^{16}_{3b}=u^{32}_{3b}=u^{54}_{3b}=\dot\iota\chi^0_{3b}\tau^0+\chi^3_{3b}\tau^3\\
&u^{21}_{3b}=u^{43}_{3b}=g_3(3\Tilde{\theta}_{\mathcal{I}})u^{65}_{3b}=g_3(3\Tilde{\theta}_{\mathcal{I}})u^{16}_{3b}\\
&u^{uu'}_{3r}=0.\\
\end{aligned}\right.
\end{equation}
\subsubsection{$w_\mathcal{I}=1$ and $w_\mathcal{T}=0$ (class UB)}
\begin{equation}\label{eq:u10_ans_1NN}
\left.\begin{aligned}
&u^{12}_{1g}=u^{34}_{1g}=u^{56}_{1g}=\dot\iota\chi^0_{1g}\tau^0+\chi^3_{1g}\tau^3\\
&u^{23}_{1g}=u^{45}_{1g}=u^{61}_{1g}=-g_3(p_\mathcal{I}\pi/3)(u^{12}_{1g})^\dagger\\
&u^{14}_{1b}=u^{36}_{1b}=u^{52}_{1b}=\dot\iota\chi^0_{1b}\tau^0+\chi^3_{1b}\tau^3,\;\;u^{14}_{1b}=-\eta g_3(p_\mathcal{I}\pi)u^{14}_{1b}\\
&u^{uu'}_{1r}=0.\\
\end{aligned}\right.
\end{equation}

\begin{equation}\label{eq:u10_ans_2NN}
\left.\begin{aligned}
&u^{uu'}_{2g}=0\\
&u^{12}_{2b}=\eta u^{34}_{2b}=\eta u^{56}_{2b}=\dot\iota\chi^0_{2b}\tau^0+\chi^3_{2b}\tau^3\\
&u^{23}_{2b}=u^{45}_{2b}=u^{61}_{2b}=-g_3(p_\mathcal{I}\pi/3)(u^{12}_{2b})^\dagger\\
&u^{14}_{2r}=\eta u^{36}_{2r}=\eta u^{52}_{2r}=\dot\iota\chi^0_{2r}\tau^0+\chi^3_{2r}\tau^3,\;\;u^{14}_{2r}=-\eta g_3(p_\mathcal{I}\pi)u^{14}_{2r}.\\
\end{aligned}\right.
\end{equation}

\begin{equation}\label{eq:u10_ans_3NN}
\left.\begin{aligned}
&u^{14}_{3g}=u^{36}_{3g}=(u^{25}_{3g})^\dagger=\dot\iota\chi^0_{3g}\tau^0+\chi^3_{3g}\tau^3,\;\;u^{14}_{3g}=-g_3(p_\mathcal{I}\pi)u^{14}_{3g}\\
&u^{16}_{3b}=u^{32}_{3b}=u^{54}_{3b}=\dot\iota\chi^0_{3b}\tau^0+\chi^3_{3b}\tau^3\\
&u^{21}_{3b}=u^{43}_{3b}=\eta u^{65}_{3b}=-\eta g_3(p_\mathcal{I}\pi/3)(u^{16}_{3b})^\dagger\\
&u^{uu'}_{3r}=0.\\
\end{aligned}\right.
\end{equation}

\subsubsection{$w_\mathcal{I}=0$ and $w_\mathcal{T}=1$ (class UC)}
In the following $\eta=g_3(\theta=n\pi)$ and $\eta_\mathcal{I}=g_3(\theta_\mathcal{I}=n_\mathcal{I}\pi)$.
\begin{equation}\label{eq:u01_ans_1NN}
\left.\begin{aligned}
&u^{12}_{1g}=\eta_\mathcal{I}u^{23}_{1g}=u^{34}_{1g}=\eta_\mathcal{I}u^{45}_{1g}=u^{56}_{1g}=\eta_\mathcal{I}u^{61}_{1g}=\chi_{1g}\tau^3\\
&u^{14}_{1b}=u^{36}_{1b}=u^{52}_{1b}=\chi_{1b}\tau^3,\;\;u^{14}_{1b}=\eta\eta_\mathcal{I}(u^{14}_{1b})^\dagger\\
&u^{13}_{1r}=u^{24}_{1r}=\eta u^{35}_{1r}=u^{46}_{1r}=\eta u^{51}_{1r}=u^{62}_{1r}=\chi_{1r}\tau^3.\\
\end{aligned}\right.
\end{equation}

\begin{equation}\label{eq:u01_ans_2NN}
\left.\begin{aligned}
&u^{13}_{2g}=u^{24}_{2g}= u^{35}_{2g}=u^{46}_{2g}= u^{51}_{2g}=u^{62}_{2g}=\chi_{2g}\tau^3\\
&u^{12}_{2b}=\eta_\mathcal{I}u^{23}_{2b}=\eta u^{34}_{2b}=\eta_\mathcal{I}u^{45}_{2b}=\eta u^{56}_{2b}=\eta_\mathcal{I}u^{61}_{2b}=\chi_{2b}\tau^3\\
&u^{14}_{2r}=\eta u^{36}_{2r}=\eta u^{52}_{2r}=\chi_{2r}\tau^3,\;\;u^{14}_{2r}=\eta\eta_\mathcal{I}(u^{14}_{2r})^\dagger.\\
\end{aligned}\right.
\end{equation}

\begin{equation}\label{eq:u01_ans_3NN}
\left.\begin{aligned}
&u^{14}_{3g}=u^{36}_{3g}=u^{25}_{3g}=\chi_{3g}\tau^3,\;\;u^{14}_{3g}=\eta_\mathcal{I}(u^{14}_{3g})^\dagger\\
&u^{16}_{3b}=\eta\eta_\mathcal{I}u^{21}_{3b}=u^{32}_{3b}=\eta\eta_\mathcal{I}u^{43}_{3b}=u^{54}_{3b}=\eta_\mathcal{I}u^{65}_{3b}=\chi_{3b}\tau^3\\
&u^{13}_{3r}=\eta u^{24}_{3r}=u^{35}_{3r}=\eta u^{46}_{3r}=u^{51}_{3r}=u^{62}_{3r}=\chi_{3r}\tau^3.\\
\end{aligned}\right.
\end{equation}

\subsubsection{$w_\mathcal{I}=1$ and $w_\mathcal{T}=1$ (class UD)}
\begin{equation}\label{eq:u11_ans_1NN}
\left.\begin{aligned}
&u^{12}_{1g}=-\eta_\mathcal{I}u^{23}_{1g}=u^{34}_{1g}=-\eta_\mathcal{I}u^{45}_{1g}=u^{56}_{1g}=-\eta_\mathcal{I}u^{61}_{1g}=\chi_{1g}\tau^3\\
&u^{14}_{1b}=u^{36}_{1b}=u^{52}_{1b}=\chi_{1b}\tau^3,\;\;u^{14}_{1b}=-\eta\eta_\mathcal{I}u^{14}_{1b}\\
&u^{13}_{1r}=-u^{24}_{1r}=\eta u^{35}_{1r}=-u^{46}_{1r}=\eta u^{51}_{1r}=-u^{62}_{1r}=\chi_{1r}\tau^3.\\
\end{aligned}\right.
\end{equation}

\begin{equation}\label{eq:u11_ans_2NN}
\left.\begin{aligned}
&u^{13}_{2g}=-u^{24}_{2g}= u^{35}_{2g}=-u^{46}_{2g}= u^{51}_{2g}=-u^{62}_{2g}=\chi_{2g}\tau^3\\
&u^{12}_{2b}=-\eta_\mathcal{I}u^{23}_{2b}=\eta u^{34}_{2b}=-\eta_\mathcal{I}u^{45}_{2b}=\eta u^{56}_{2b}=-\eta_\mathcal{I}u^{61}_{2b}=\chi_{2b}\tau^3\\
&u^{14}_{2r}=\eta u^{36}_{2r}=\eta u^{52}_{2r}=\chi_{2r}\tau^3,\;\;u^{14}_{2r}=-\eta\eta_\mathcal{I}u^{14}_{2r}.\\
\end{aligned}\right.
\end{equation}

\begin{equation}\label{eq:u11_ans_3NN}
\left.\begin{aligned}
&u^{14}_{3g}=u^{36}_{3g}=u^{25}_{3g}=\chi_{3g}\tau^3,\;\;u^{14}_{3g}=-\eta_\mathcal{I}u^{14}_{3g}\\
&u^{16}_{3b}=-\eta\eta_\mathcal{I}u^{21}_{3b}= u^{32}_{3b}=-\eta\eta_\mathcal{I}u^{43}_{3b}=u^{54}_{3b}=- \eta_\mathcal{I}u^{65}_{3b}=\chi_{3b}\tau^3\\
&u^{13}_{3r}=-\eta u^{24}_{3r}= u^{35}_{3r}=-\eta u^{46}_{3r}=u^{51}_{3r}=- u^{62}_{3r}=\chi_{3r}\tau^3.\\
\end{aligned}\right.
\end{equation}

The symmetry allowed mean-field amplitudes on the reference bonds up to third nearest-neighbor are tabulated in Table~\ref{table:u1_ansatz_3nn}.

\begin{table*}
\caption{Spatial constraints for symmetric $U(1)$ mean-field \textit{Ans\"atze} up to third nearest neighbours. The parameters $(b,c,d)_{1g}$, $(b,c,d)_{1r}$, $(b,c,d)_{2g}$,  $(b,c,d)_{2b}$, $(b,c,d)_{3b}$, $(b,c,d)_{3r}$ are not spatially constrained and hence not listed. Note that time-reversal constraints (namely, when $w_{\mathcal{T}}=0$ all bonds with ``$1r$'', ``$2g$'', ``$3r$'' vanish) are not listed in this table.}
\label{table:u1_ansatz_3nn_with_triplets}
\begin{ruledtabular}
\begin{tabular}{ccccc}
\multirow{2}{*}{$\{w_\mathcal{I},w_\mathcal{T}\}$} & \multirow{2}{*}{$\{n,p_\mathcal{I}\}$} & {1NN} & %
    {2NN} & {3NN}\\
\cline{3-5}
& & $(a,b,c,d)_{1b}$ & $(a,b,c,d)_{2r}$ & $(a,b,c,d)_{3g}$ \\
\hline
$\{1,0\}$ & $\{0,0\}$ & \multirow{8}{*}{$=(-a,c,b,-d)_{1b} e^{-\dot\iota (n+p_{\mathcal{I}})\pi}$} & \multirow{8}{*}{$=(-a,c,b,-d)_{2r} e^{-\dot\iota (n+p_{\mathcal{I}})\pi}$}& \multirow{8}{*}{$=(-a,c,b,-d)_{1b} e^{-\dot\iota p_{\mathcal{I}}\pi}$} \\
$\{1,0\}$ & $\{0,1\}$ &&& \\
$\{1,0\}$ & $\{0,2\}$ &&& \\
$\{1,0\}$ & $\{0,3\}$ &&& \\
$\{1,0\}$ & $\{1,0\}$ &&& \\
$\{1,0\}$ & $\{1,1\}$ &&& \\
$\{1,0\}$ & $\{1,2\}$ &&& \\
$\{1,0\}$ & $\{1,3\}$ &&& \\
\hline
$\{w_\mathcal{I},w_\mathcal{T}\}$ & $\{n,n_\mathcal{I}\}$ & $(a,b,c,d)_{1b}$ & $(a,b,c,d)_{2r}$ & $(a,b,c,d)_{3g}$ \\
\hline
$\{0,1\}$ & $\{0,0\}$ & \multirow{4}{*}{$=(-a^*,c^*,b^*,-d^*)_{1b} (-1)^{n+n_{\mathcal{I}}}$} & \multirow{4}{*}{$=(-a^*,c^*,b^*,-d^*)_{2r} (-1)^{n+n_{\mathcal{I}}}$}& \multirow{4}{*}{$=(-a^*,c^*,b^*,-d^*)_{1b} (-1)^{n_{\mathcal{I}}}$} \\
$\{0,1\}$ & $\{0,1\}$ &&& \\
$\{0,1\}$ & $\{1,0\}$ &&& \\
$\{0,1\}$ & $\{1,1\}$ &&& \\
\hline
$\{1,1\}$ & $\{0,0\}$ & \multirow{4}{*}{$=(-a,c,b,-d)_{1b} (-1)^{n+n_{\mathcal{I}}}$} & \multirow{4}{*}{$=(-a,c,b,-d)_{2r} (-1)^{n+n_{\mathcal{I}}}$}& \multirow{4}{*}{$=(-a,c,b,-d)_{1b} (-1)^{n_{\mathcal{I}}}$} \\
$\{1,1\}$ & $\{0,1\}$ &&& \\
$\{1,1\}$ & $\{1,0\}$ &&& \\
$\{1,1\}$ & $\{1,1\}$ &&& \\
\hline
$\{w_\mathcal{I},w_\mathcal{T}\}$ & $\{\Tilde{\theta}_{\mathcal{I}}\}$ & $(a,b,c,d)_{1b}$ & $(a,b,c,d)_{2r}$ & $(a,b,c,d)_{3g}$ \\
\hline
$\{0,0\}$ & $\{m\pi/n\}$ & $=(-a^*,c^*,b^*,-d^*)_{1b} e^{-\dot\iota\Tilde{\theta}_{\mathcal{I}}}$ & $=(-a^*,c^*,b^*,-d^*)_{2r} e^{\dot\iota\Tilde{\theta}_{\mathcal{I}}}$ & $=(-a^*,c^*,b^*,-d^*)_{3g}$  \\
\end{tabular}
\end{ruledtabular}
\end{table*}

\begin{table*}
\caption{Spatial constraints for symmetric $\mathbb{Z}_2$ mean-field \textit{Ans\"atze} up to third nearest neighbours. The parameters $(b,c,d)_{1g}$, $(b,c,d)_{1r}$, $(b,c,d)_{2g}$,  $(b,c,d)_{2b}$, $(b,c,d)_{3b}$, $(b,c,d)_{3r}$ are not spatially constrained (but are still constrained by time-reversal symmetry) and hence not listed. Note that time- reversal constraints are not listed in this table.}
\label{table:z2_ansatz_with_triplets}
\begin{ruledtabular}
\begin{tabular}{cccc}
\multirow{2}{*}{$\{\eta,\eta_{\mathcal{I}}\}$}  & {1NN} & %
    {2NN} & {3NN}\\
\cline{2-4}
 & $(a_h,b_h,c_h,d_h,a_p,b_p,c_p,d_p)_{1b}$ & $(a_h,b_h,c_h,d_h,a_p,b_p,c_p,d_p)_{2r}$ & $(a_h,b_h,c_h,d_h,a_p,b_p,c_p,d_p)_{3g}$ \\
\hline 
\multirow{2}{*}{$\{\pm,\pm\}$} & {$= \eta \eta_{\mathcal{I}} \times$} &
{$=\eta \eta_{\mathcal{I}}\times$}  & 
{$=\eta_{\mathcal{I}}\times$} \\
& 
{$(-a_h^*,c_h^*,b_h^*,-d_h^*,a_p,-c_p,-b_p,d_p)_{1b}$}&
{$(-a_h^*,c_h^*,b_h^*,-d_h^*,a_p,-c_p,-b_p,d_p)_{2r}$}&
{$(-a_h^*,c_h^*,b_h^*,-d_h^*,a_p,-c_p,-b_p,d_p)_{3g}$}\\
\end{tabular}
\end{ruledtabular}
\end{table*}

\section{Symmetric mean-field \textit{Ans\"atze} including triplet terms up to 3NN}

The most general mean-field Hamiltonian for fermionic spinons is written as
\begin{equation}
H = \sum_{i = 0,x,y,z} H^i
\end{equation}
with
\begin{equation}
\begin{aligned}
H^i &= \sum_{ \mathbf{r}_\alpha,\mathbf{r}'_\beta} H^i_{\mathbf{r}_\alpha,\mathbf{r}'_\beta},\\
H^i_{\mathbf{r}_\alpha,\mathbf{r}'_\beta} &= \mathrm{Tr}[\tau^\alpha \Psi_{\mathbf{r}_\alpha}u^{(i)}_{\mathbf{r}_\alpha,\mathbf{r}'_\beta}\Psi^\dag_{\mathbf{r}'_\beta}],
\end{aligned}
\end{equation}
where ${\hat\Psi}_{\mathbf{r}_\alpha}= \begin{pmatrix}{\hat f}_{\mathbf{r}_\alpha,\uparrow}&{\hat f}^\dag_{\mathbf{r}_\alpha,\downarrow}\\ {\hat f}_{\mathbf{r}_\alpha,\downarrow}&-{\hat f}_{\mathbf{r}_\alpha,\uparrow}^\dag \end{pmatrix}$.
For the bond $\mathbf{r}_\alpha\leftarrow \mathbf{r}'_\beta$, we use eight complex numbers $a_h,b_h,c_h,d_h,a_p,b_p,c_p,d_p$ to parametrize the 16 real parameters in $u^{(i)}_{\mathbf{r}_\alpha,\mathbf{r}'_\beta}$:

\begin{equation}
\begin{aligned}
u^{(0)}_{\mathbf{r}_\alpha,\mathbf{r}'_\beta}
&=\dot\iota\mathrm{Re}a_h \tau^0 - \mathrm{Re}a_p \tau^1-\mathrm{Im}a_p \tau^2-\mathrm{Im}a_h \tau^3,\\
u^{(x)}_{\mathbf{r}_\alpha,\mathbf{r}'_\beta}
&=\mathrm{Re}b_h \tau^0 +\dot\iota( \mathrm{Re}b_p \tau^1-\mathrm{Im}b_p \tau^2+\mathrm{Im}b_h \tau^3),\\
u^{(y)}_{\mathbf{r}_\alpha,\mathbf{r}'_\beta}
&=\mathrm{Re}c_h \tau^0 +\dot\iota( \mathrm{Re}c_p \tau^1-\mathrm{Im}c_p \tau^2+\mathrm{Im}c_h \tau^3),\\
u^{(z)}_{\mathbf{r}_\alpha,\mathbf{r}'_\beta}
&=\mathrm{Re}d_h \tau^0 +\dot\iota( \mathrm{Re}d_p \tau^1-\mathrm{Im}d_p \tau^2+\mathrm{Im}d_h \tau^3).
\end{aligned}
\end{equation}
More explicitly, we have
\begin{equation}\label{simpH}
\begin{aligned}
H^0_{\mathbf{r}_\alpha,\mathbf{r}'_\beta}  
=&
\dot\iota a_h^*({\hat f}^\dag_{\mathbf{r}_\alpha,\uparrow} {\hat f}_{\mathbf{r}'_\beta,\uparrow}+{\hat f}^\dag_{\mathbf{r}_\alpha,\downarrow}{\hat f}_{\mathbf{r}'_\beta,\downarrow})\\
&+a_p ({\hat f}^\dag_{\mathbf{r}_\alpha,\uparrow}{\hat f}^\dag_{\mathbf{r}'_\beta,\downarrow}-{\hat f}^\dag_{\mathbf{r}_\alpha,\downarrow} {\hat f}^\dag_{\mathbf{r}'_\beta,\uparrow})+h.c.,\\
H^x_{\mathbf{r}_\alpha,\mathbf{r}'_\beta}  
=&
-b_h^*({\hat f}^\dag_{\mathbf{r}_\alpha,\downarrow}{\hat f}_{\mathbf{r}'_\beta,\uparrow}+{\hat f}^\dag_{\mathbf{r}_\alpha,\uparrow} {\hat f}_{\mathbf{r}'_\beta,\downarrow})\\
&-\dot\iota b^*_p({\hat f}^\dag_{\mathbf{r}_\alpha,\uparrow}{\hat f}^\dag_{\mathbf{r}'_\beta,\uparrow}-{\hat f}^\dag_{\mathbf{r}_\alpha,\downarrow}{\hat f}^\dag_{\mathbf{r}'_\beta,\downarrow})
+h.c.,\\
H^y_{\mathbf{r}_\alpha,\mathbf{r}'_\beta}  
=&
-\dot\iota c^*_h({\hat f}^\dag_{\mathbf{r}_\alpha,\downarrow}{\hat f}_{\mathbf{r}'_\beta,\uparrow}-{\hat f}^\dag_{\mathbf{r}_\alpha,\uparrow}{\hat f}_{\mathbf{r}'_\beta,\downarrow})\\
&-c^*_p({\hat f}^\dag_{\mathbf{r}_\alpha,\uparrow}{\hat f}^\dag_{\mathbf{r}'_\beta,\uparrow}+
{\hat f}^\dag_{\mathbf{r}_\alpha,\downarrow}{\hat f}^\dag_{\mathbf{r}'_\beta,\downarrow})+h.c.,\\
H^z_{\mathbf{r}_\alpha,\mathbf{r}'_\beta}  
=&
-d_h^*({\hat f}^\dag_{\mathbf{r}_\alpha,\uparrow}{\hat f}_{\mathbf{r}'_\beta,\uparrow}-{\hat f}^\dag_{\mathbf{r}_\alpha,\downarrow}{\hat f}_{\mathbf{r}'_\beta,\downarrow})\\
&+\dot\iota d^*_p({\hat f}^\dag_{\mathbf{r}_\alpha,\uparrow}{\hat f}^\dag_{\mathbf{r}'_\beta,\downarrow}+{\hat f}^\dag_{\mathbf{r}_\alpha,\downarrow}{\hat f}^\dag_{\mathbf{r}'_\beta,\uparrow})+h.c.,
\end{aligned}
\end{equation}

We define in Table \ref{table:bond_convention} the bond parameters for the representative bonds up to 3NN. All other bonds can then be obtained by performing certain PSG operation from these bonds. Table \ref{table:bond_convention} serves as the reference to map the terms and parameters in Tables \ref{table:u1_ansatz_3nn}, \ref{table:z2_ansatz}, and \ref{table:u1_ansatz_3nn_with_triplets}, \ref{table:z2_ansatz_with_triplets}.

Note that for onsite bond, we only have four complex parameters that are possibly nonzero, $\alpha_h$, $\beta_p$, $\gamma_p$, $\delta_p$, due to fermion anticommutativity and hermiticity of the Hamiltonian.

\subsection{$U(1)$ Ans\"{a}tze}

Note that for $U(1)$ PSG \textit{Ans\"atze}, we only have hopping bilinears and no pairing, therefore the parameters with subscript ``$p$" (hence the $\tau^1$ and $\tau^2$ terms) vanish. We then simplify the notation of the hopping parameters by omitting the subscript ``$h$" (i.e., $(a,b,c,d):=(a_h,b_h,c_h,d_h)$) and write
\begin{equation}
\begin{aligned}
u^{(0)}_{\mathbf{r}_\alpha,\mathbf{r}'_\beta} &= \dot\iota\mathrm{Re}a \tau^0-\mathrm{Im}a \tau^3,\\
u^{(x)}_{\mathbf{r}_\alpha,\mathbf{r}'_\beta} &= \mathrm{Re}b \tau^0+\dot\iota\mathrm{Im}b \tau^3,\\
u^{(y)}_{\mathbf{r}_\alpha,\mathbf{r}'_\beta} &= \mathrm{Re}c \tau^0+\dot\iota\mathrm{Im}c \tau^3,\\
u^{(z)}_{\mathbf{r}_\alpha,\mathbf{r}'_\beta} &= \mathrm{Re}d \tau^0+\dot\iota\mathrm{Im}d \tau^3.
\end{aligned}
\end{equation}

The spatial constraints for these parameters are summarized in Table \ref{table:u1_ansatz_3nn_with_triplets}.

Effect of time-reversal for $U(1)$ Ans\"{a}tze:
\begin{itemize} 
\item When $w_{\mathcal{T}}=0$: TRS forbids bonds connecting sublattices with same sublattice parity, therefore all bonds with ``$1r$'', ``$2g$'', and ``$3r$'' are constrained to vanish by the PSG classes;
\item When $w_{\mathcal{T}}=1$: TRS forbids the appearance of $\dot\iota \tau^0$ in all bonds. This means that we have the TRS constraints $(\mathrm{Re}a,\mathrm{Re}b,\mathrm{Re}c,\mathrm{Re}d)=(0,0,0,0)$ while $(\mathrm{Im}a,\mathrm{Im}b,\mathrm{Im}c,\mathrm{Im}d)$ are not constrained by TRS.
\end{itemize}

\subsection{$\mathbb{Z}_2$ Ans\"{a}tze}

The spatial constraints for the parameters of the $\mathbb{Z}_2$ Ans\"{a}tze are summarized in Table \ref{table:u1_ansatz_3nn_with_triplets}.

Effect of time reversal for $\mathbb{Z}_2$ Ans\"{a}tze:

\begin{itemize}
\item When $\{\eta_{\mathcal{T}},g_{\mathcal{T}}\} = \{+,\dot\iota \tau^2\}$, the constraints of TRS is the same across all bond types: coefficient in front of $\tau^{0,2}$ vanish, meaning $(\mathrm{Re} a_h,\mathrm{Re} b_h,\mathrm{Re} c_h,\mathrm{Re} d_h,\mathrm{Im}a_p,\mathrm{Im}b_p,\mathrm{Im}c_p,\mathrm{Im}d_p)=(0,0,0,0,0,0,0,0)$ while the other eight real components, $(\mathrm{Im} a_h,\mathrm{Im} b_h,\mathrm{Im} c_h,\mathrm{Im} d_h,\mathrm{Re}a_p,\mathrm{Re}b_p,\mathrm{Re}c_p,\mathrm{Re}d_p)$, are not constrained by TRS.
\item When $\{\eta_{\mathcal{T}},g_{\mathcal{T}}\} = \{-,\dot\iota \tau^2\}$: the constraints of TRS is no more the same across all bond types:
\begin{itemize}
\item For the Onsite bonds, NN bond type ``$r$'', 2nd NN bond type ``$g$'', and 3nd NN bond type ``$r$'': coefficient in front of $\tau^{0,2}$ vanish, meaning $(\mathrm{Re} a_h,\mathrm{Re} b_h,\mathrm{Re} c_h,\mathrm{Re} d_h,\mathrm{Im}a_p,\mathrm{Im}b_p,\mathrm{Im}c_p,\mathrm{Im}d_p)=(0,0,0,0,0,0,0,0)$ while the other eight real components, $(\mathrm{Im} a_h,\mathrm{Im} b_h,\mathrm{Im} c_h,\mathrm{Im} d_h,\mathrm{Re}a_p,\mathrm{Re}b_p,\mathrm{Re}c_p,\mathrm{Re}d_p)$, are not constrained by TRS. 
\item For the NN bond types ``$g$'', ``$b$'', the 2nd NN bond types ``$b$'', ``$r$'', and 3nd NN bond types ``$g$'', ``$b$'', the coefficient in front of $\tau^{1,3}$ vanish, meaning $(\mathrm{Im} a_h,\mathrm{Im} b_h,\mathrm{Im} c_h,\mathrm{Im} d_h,\mathrm{Re}a_p,\mathrm{Re}b_p,\mathrm{Re}c_p,\mathrm{Re}d_p)=(0,0,0,0,0,0,0,0)$ while the other eight real components, $(\mathrm{Re} a_h,\mathrm{Re} b_h,\mathrm{Re} c_h,\mathrm{Re} d_h,\mathrm{Im}a_p,\mathrm{Im}b_p,\mathrm{Im}c_p,\mathrm{Im}d_p)$, are not constrained by TRS.
\end{itemize}
\end{itemize}

\section{Spin Structure Factor}
\label{app:calculation_dsf}
The dynamical spin structure factor is defined as
\begin{equation}\label{eq:dsf}
	\left.\begin{aligned}
\mathcal{S}^{\lambda\lambda'}(\mathbf{q},\omega)=\int^{+\infty}_{-\infty}\frac{d\tau e^{\dot{\iota}\omega\tau}}{2\pi \mathcal{N}}&\sum_{i,j}e^{\dot{\iota}\mathbf{q}\cdot\mathbf{r}_{ij}}\langle \hat{S}^\lambda_{i}(\tau)\hat{S}^{\lambda'}_{j}(0)\rangle\\
	\end{aligned}\right.
\end{equation}
where $\lambda,\lambda'\in\{x,y,z\}$, $\mathbf{r}_{ij}=\mathbf{r}_i-\mathbf{r}_j$. and $\hat{S}^z_{i}(\tau)=e^{\dot{\iota}\hat{H}\tau}\hat{S}^z_{i}e^{-\dot{\iota}\hat{H}\tau}$. Due to the presence of the spin-rotational symmetry, it is sufficient to consider the longitudinal components only, i.e., 
\begin{equation}\label{eq:dsf_zz}
	\left.\begin{aligned}
\mathcal{S}^{zz}(\mathbf{q},\omega)&=\int^{+\infty}_{-\infty}\frac{d\tau e^{\dot{\iota}\omega\tau}}{2\pi \mathcal{N}}\sum_{i,j}e^{\dot{\iota}\mathbf{q}\cdot\mathbf{r}_{ij}}\langle e^{\dot{\iota}\hat{H}\tau}\hat{S}^z_{i}e^{-\dot{\iota}\hat{H}\tau}\hat{S}^z_{j}\rangle.\\
	\end{aligned}\right.
\end{equation}
This, in terms of fermion operators reads as
\begin{equation}\label{eq:dsf_zz_f}
	\left.\begin{aligned}
\mathcal{S}^{zz}(\mathbf{q},\omega)&=\int^{+\infty}_{-\infty}\frac{d\tau e^{\dot{\iota}\omega\tau}}{8\pi \mathcal{N}}\sum_{i,j}e^{\dot{\iota}\mathbf{q}\cdot\mathbf{r}_{ij}}\sigma^z_{\alpha\alpha}\sigma^z_{\beta\beta}\\
&\times\sum_{\alpha,\beta}\langle e^{\dot{\iota}\hat{H}\tau}\hat{f}^\dagger_{i,\alpha}\hat{f}_{i,\alpha}e^{-\dot{\iota}\hat{H}\tau}\hat{f}^\dagger_{j,\beta}\hat{f}_{j,\beta}\rangle.\\
	\end{aligned}\right.
\end{equation}
For the $U(1)$ \textit{Ans\"atze}, $\uparrow$ and $\downarrow$ sectors are decoupled. As a result, the basis contains only annihilation operators in each sector. Consider a unitary matrix $U$ such that $U^\dagger\hat{H}U=\text{diag}(\epsilon_1,\epsilon_2,\dots,\epsilon_{\mathcal{N}}$), where $\mathcal{N}$ is the total number of sites. Consequently, the basis vectors will transform as $\hat{f}_{i,\alpha}=U_{i\mu}\hat{\xi}_{\mu,\alpha}$ and Eq.~\eqref{eq:dsf_zz_f} can be recast as
\begin{equation}\label{eq:dsf_u_1}
	\left.\begin{aligned}
\mathcal{S}^{zz}(\mathbf{q},\omega)&=\int^{+\infty}_{-\infty}\frac{dte^{\dot{\iota}\omega\tau}}{8\pi \mathcal{N}}\sum_{i,j,\mu,\mu'\nu,\nu'}e^{\dot{\iota}\mathbf{q}\cdot \mathbf{r}_{ij}}\sigma^z_{\alpha\alpha}\sigma^z_{\beta\beta}\\
&\times U^*_{i,\mu}U_{i,\mu'}U^*_{j,\nu}U_{j,\nu'}\\
&\times\sum_{\alpha,\beta}\langle e^{\dot{\iota}\hat{H}\tau}\hat{\xi}^\dagger_{\mu,\alpha}\hat{\xi}_{\mu',\alpha}e^{-\dot{\iota}\hat{H}\tau}\hat{\xi}^\dagger_{\nu,\beta}\hat{\xi}_{\nu',\beta}\rangle.\\
	\end{aligned}\right.
\end{equation}
The scattering mechanism is as follows. At time $\tau=0$, a pair of excitations is created by removing a fermion with a state $(\nu',\beta)$ from the filled bands (i.e., bellow the fermi level) and creating a fermion with a state $(\nu,\beta)$ at the empty bands (i.e., above the fermi energy) followed by the annihilation of the pair of excitations at time $\tau$. Thus, $\langle e^{\dot{\iota}\hat{H}\tau}\hat{\xi}^\dagger_{\mu,\alpha}\hat{\xi}_{\mu',\alpha}e^{-\dot{\iota}\hat{H}\tau}\hat{\xi}^\dagger_{\nu,\beta}\hat{\xi}_{\nu',\beta}\rangle$ gives
\begin{equation}\label{eq:dsf_u_2}
	\left.\begin{aligned}
&e^{-\dot{\iota}(\epsilon_{\nu}-\epsilon_{\nu'})\tau}\times\delta(\nu',\mu)\delta_{\mu',\nu}\delta_{\alpha,\beta}.\\
	\end{aligned}\right.  
\end{equation}
Note that $\epsilon_\nu$ is independent of spin index because of spin symmetry. Substitution of the Eq.~\eqref{eq:dsf_u_2} in Eq.~\eqref{eq:dsf_u_1}, yields
\begin{equation}\label{eq:dsf_u_3}
	\left.\begin{aligned}
\mathcal{S}^{zz}(\mathbf{q},\omega)&=\frac{1}{2\mathcal{N}}\sum_{i,j,\mu,\nu}e^{\dot{\iota}\mathbf{q}\cdot \mathbf{r}_{ij}}\delta(\omega-\epsilon_{\nu}+\epsilon_{\mu})\\
&\times U^*_{i,\mu}U_{i,\nu}U^*_{j,\nu}U_{j,\mu} n_{\mu}(1-n_{\nu}).\\
	\end{aligned}\right.
\end{equation}
Here, $n_{\gamma}=\frac{1}{e^{\beta(\epsilon_{\gamma}-\epsilon_F)}+1}$ with Fermi energy $\epsilon_F$. At absolute zero temperature, i.e., $\beta=\infty$, Eq.~\eqref{eq:dsf_u_3} can be written using a step function $\theta(x)$ as follows
\begin{equation}\label{eq:dsf_u_4}
	\left.\begin{aligned}
\mathcal{S}^{zz}(\mathbf{q},\omega)&=\frac{1}{2\mathcal{N}}\sum_{i,j,\mu,\nu}e^{\dot{\iota}\mathbf{q}\cdot \mathbf{r}_{ij}}\delta(\omega-\epsilon_{\nu}+\epsilon_{\mu})\\
&\times U^*_{i,\mu}U_{i,\nu}U^*_{j,\nu}U_{j,\mu}\theta(\epsilon_F-\epsilon_{\mu})\theta(\epsilon_{\nu}-\epsilon_F).\\
	\end{aligned}\right.
\end{equation}
Now, the equal-time momentum resolved spin-spin correlation function can be calculated from the above equation as $\mathcal{S}^{zz}_{eqt}(\mathbf{q})=\sum_{\omega}\mathcal{S}^{zz}(\mathbf{q},\omega)$. Thus,
\begin{equation}\label{eq:ssf}
	\left.\begin{aligned}
\mathcal{S}^{zz}_{\rm eqt}(\mathbf{q})=\frac{1}{2\mathcal{N}}\sum_{i,j,\mu,\nu}&e^{\dot{\iota}\mathbf{q}\cdot \mathbf{r}_{ij}}U^*_{i,\mu}U_{i,\nu}U^*_{j,\nu}U_{j,\mu}\\
&\times \theta(\epsilon_F-\epsilon_{\mu})\theta(\epsilon_{\nu}-\epsilon_F).\\
	\end{aligned}\right.
\end{equation}
\vspace{7mm}

%

\end{document}